%% file: main.tex
\documentclass[11pt, a4paper, logo, copyright]{googledeepmind}

\usepackage[utf8]{inputenc} 
\usepackage[T1]{fontenc}    

\usepackage[table]{xcolor}  
\usepackage{booktabs}       
\usepackage{multirow}
\usepackage{makecell}
\usepackage{tabularx}
\usepackage{longtable}
\usepackage{threeparttable}
\usepackage{colortbl}
\usepackage{tcolorbox}
\tcbuselibrary{breakable}
\usepackage{booktabs}
\usepackage{tabularx}
\usepackage{array}
\usepackage{xcolor}

\definecolor{tablegray}{RGB}{242,244,247}

\usepackage{array}
\usepackage{booktabs}
\usepackage{longtable}
\usepackage{pdflscape}
\usepackage{caption}
\usepackage{titletoc}
\usepackage{lipsum}

\tcbuselibrary{most}

\newcolumntype{L}[1]{%
    >{\raggedright\arraybackslash}p{#1}%
}

\usepackage{graphicx}
\usepackage{subcaption}
\usepackage{caption}
\usepackage{float}
\usepackage{adjustbox}
\usepackage{wrapfig}
\usepackage[inkscapelatex=false]{svg}

\usepackage{amsmath}
\usepackage{amssymb}
\usepackage{amsfonts}
\usepackage{nicefrac}
\usepackage{pifont}         

\usepackage{algorithm}
\usepackage{algpseudocode}

\usepackage{geometry}
\usepackage{microtype}      
\usepackage{url}            
\usepackage{hyperref}       
\usepackage{xspace}         
\usepackage{enumitem}
\usepackage{array}
\usepackage{natbib}
\usepackage[normalem]{ulem} 
\usepackage{soul}
\usepackage{lipsum}

\usepackage{tikz}
\usetikzlibrary{shapes, arrows.meta, positioning, patterns, shapes.geometric}

\definecolor{clsColor}{RGB}{252,146,114}      
\definecolor{aucColor}{RGB}{222,45,38}        

\newcommand{\CLSLabel}{%
    \tikz[baseline=(char.base)]\node[
        shape=rectangle,
        fill=clsColor,
        text=white,
        font=\small\sffamily\bfseries,
        rounded corners=3pt,
        inner sep=2.5pt
    ] (char) {CLS};\xspace
}

\newcommand{\AUCLabel}{%
    \tikz[baseline=(char.base)]\node[
        shape=rectangle,
        fill=aucColor,
        text=white,
        font=\small\sffamily\bfseries,
        rounded corners=3pt,
        inner sep=2.5pt
    ] (char) {AUC};\xspace
}

\definecolor{ForColor}{RGB}{8,81,156}       
\definecolor{pearsonColor}{RGB}{8,81,156}   
\definecolor{rmseColor}{RGB}{49,130,189}     
\definecolor{mseColor}{RGB}{107,174,214}    
\definecolor{maeColor}{RGB}{198,219,239}    
\definecolor{rsquaredColor}{RGB}{14,140,205} 

\newcommand{\ForLabel}{%
    \tikz[baseline=(char.base)]\node[
        shape=rectangle,
        fill=ForColor,
        text=white,
        font=\small\sffamily\bfseries,
        rounded corners=3pt,
        inner sep=2.5pt
    ] (char) {FOR};\xspace
}

\newcommand{\PearsonLabel}{%
    \tikz[baseline=(char.base)]\node[
        shape=rectangle,
        fill=pearsonColor,
        text=white,
        font=\small\sffamily\bfseries,
        rounded corners=3pt,
        inner sep=2.5pt
    ] (char) {$r$};\xspace
}

\newcommand{\RSquaredLabel}{%
    \tikz[baseline=(char.base)]\node[
        shape=rectangle,
        fill=rsquaredColor,
        text=white,
        font=\small\sffamily\bfseries,
        rounded corners=3pt,
        inner sep=2.5pt
    ] (char) {$r^2$};\xspace
}

\definecolor{demColor}{RGB}{110,110,110}
\definecolor{crdColor}{RGB}{220,38,38}
\definecolor{plmColor}{RGB}{37,99,235}
\definecolor{slpColor}{RGB}{124,58,237}
\definecolor{mtnColor}{RGB}{16,185,129}
\definecolor{sknColor}{RGB}{245,158,11}
\definecolor{varColor}{RGB}{14,116,144}
\definecolor{strColor}{RGB}{168,85,247}

\definecolor{stateColor}{RGB}{107,114,128}
\definecolor{actionColor}{RGB}{34,197,94}

\newcommand{\Badge}[2]{%
    \tikz[baseline=(char.base)]\node[
        shape=rectangle,
        fill=#1,
        text=white,
        font=\small\sffamily\bfseries,
        rounded corners=3pt,
        inner sep=2.5pt
    ] (char) {#2};\xspace
}

\newcommand{\DEMLabel}{\Badge{demColor}{DEM}}
\newcommand{\CRDLabel}{\Badge{crdColor}{CRD}}
\newcommand{\PLMLabel}{\Badge{plmColor}{PLM}}
\newcommand{\SLPLabel}{\Badge{slpColor}{SLP}}
\newcommand{\MTNLabel}{\Badge{mtnColor}{MTN}}

\newcommand{\VARLabel}{\Badge{varColor}{VAR}}

\newcommand{\StateLabel}{\Badge{stateColor}{State}}
\newcommand{\ActionLabel}{\Badge{actionColor}{Action}}

\definecolor{slfColor}{RGB}{103,121,144}   
\definecolor{labColor}{RGB}{201,171,85}    

\newcommand{\SourceLabel}[2]{%
    \tikz[baseline=(char.base)]\node[
        shape=rectangle,
        fill=#1,
        text=white,
        font=\small\sffamily\bfseries,
        rounded corners=3pt,
        inner sep=2.5pt
    ] (char) {#2};\xspace
}

\newcommand{\SLFLabel}{\SourceLabel{slfColor}{SLF}}
\newcommand{\LABLabel}{\SourceLabel{labColor}{LAB}}

\definecolor{lightpurple}{RGB}{245, 242, 255}

\colorlet{soulblue}{blue!20}

\newcommand{\cmark}{\textcolor{green!70!black}{\ding{51}}}
\newcommand{\xmark}{\textcolor{red!70!black}{\ding{55}}}

\definecolor{tablegreen}{RGB}{46, 125, 50}
\definecolor{tablered}{RGB}{198, 40, 40}
\definecolor{tableorange}{RGB}{249, 168, 37}
\definecolor{highlightgray}{RGB}{242, 242, 242}

\title{Personalized coaching via coachable health age and counterfactual wearable trajectories}

\usepackage{booktabs}   
\usepackage{amssymb}    
\usepackage{pifont}     
\usepackage{xcolor}     
\usepackage{colortbl}   
\usepackage{makecell}   

\definecolor{tablegreen}{RGB}{46, 125, 50}     
\definecolor{tablered}{RGB}{198, 40, 40}       
\definecolor{tableorange}{RGB}{249, 168, 37}   
\definecolor{highlightgray}{RGB}{242, 242, 242} 

\newcommand{\pmark}{\textcolor{tableorange}{\ensuremath{\circ}}} 

\reportnumber{}

\author[1,2,*,$\dagger$]{Simon A. Lee}
\author[1,*]{Zeinab Esmaeilpour}
\author[1]{Menglian Zhou}
\author[1]{Marius Guerard}
\author[1]{Sohrab Sidhu}
\author[1]{Alexandru Solot}
\author[1]{Hamid Palangi}
\author[1]{Pramod Rudrapatna}
\author[1]{Ming-Zher Poh}
\author[1]{Max A. Xu}
\author[1]{Daniel McDuff}
\author[1]{Dimitrios Antos}
\author[1]{Yuzhe Yang}
\author[2]{Jeffrey N. Chiang}
\author[1]{Anthony Z. Faranesh}
\author[1]{Lindsey Sunden}
\author[1]{Mark Malhotra}
\author[1]{Shwetak Patel}
\author[1,$\ddagger$]{A. Ali Heydari}
\author[1,$\ddagger$]{Ahmed A. Metwally}

\affil[1]{Google Research}
\affil[2]{University of California, Los Angeles}
\affil[*]{Co-first}
\affil[$\dagger$]{Work Completed During Research Internship} 
\affil[$\ddagger$]{Equal Leadership}

\correspondingauthor{Correspondence to: \{aliheydari, aametwally\}@google.com.}

\begin{abstract}
Personalized health coaching requires moving beyond retrospective wearable summaries toward quantitative, actionable behavioral prescriptions. Here, we introduce Coachable Health Age (CHA), an interpretable health-age metric constructed from modifiable physiological and behavioral signals passively captured by consumer wearables. Across the WEAR-ME ($N = 2{,}780$), \textit{All of Us} ($N = 22{,}748$), and longitudinal Fitbit ($N = 1{,}464$) cohorts, CHA aligns with established cardiometabolic health and risk scores and it improves prognostic discrimination for incident cardiovascular and metabolic disease beyond age, sex, and body mass index. To enable prospective behavioral optimization, we developed Sensor World Model (SensorWM), an action-conditioned wearable world model pretrained on over 2.2~million person-weeks of longitudinal sensor data from approximately 100{,}000 participants. SensorWM forecasts multi-week physiological trajectories conditioned on behavioral actions ($R^2$ = 0.664), outperforming a general-purpose time-series foundation model while preserving clinically relevant information for downstream prediction tasks, including diabetes and hypertension. Finally, we integrate CHA and SensorWM into an agentic counterfactual coaching system to synthesize personalized 12-week coaching plans. In blinded paired reviews, certified lifestyle coaches preferred counterfactual-grounded plans over plans grounded in wearable history alone for overall quality (71.4\%, \(P<0.001\)) and anticipated adherence (71.0\%, \(P<0.001\)). Board-certified clinicians endorsed safe exercise-intensity limits in 99.2\% of ratings of counterfactual-grounded plans and 97.6\% of ratings of comparison plans. Automated evaluations across the full cohort also favored counterfactual-grounded plans. These findings establish a framework integrating longitudinal wearable tracking, action-conditioned physiological forecasting, and personalized coaching, supporting prospective evaluation of its effects on behavioral adherence and cardiometabolic health.

\end{abstract}
\begin{document}

\maketitle

\section{Main}
\label{intro}
A substantial proportion of the global burden of chronic disease, particularly cardiometabolic disease, is attributable to modifiable behaviors such as physical inactivity, poor sleep, and unhealthy diet, and is therefore preventable \cite{lloyd2022life,Vaduganathan2022GBDCardiovascular}. Translating this preventive potential into sustained behavior change requires measures of health that are personalized, intuitive, and responsive to changes in daily behavior. Although chronological age is among the strongest predictors of mortality and chronic disease, it fails to capture the substantial heterogeneity in physiological function among individuals of the same age. Biological or health age measures have therefore been developed from molecular, clinical and wearable-derived features to quantify deviations from expected age-related physiology \cite{klemera2006new,levine2018epigenetic,miller2025wearable}. 
By expressing multidimensional physiological information on a familiar age scale, these measures can make health status more interpretable and personally salient \cite{spiegelhalter2016how}. Related constructs such as ``heart age'' and ``fitness age'' have been used to communicate risk in accessible terms\cite{bonner2021interventions} and in some trials such framing has improved risk perception and health behaviors, including smoking cessation \cite{Parkes2008Step2quit}, although effects have been inconsistent \cite{bonner2021interventions,Bakhit2024RiskCommunication}. Tracked longitudinally, health age could additionally serve as a feedback signal linking daily behaviors to longer-term changes in physiological state.

Most existing approaches derive health age by training machine-learning models to predict chronological age from multiple clinical biomarkers, or continuous wearable data \cite{horvath2013dna,pyrkov2018quantitative,miller2025wearable}, with a predicted age below chronological age, a negative age delta, interpreted as evidence of biological youthfulness or more favorable functional health. Second generation clocks trained on mortality or phenotypic outcomes improve risk predictions \cite{levine2018epigenetic, mccrory2021grimage}, but like age trained models, they are optimized to capture risk rather than modifiability. Chronological age prediction and health coaching are not completely aligned objectives. A model optimized to predict chronological age or mortality risk may rely on physiological characteristics that are strongly age-associated but fixed, difficult to modify, or only weakly informative about behaviors an individual can act upon. They are also typically opaque; post hoc attribution methods can indicate which inputs influenced an estimate, but not how that estimate would change if a person altered their behavior\cite{ghassemi2021false}. A health-age measure intended for coaching should therefore be built from interpretable, repeatedly measurable characteristics that can serve as practical targets for behavioral feedback \cite{michie2009effective,michie2013behavior,bandura2004health}.

A coachable health-age measure describes an individual's current state. Prospective coaching additionally requires understanding how that state may evolve by comparing health trajectories under alternative behavioral scenarios \cite{hood2012personal}. 
The density and continuity of consumer-wearable signals make them well suited to this task. Recurrent and transformer models have been used to model physiological states and incident risk from longitudinal health records and wearable data \cite{rajkomar2018scalable, miller2025wearable}. This work has culminated in large-scale wearable ``foundation models'' trained using self-supervised objectives across thousands to millions of participants, which learn general-purpose physiological representations \cite{abbaspourazadlarge, narayanswamy2026towards, lee2026himae, zhou2026physiology, cao2026wavesfm, jing2026one, li2026glucofm} and support imputation and interpolation of sensor values over time \cite{narayanswamy2024scaling}. In parallel, the digital-twin paradigm proposes individualized virtual replicas that simulate personal health trajectories to enable anticipatory care \cite{katsoulakis2024digital, lee2026forecasting}. These approaches are predominantly observational. They estimate the future associated with observed or persisting behavioral patterns. Related longitudinal outcome-modeling methods have considered predictions under alternative treatment sequences \cite{lim2018forecasting,bica2020estimating,melnychuk2022causal, urbut2026bayesian}, while action-conditioned world models provide a general computational framework for representing how predicted future states depend on specified actions \cite{ha2018world, lecun2022path}. For wearable health coaching, however, the relevant actions are not primarily discrete clinical treatments, but continuously varying behaviors such as physical activity, sleep, and cardiovascular exercise. This motivates an action-conditioned wearable sensor model that can generate individualized physiological projections under feasible behavioral scenarios and summarize those projections using a common, coachable outcome.

Translating such projections into guidance that people can follow is a further challenge. Large language models (LLMs) can deliver personalized, conversational health coaching and have shown expert-level performance in interpreting wearable sleep and fitness data \cite{ merrill2026transforming, khasentino2025personal, heydari2025anatomy}. However, current LLM-based health coaches are typically grounded in retrospective summaries of an individual's data rather than in quantitative projections of how their physiology would respond to specific behavioral changes, yielding recommendations that rely on qualitative population heuristics rather than calibrated, individualized dose-response estimates\cite{khasentino2025personal,merrill2026transforming}. Moreover, LLMs can generate plausible but unsupported claims \cite{thirunavukarasu2023large}, and despite recent benchmarks in wearable-based coaching \cite{khasentino2025personal,heydari2025anatomy}, few systems have been rigorously evaluated by both lifestyle-coaching and clinical experts \cite{bedi2025testing}. Coupling LLM coaching with an explicit model of physiological response and an interpretable outcome measure could make recommendations both personalized and directly traceable to their projected health benefits.

Here, we introduce \textbf{Coachable Health Age (CHA)}, an interpretable health-age measure built from physiological and behavioral signals that are passively and repeatedly measured by consumer wearables, including resting heart rate, sleep duration and physical activity, and that reflect modifiable aspects of everyday health. We embed CHA within an architecture that couples an action-conditioned wearable \textbf{Sensor World Model (SensorWM)} with an agentic coaching system. The world model learns predictive physiological dynamics from longitudinal wearable trajectories and generates model-implied future states conditioned on alternative patterns of daily activity, sleep, and structured exercise. The coaching agent integrates these counterfactual projections with a feasibility-constrained optimization procedure to produce personalized behavioral plans grounded in projected changes in CHA. We evaluate each component of the framework, assessing the clinical, prognostic, and longitudinal validity of CHA, the forecasting fidelity of SensorWM in driving the coaching simulation engine, the behavior of its scenario-conditioned projections, and the quality of coaching plans generated with and without simulation-derived information. In blinded evaluations, lifestyle coaches consistently preferred plans informed by CHA and SensorWM projections over plans grounded in baseline wearable data alone for overall plan quality and anticipated long-term adherence, and clinicians rated both sets of plans similarly on exercise-intensity safety and physiological accuracy.

\section{Results}

\subsection{Study design and dataset characteristics}

\begin{figure}[t!]
    \centering
 \includegraphics[width=\linewidth, height=0.8\textheight, keepaspectratio]{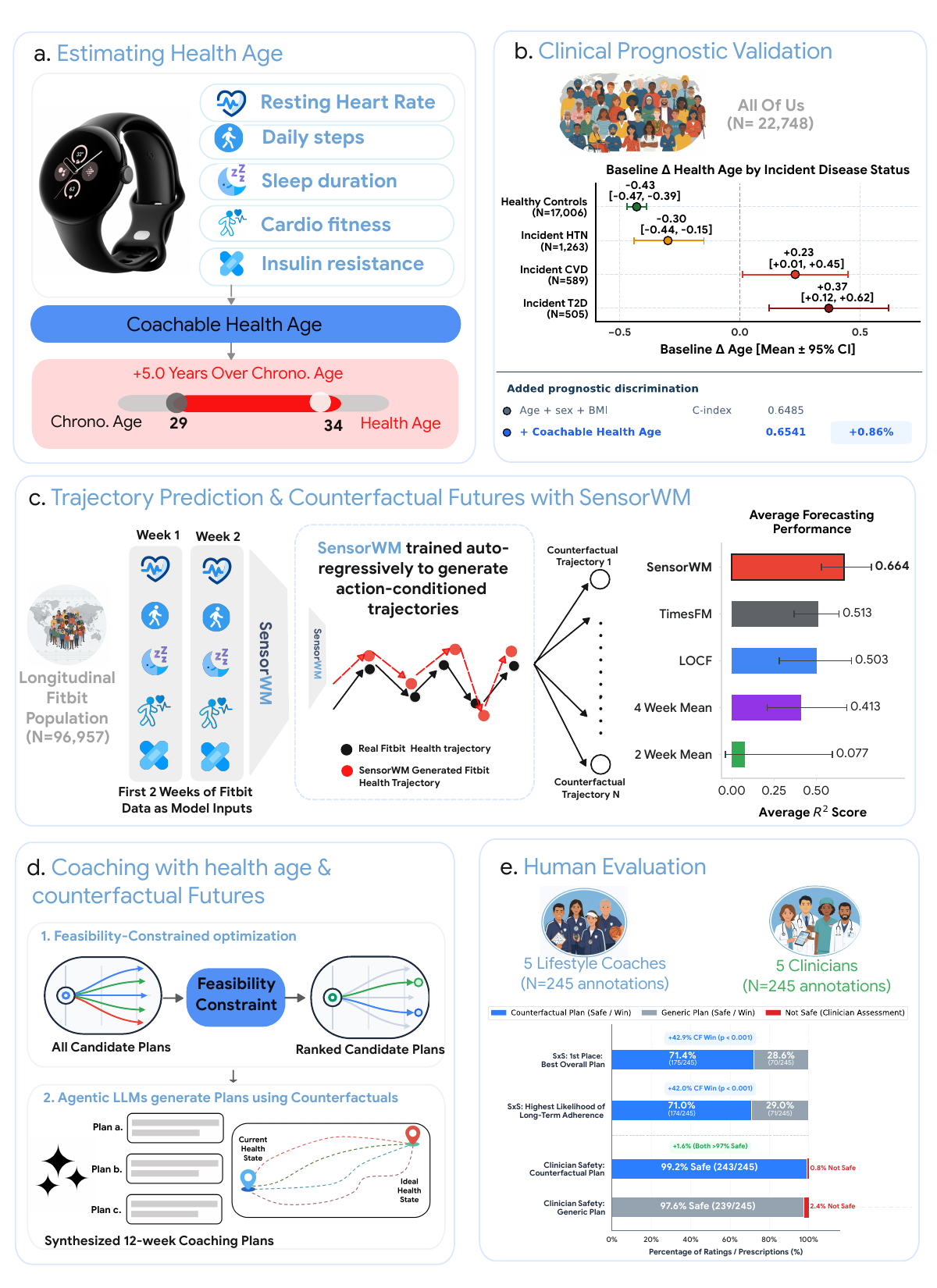}
    \caption{\small\textbf{Coachable Health Age (CHA) framework for personalized health optimization.} \textbf{(a)} Wearable sensor physiological and behavioral metrics are integrated to estimate CHA relative to chronological age. \textbf{(b)} Clinical prognosis across incident cardiometabolic outcomes in All of Us. \textbf{(c)} SensorWM auto-regressively simulates action-conditioned counterfactual lifestyle trajectories over a 12-week horizon. \textbf{(d)} Feasibility-constrained optimization balances forecasted Health Age reduction against historical effort penalties and clinical safety gates to synthesize a personalized 12-week coaching plan. \textbf{(e)} Dual-track human evaluation by lifestyle coaches (10 behavioral rubrics and side-by-side preference) and clinicians (clinical safety gate and 6 clinical rubrics).}
    \label{fig:main1}
\end{figure}

We designed the study as three linked analyses evaluating the core components of our end-to-end CHA and counterfactual coaching framework (Figure~\ref{fig:main1}; cohort demographics and baseline wearable characteristics in Table~\ref{tab:cohort_characteristics_aou}). First, we constructed and evaluated CHA, an age-equivalent summary of wearable-derived physiological and behavioral health built from repeatedly measurable, modifiable signals (Figure~\ref{fig:main1}a). We assessed its alignment against established cardiovascular and metabolic health and risk scores (modified Life's Essential 8 and Metabolic Syndrome Severity) and self-reported health in the WEAR-ME cohort, and characterized its within-person longitudinal modifiability across repeated 3-month windows over two years in the longitudinal Fitbit cohort. Independent prospective clinical validation was then performed in the \textit{All of Us} Research Program by relating baseline and longitudinal $\Delta\text{CHA}$ ($\Delta\text{CHA} = \text{CHA} - \text{chronological age}$) to incident cardiometabolic disease, specifically hypertension, type 2 diabetes, and cardiovascular disease as well as incremental prognostic discrimination beyond age, sex, and body mass index (Figure~\ref{fig:main1}b).

Second, we developed SensorWM, an action-conditioned latent world model for
longitudinal wearable data (Figure~\ref{fig:main1}c). SensorWM was pretrained on real world
longitudinal wearable trajectories from 96,957 participants (approximately 100,000), comprising 2.28 million total person-weeks of wearable data (182,797 valid 12-week windows; 42,039 person-years). The model was
trained to represent physiological state in a latent space and to predict its
future evolution conditional on both recent physiological trajectory and
observed behavioral actions. Model fidelity was evaluated by comparing
generated and observed physiological trajectories, feature-level forecasting
accuracy, and preservation of clinically relevant information on downstream
health prediction tasks in a held-out WEAR-ME cohort.

Third, we evaluated whether grounding an LLM coaching agent in CHA and SensorWM counterfactual projections improved personalized 12-week behavioral prescriptions (Figure~\ref{fig:main1}d,e). For each eligible WEAR-ME participant, SensorWM was initialized from recent wearable history to simulate future physiological trajectories under candidate modifications in daily steps, sleep duration, and cardiorespiratory exercise. Candidate trajectories were mapped to projected changes in CHA and ranked using a feasibility-constrained optimization objective that balances projected CHA reduction against participant-specific behavioral effort and safety bounds (Figure~\ref{fig:main1}d). We then compared two matched LLM coaching conditions that shared the same underlying language model, participant wearable history, and plan structure, differing only in access to simulation-derived information: (1) \textbf{Counterfactual Coaching}, which received the participant's baseline CHA, projected status-quo trajectory, and feasibility-ranked counterfactual pathways, and (2) \textbf{Generic Coaching}, which generated recommendations from observed baseline wearable history alone. Plan quality, clinical safety, physiological plausibility, and anticipated long-term adherence were evaluated through a blinded, randomized dual-track human expert study with five certified lifestyle coaches and five board-certified clinicians across stratified WEAR-ME personas, complemented by a full-cohort automated evaluation across the WEAR-ME dataset using rubric-aligned LLM autoraters (Figure~\ref{fig:main1}e). All plan generation and expert evaluations were conducted offline using retrospective participant profiles prior to prospective deployment.

\subsection{Coachable Health Age Framework}
To serve as both a continuous summary of everyday health and a quantitative target for prospective behavioral coaching, a health-age measure must satisfy seven core design criteria (Supplementary Table~\ref{tab:modeling_comparison_5row}): it must be (1) \textit{predictable from consumer wearables} via passive, high-frequency sensing; (2) \textit{explainable} at the level of individual inputs; (3) \textit{flexible and extensible} to additional features; (4) \textit{engageable} for longitudinal user communication; (5) \textit{coachable}, mapping directly to modifiable daily habits; (6) \textit{movable} within individuals over coaching relevant timescales; and (7) \textit{grounded} in clinical outcomes. Existing age formulations satisfy only subsets of these requirements. Chronological age provides an intuitive reference scale but cannot capture physiological heterogeneity among peers of the same age. Machine learning clocks trained to predict chronological age from wearable streams (such as foundation-model embeddings) capture age correlated sensor variation, but their training objective does not prioritize modifiable behaviors or present cardiometabolic health, and their black-box representations limit actionable attribution. Conversely, clinical biological clocks such as PhenoAge are grounded in mortality-associated blood biomarkers, but rely on invasive, episodic laboratory draws that cannot be tracked continuously by consumer wearables. Finally, directly prompting large language models (LLMs) to estimate a numerical health age from wearable summaries introduces stochastic variation across inference settings and lacks deterministic epidemiological calibration (evaluated separately in Supplementary Information~\ref{llm}: LLM Health Age \& Consistency).

We therefore formulated Coachable Health Age (CHA) by mapping repeatedly measurable, modifiable wearable signals, core behaviors and physiology comprising resting heart rate, daily steps, sleep duration, and weekly Active Zone Minutes (AZM)---onto an intuitive age-equivalent scale using collinearity-adjusted epidemiological hazard ratios (Figure~\ref{fig:main2}a; Methods). Two architectural choices distinguish this framework. First, rather than parameterizing risk solely against late-life all-cause mortality, CHA anchors its core wearable domains to published epidemiological associations with cardiometabolic outcomes, including type 2 diabetes, heart failure, and cardiovascular disease, emphasizing risk factors that are measurable through consumer wearables and potentially responsive to behavioral modification. Second, because CHA aggregates risk additively in log-hazard space rather than through a monolithic neural network, the framework is inherently modular and expandable: optional wearable or clinical signals such as cardiorespiratory fitness ($\mathrm{VO}_2\max$), sleep midpoint regularity, and wearable-estimated insulin resistance, or emerging sensing modalities can be seamlessly incorporated when available without retraining the underlying formulation. Together, these design characteristics motivate CHA as an interpretable, cardiometabolic grounded outcome metric (Supplementary Table~\ref{tab:modeling_comparison_5row}) and reflect design considerations rather than an empirical ranking of alternative age measures.

\begin{table*}[t!]
\centering
\caption{
\textbf{Demographic, longitudinal wearable, and clinical characteristics of the study cohorts.}
Categorical variables are reported as $n$ (\%) and continuous variables as
mean $\pm$ standard deviation, unless otherwise specified.
The disease-free baseline cohort excludes participants with a history of
hypertension, type 2 diabetes, or cardiovascular disease on or before the
end of their 90-day baseline wearable window.
}
\label{tab:cohort_characteristics_aou}

\scriptsize
\renewcommand{\arraystretch}{1.18}
\setlength{\tabcolsep}{3.5pt}

\begin{adjustbox}{max width=\textwidth,center}
\begin{tabular}{@{}lccccc@{}}
\toprule[1.2pt]
& \multicolumn{3}{c}{\textbf{CHA AoU Analysis}}
& \multicolumn{2}{c}{\textbf{SensorWM / Expert Evaluation}} \\
\cmidrule(lr){2-4}
\cmidrule(l){5-6}

\textbf{Characteristic}
& \textbf{Total Wearable}
& \textbf{Disease-Free}
& \textbf{Cox / C-Index}
& \textbf{Pre-training}
& \textbf{WEAR-ME} \\

& \textbf{Cohort}
& \textbf{Baseline (KM)}
& \textbf{Cohort}
& \textbf{Cohort}
& \textbf{Evaluation} \\
\midrule

\rowcolor{gray!12}
\multicolumn{6}{@{}l}{\textbf{Cohort Size and Wearable Observation}} \\

\quad Participants, $N$
& 22,748 & 18,423 & 14,663 & 96,957 & 2,780 \\





\quad Total valid 3-month wearable windows, $n$
& 276,378 & 227,557 & 188,033 & 182,797 & 3,289 \\

\quad Valid 3-month windows per participant
& $12.1 \pm 8.3$ & $12.4 \pm 8.4$ & $12.8 \pm 8.5$ & $1.9 \pm 1.5$ & $1.2 \pm 0.4$ \\

\quad Active wearable tracking, person-years
& 68,101 & 56,071 & 46,332 & 42,039 & 756 \\

\quad Follow-up span, years
& $3.59 \pm 2.28$ & $3.65 \pm 2.30$ & $3.80 \pm 2.31$ & $0.43 \pm 0.35$ & $0.38 \pm 0.08$ \\

\addlinespace[3pt]
\rowcolor{gray!12}
\multicolumn{6}{@{}l}{\textbf{Baseline Demographics}} \\

\quad Age, years
& $50.2 \pm 15.5$ & $48.2 \pm 15.4$ & $48.7 \pm 15.5$ & $40.7 \pm 16.6$ & $48.5 \pm 12.0$ \\

\quad Age range, years
& 13--85 & 13--84 & 13--84 & 18–100 & 23--100 \\




\addlinespace[2pt]
\quad Sex, $n$ (\%) & & & & & \\

\quad\quad Female
& 16,031 (70.5) & 13,164 (71.5) & 10,446 (71.2)
& 47,651 (49.2) & 1,652 (59.4) \\

\quad\quad Male
& 6,209 (27.3) & 4,814 (26.1) & 3,872 (26.4)
& 49,294 (50.8) & 1,123 (40.4) \\

\quad\quad Other / Unknown / Not reported
& 508 (2.2) & 445 (2.4) & 345 (2.4)
& 12 (0.0) & 5 (0.2) \\

\addlinespace[3pt]
\rowcolor{gray!12}
\multicolumn{6}{@{}l}{\textbf{Body Mass Index (BMI)}} \\

\quad BMI, $\mathrm{kg/m^2}$
& $29.61 \pm 7.18^{a}$ & $28.95 \pm 6.88^{b}$
& $28.95 \pm 6.88$ & $26.21 \pm 6.46$ & $29.52 \pm 6.85$ \\





\addlinespace[3pt]
\rowcolor{gray!12}
\multicolumn{6}{@{}l}{\textbf{Baseline Wearable and Physiological Characteristics}} \\

\quad Daily steps, steps/day
& $7{,}846 \pm 3{,}574$ & $8{,}040 \pm 3{,}560$
& $8{,}131 \pm 3{,}511$ & $5,604 \pm 3,601$ & $7,405 \pm 3,519$ \\

\quad Resting heart rate proxy, bpm
& $55.2 \pm 7.3$ & $55.1 \pm 7.2$
& $55.2 \pm 7.1$ & $69.1 \pm 8.8$ & $68.2 \pm 8.6$ \\

\quad Main sleep duration, hours/night
& $6.36 \pm 1.20$ & $6.41 \pm 1.17$
& $6.43 \pm 1.16$ & $6.75 \pm 1.01$ & $7.12 \pm 1.15$ \\

\quad Active Zone Minutes, minutes/week
& $373.7 \pm 374.5$ & $381.3 \pm 375.7$
& $388.1 \pm 377.2$ & $302.4 \pm 303.8$ & $198.4 \pm 142.6$ \\

\addlinespace[3pt]
\rowcolor{gray!12}
\multicolumn{6}{@{}l}{\textbf{Coachable Health Age}} \\

\quad Coachable Health Age, years
& $49.97 \pm 15.84$ & $47.77 \pm 15.64$
& $48.18 \pm 15.69$ & $44.10 \pm 17.13$ & $48.46 \pm 12.02$ \\

\quad Health Age Delta, $\Delta$ years
& $-0.26 \pm 2.67$ & $-0.42 \pm 2.60$
& $-0.51 \pm 2.56$ & $+3.40 \pm 3.80$ & $-0.04 \pm 3.00$ \\

\addlinespace[3pt]
\rowcolor{gray!12}
\multicolumn{6}{@{}l}{\textbf{Cardiometabolic Disease Status}} \\

\quad Pre-existing cardiometabolic condition, $n$ (\%)
& 4,325 (19.0) & 0 (0.0) & 0 (0.0) & --- & --- \\

\quad Incident composite diagnosis, $n$ (\%)$^{c}$
& 1,417 (6.2) & 1,417 (7.7) & 1,402 (9.5) & --- & --- \\

\quad\quad Incident hypertension
& 1,263 (5.6) & 1,138 (6.2) & 1,125 (7.7) & --- & --- \\

\quad\quad Incident type 2 diabetes
& 505 (2.2) & 315 (1.7) & 309 (2.1) & --- & --- \\

\quad\quad Incident cardiovascular disease
& 589 (2.6) & 309 (1.7) & 304 (2.1) & --- & --- \\

\bottomrule[1.2pt]
\end{tabular}
\end{adjustbox}

\vspace{3pt}
\begin{minipage}{\textwidth}
\raggedright
\scriptsize
\textit{Notes.}
Values are $n$ (\%) or mean $\pm$ standard deviation unless otherwise
specified. Percentages are calculated using the corresponding cohort
size and may not sum to 100\% because of rounding.
An em dash (---) indicates that a characteristic was not available for the corresponding cohort.
Observation duration in the pre-training and WEAR-ME evaluation
cohorts is measured in weeks, whereas longitudinal clinical follow-up
is summarized using valid 3-month wearable windows and years. In the \textit{All of Us} cohort, all participants were $\ge 18$ years of age at study enrollment; baseline age reflects age at the first valid 90-day wearable window, which includes retrospective Bring Your Own Device (BYOD) Fitbit records collected from age 13 onward.
The Cox / C-index cohort is restricted to disease-free participants
with available EHR-derived BMI.
$^{a}$ BMI available for $N=18{,}757$ participants in the total
wearable cohort.
$^{b}$ BMI available for $N=14{,}663$ participants in the
disease-free cohort.
$^{c}$ Individual incident disease counts are not mutually
exclusive, as participants may develop more than one condition during follow-up; in the total wearable cohort, condition-specific incident counts exclude only baseline history of that respective condition, whereas in the disease-free and Cox cohorts all participants are free of all three conditions at baseline. In the pre-training cohort, 2.28 million total person-weeks of observations yielded 182,797 complete 12-week (84-day) windows (2.19 million windowed person-weeks; 42,039 person-years).
\end{minipage}

\end{table*}

\subsection{Cardiometabolic prognosis and longitudinal modifiability of CHA}

Because CHA incorporates modifiable lifestyle and physiological domains emphasized in preventive cardiometabolic disease, we first verified its alignment against established clinical health and risk scores namely modified Life's Essential 8 (mod-LE8) and Metabolic Syndrome Severity Score (MSSS) as well as independent self-reported health status in the WEAR-ME cohort (Supplementary Figure~\ref{fig:le8}). Whereas wearable models trained to predict chronological age showed weak concordance with current cardiometabolic health (mod-LE8 $|r|=0.141$; MSSS $|r|\le 0.048$) and direct LLM-based health-age estimates were poorly calibrated ($|r|=0.110\text{--}0.214$), CHA preserved strong alignment with both mod-LE8 ($|r|=0.512$) and MSSS ($|r|=0.355$), exceeding blood-based PhenoAge ($|r|=0.406$ and $|r|=0.319$, respectively) and varying monotonically across self-reported health categories from excellent to poor. Thus, unlike wearable clocks, CHA directly reflects an individual's current cardiometabolic and perceived health.
 
\begin{figure*}[t!]
    \centering
    \includegraphics[width=\linewidth]{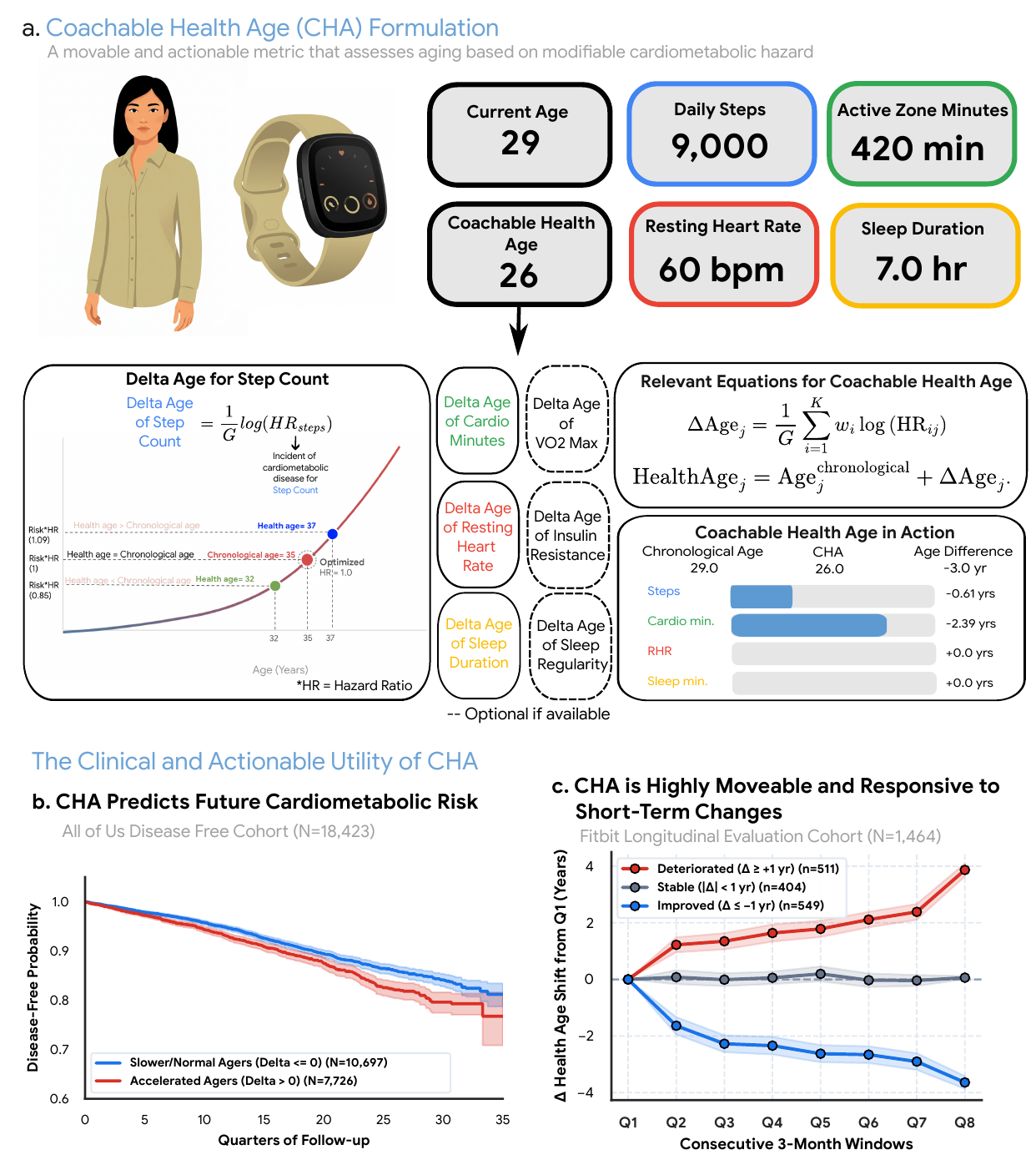}
    \caption{
\small\textbf{Formulation, cardiometabolic prognosis, and short-term modifiability of CHA.}
\textbf{(a)} Overview of the CHA formulation, mapping core wearable signals (resting heart rate, sleep duration, cardiorespiratory minutes, and step count) and optional metabolic or fitness markers (insulin resistance, sleep regularity, and $\text{VO}_2\text{ max}$) onto a Gompertz cardiometabolic hazard curve (\textit{left}), alongside the governing equations and an illustrative four-component calculation (\textit{right}). The hypothetical example assumes a chronological age of 29 years, 9,000 daily steps, 420 weekly Active Zone Minutes, a resting heart rate of 60 beats per minute, and 7 hours of nightly sleep. Optional components are omitted.
\textbf{(b)} Kaplan--Meier estimates of remaining free of incident cardiometabolic disease in the \textit{All of Us} cohort ($N = 18{,}423$), stratified by baseline Health Age Delta ($\Delta\text{CHA} \le 0$, $n = 10{,}697$ versus $\Delta\text{CHA} > 0$, $n = 7{,}726$).
\textbf{(c)} Mean $\Delta\text{CHA}$ shift from baseline (Q1) across eight consecutive 3-month windows (2 years) among deteriorated ($\Delta \ge +1\text{ yr}$, $n = 511$), stable ($|\Delta| < 1\text{ yr}$, $n = 404$), and improved ($\Delta \le -1\text{ yr}$, $n = 549$) trajectory groups ($N = 1{,}464$). Shaded bands indicate 95\% confidence intervals.
    }
    \label{fig:main2}
\end{figure*}

To independently validate the prognostic utility of CHA, we evaluated longitudinal incident cardiometabolic disease in the \textit{All of Us} Research Program wearable cohort ($N=22{,}748$, of whom $N=18{,}423$ were free of all three conditions at baseline and $N=4{,}325$ had pre-existing baseline disease; Figure~\ref{fig:main1}b). Compared with healthy controls who remained disease-free throughout follow-up ($N=17{,}006$; mean $\Delta\text{CHA} = -0.43$ years, $95\%\text{ CI}: [-0.47, -0.39]$), baseline $\Delta\text{CHA}$ was higher among participants who subsequently developed incident hypertension ($N=1{,}263$; $-0.30$ years, $[-0.44, -0.15]$), cardiovascular disease ($N=589$; $+0.23$ years, $[+0.01, +0.45]$), or type 2 diabetes ($N=505$; $+0.37$ years, $[+0.12, +0.62]$; Figure~\ref{fig:main1}b).
Over up to 36 quarters of follow-up, Kaplan--Meier analyses confirmed that baseline accelerated agers ($\Delta\text{CHA} > 0$, $n=7{,}726$) experienced significantly lower cardiometabolic disease-free survival than slower or normal agers ($\Delta\text{CHA} \le 0$, $n=10{,}697$; 8-year survival $79.28\%$ versus $82.41\%$, log-rank $P = 7.09 \times 10^{-6}$; Figure~\ref{fig:main2}b and Extended Data Table~\ref{tab:km_quarters}). In multivariable Cox proportional hazards models adjusted for baseline chronological age, sex, and BMI, each 1-year increase in $\Delta\text{CHA}$ remained independently associated with higher incident cardiometabolic risk both at baseline ($\mathrm{HR} = 1.02$, $95\%\text{ CI}: [1.01, 1.04]$, $P = 2.8 \times 10^{-4}$; $N=14{,}663$, $1{,}402$ events) and in lagged time-varying models where $\Delta\text{CHA}_{t-1}$ predicted incident diagnosis in the subsequent quarter ($\mathrm{HR} = 1.04$, $95\%\text{ CI}: [1.02, 1.05]$, $P = 6.0 \times 10^{-7}$; $N=13{,}842$, $164{,}336$ person-quarters; Extended Data Table~\ref{tab:cox_models}). Crucially, adding CHA to the clinical baseline of age, sex, and BMI significantly improved prognostic discrimination across the full cohort (Harrell's C-index $0.6541$ versus $0.6485$, $\Delta\text{C} = +0.0056$, $P=0.005$), in 5-fold cross-validation ($\Delta\text{C} = +0.0053$, $P=0.016$), and most pronouncedly in adults under 50 years of age ($N=7{,}399$; C-index $0.6762$ versus $0.6648$, $\Delta\text{C} = +0.0114$, $P < 0.001$; Extended Data Table~\ref{tab:c_index_comparison} and Supplementary Figure~\ref{fig:c-index}).


\subsection{Sensor world models learn physiological health trajectories}

\begin{figure*}[t!]
    \centering
    \includegraphics[width=0.93\linewidth]{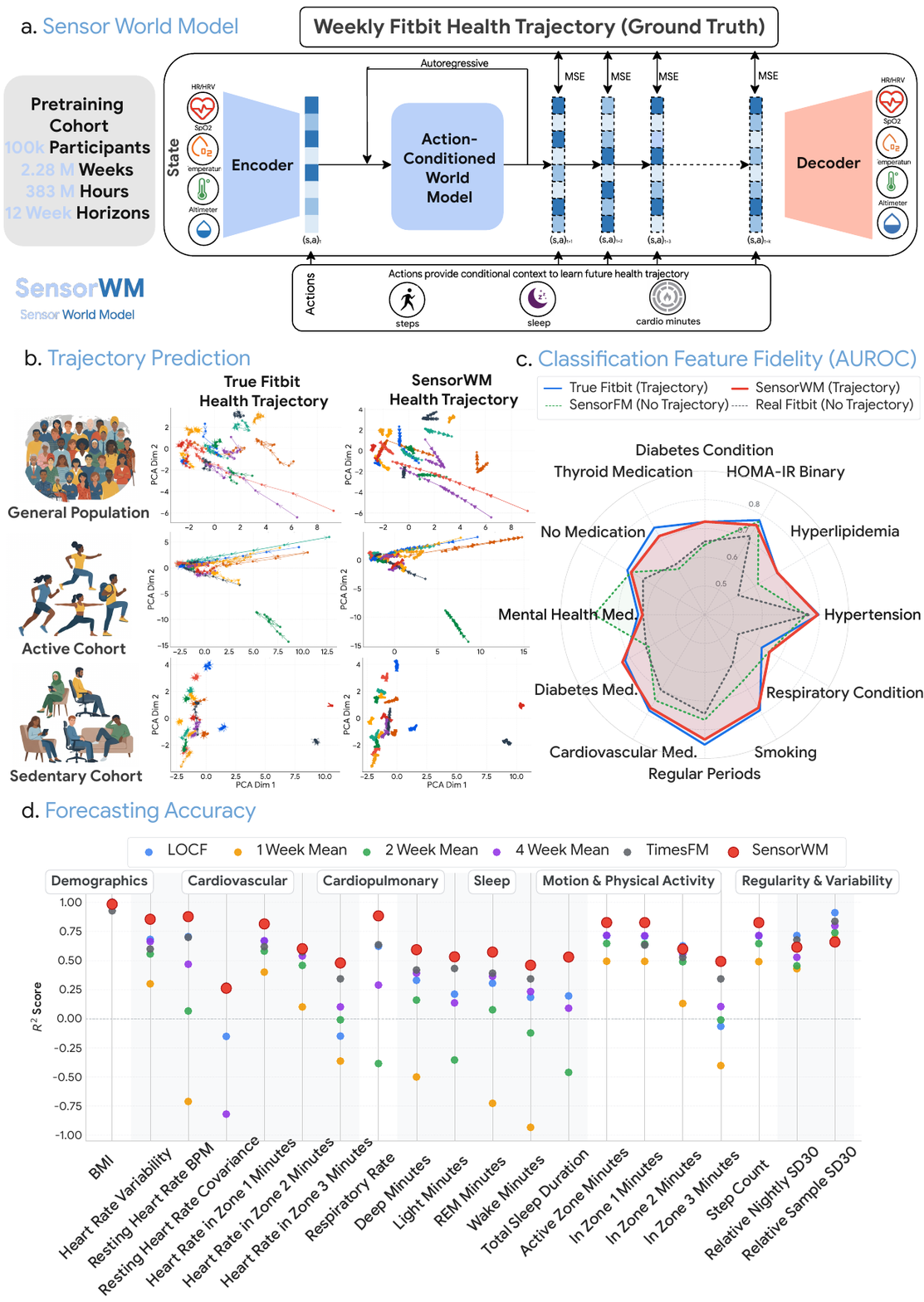}
    \caption{\small\textbf{Workflow and Validation of the Sensor World Model.} \textbf{(a)} Sensor world model architecture for action-conditioned prediction of future health trajectories from wearable sensor data. \textbf{(b)} PCA projections of ground-truth (left) and SensorWM-predicted (right) health trajectories across general, active, and sedentary cohorts. \textbf{(c)} Downstream classification performance (AUROC) comparing real and SensorWM-generated features, with and without trajectory information. \textbf{(d)} Forecasting accuracy ($R^2$) across 20 wearable health features, comparing SensorWM against baseline models.}
    \label{fig:main3}
\end{figure*}

Estimating an actionable Health Age requires more than measuring an individual’s present physiological state. It requires a model of how that state evolves over time and how predicted trajectories vary with observed behavioral patterns. We therefore developed SensorWM, an action-conditioned sensor world model trained to reconstruct and forecast longitudinal wearable measurements (Figure \ref{fig:main3}a). SensorWM was pretrained on trajectories from 96,957 research-consented participants, comprising 2.28 million total person-weeks. Each participant was represented as a weekly sequence of physiological states and observable actions. State inputs included cardiovascular (e.g., Heart Rate Variability), cardiopulmonary (e.g., Respiratory Rate), sleep (e.g., Deep sleep minutes), and variability-related (e.g. dispersion) measurements (e.g., Relative Sample SD30), whereas action inputs captured modifiable behaviors, including step count, sleep duration and time spent in cardio-intensity activity (Full Decomposition Breakdowns of Feature Splits in Supplementary Table \ref{tab:state_action_features}). 

SensorWM consists of a physiological state encoder and an action-conditioned transition model. A multilayer perceptron first encodes physiological measurements into a 128-dimensional latent representation. The transition model then predicts how this representation changes over time based on the current physiological state, recent changes in that state, and observed behavioral actions. Specifically, it learns a linear transition operator that updates the latent representation using its recent trajectory. Both components are trained jointly using a self-supervised joint-embedding predictive architecture (JEPA), closely following the training objective of LeWorldModel which models trajectory fidelity and a regularization term \cite{maes2026leworldmodel}.

This state–action formulation enabled SensorWM to predict future physiological states based on both observed behaviors and recent physiological changes, allowing the model to capture not only an individual's current health status but also the direction and trajectory of their physiological evolution. The resulting architecture supports autoregressive forecasting and the simulation of alternative behavioral scenarios over extended horizons, providing the physiological projections used in downstream counterfactual coaching. Additional details on model architecture, training objectives, and recursive inference are provided in Methods.

We next tested whether generated trajectories retained clinically relevant information. This evaluation is more stringent than minimizing point-wise reconstruction error; a forecast can be numerically close to the observed sequence while erasing weak but medically informative covariation across its learned latent representation. We therefore trained matched downstream classifiers using two settings: (1) observed wearable trajectories, (2) SensorWM-generated trajectories. We then evaluated the two settings across health conditions, laboratory-derived metabolic phenotypes, medication use, behavioral activity and women's health prediction tasks. Across most downstream prediction tasks, SensorWM-generated trajectories outperformed the pretrained Sensor Foundation Model (SensorFM) and Fitbit features without longitudinal trajectory information (Figure \ref{fig:main3}c). For example, generated trajectories yielded an AUROC of 0.724 for diabetes compared with 0.643 using SensorFM ($p =1.85 \times10^{-5}$), 0.692 versus 0.614 for hyperlipidemia ($p =9.64 \times10^{-10}$), and 0.793 versus 0.753 for hypertension ($p =5.85 \times10^{-4}$). We next evaluated whether SensorWM preserved the clinical information contained in the true longitudinal Fitbit trajectories. Across these tasks, classification performance from generated trajectories closely followed that obtained from the observed data (Figure \ref{fig:main3}c), with most comparisons showing no statistically significant differences. For example, generated trajectories yielded an AUROC of 0.724 for diabetes compared with 0.723 using Fitbit observed trajectories ($p$=0.7037), 0.692 versus 0.691 for hyperlipidemia, and 0.793 versus 0.796 for hypertension ($p$=0.7784). Performance was likewise preserved for HOMA-IR-defined insulin resistance, with an AUROC of 0.760 from generated trajectories compared with 0.780 from observed data ($p$=0.2782). These results suggest that SensorWM captures clinically informative longitudinal dynamics while preserving much of the predictive information present in observed wearable trajectories.


We then quantified forecasting accuracy across wearable domains and compared SensorWM with persistence and temporal baselines, including the mean of the preceding one, two, or four weeks, last observation carried forward (LOCF), and the time-series foundation model TimesFM \cite{das2024decoderonlyfoundationmodeltimeseries}. SensorWM was initialized using the first two observed physiological states and subsequently propagated autoregressively, with each physiological state transition conditioned on the corresponding behavioral action vector.
The largest improvements occurred for measurements with substantial temporal variation and cross-modal dependence. SensorWM attained \(R^2=0.877\) for resting heart rate, compared with 0.705 for last observation carried forward and 0.686 for TimesFM. For heart-rate variability, SensorWM reached \(R^2=0.855\), compared with 0.682 and 0.605, respectively. Respiratory rate was forecast with \(R^2=0.884\), substantially exceeding last observation carried forward \((R^2=0.623)\) and TimesFM \((R^2=0.503)\). SensorWM also performed strongly for time spent in lower-intensity heart-rate zones, achieving \(R^2=0.814\) for heart-rate zone 1 minutes and \(R^2=0.826\) for activity zone 1 minutes. These features reflect both behavior and physiological response; predicting them therefore requires the model to integrate activity patterns with individual cardiovascular dynamics rather than rely solely on univariate persistence.

\subsection{SensorWM enables individualized counterfactual simulation and feasibility-constrained coaching}
\label{sec:counterfactual_simulation}

\begin{figure}[t!]
    \centering
    \includegraphics[width=\linewidth]{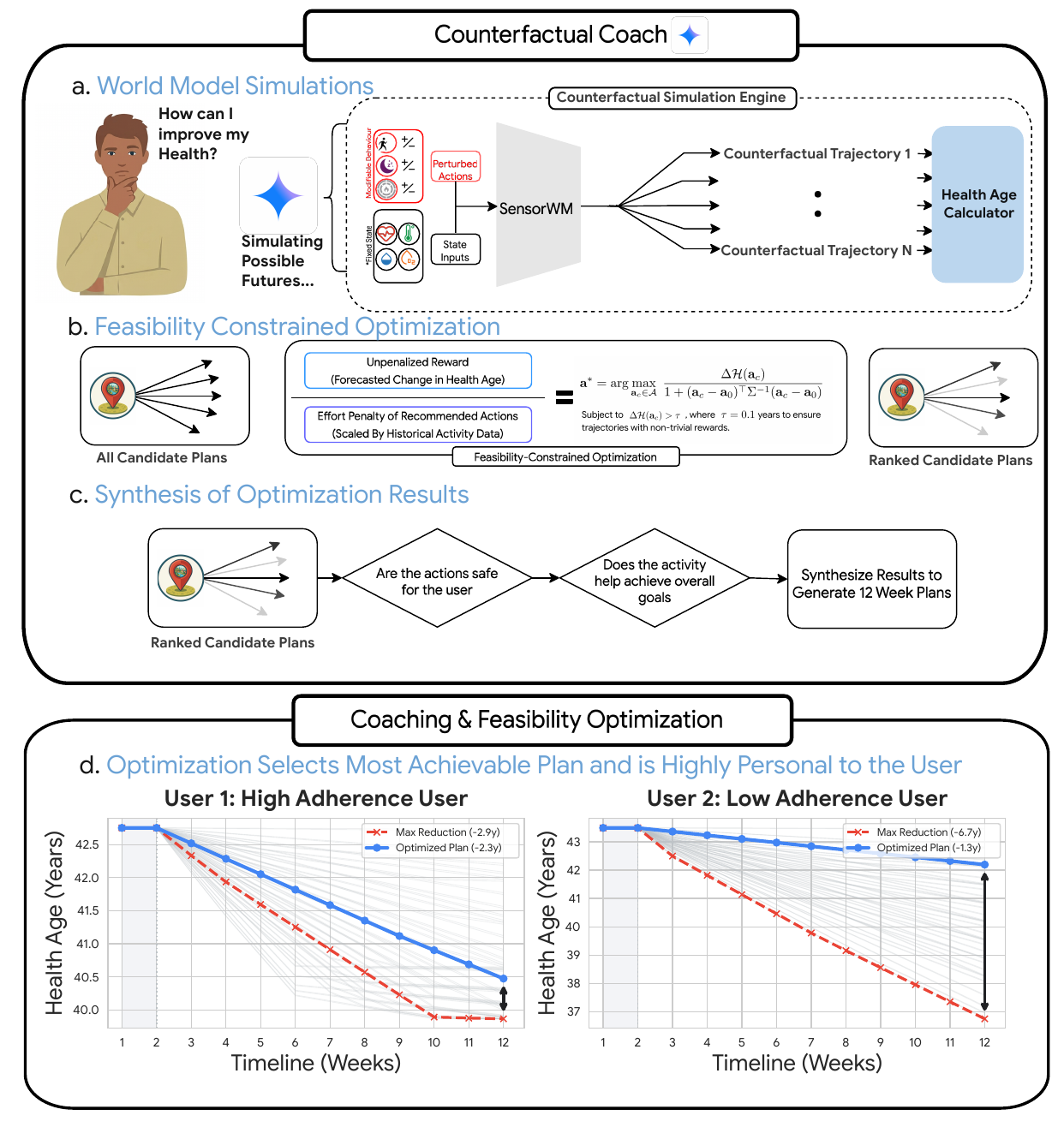}
    \caption{\small\textbf{SensorWM Generates and Prioritizes Personalized Behavioral Plans for Health Coaching.} \textbf{(a)} Counterfactual simulation of 12-week health trajectories under alternative behavioral plans using SensorWM. \textbf{(b)} Feasibility-constrained optimization ranks candidate plans by projected Health Age reduction and individualized behavioral effort. \textbf{(c)} Ranked plans are screened for safety and behavioral feasibility before synthesis into 12-week coaching plans. \textbf{(d)}  Representative trajectories illustrate how optimization balances achievable Health Age improvements with participant-specific historical behavioral patterns.}
    \label{fig:counterfactual}
\end{figure}

We next used SensorWM to project how individual physiological trajectories may change under alternative behavioral scenarios. For each participant, SensorWM was initialized from the same recent physiological history and rolled forward for 12 weeks under candidate changes in daily steps, sleep duration, and cardio-intensity activity. Each projected trajectory was decoded into wearable-derived physiological measures and mapped to CHA, allowing alternative model-implied futures to be compared from a common starting state (Figure~\ref{fig:counterfactual}a). 

To translate these projections into actionable coaching options, we coupled SensorWM with a feasibility-constrained optimization procedure (Figure~\ref{fig:counterfactual}b, Methods). For each participant, candidate behavioral plans were evaluated according to their projected reduction in CHA and the behavioral effort required to achieve them. Effort was defined relative to the participant's own recent behavioral variability, such that changes consistent with previously observed behavior were favored over larger departures from their usual range. The resulting candidate plans were ranked to prioritize projected physiological improvement while accounting for participant-specific feasibility.

The coaching engine then applied additional constraints before plan generation (Figure~\ref{fig:counterfactual}c). Only directly modifiable behaviors (daily steps, sleep duration, and cardio-intensity activity) were treated as intervention targets, whereas resting heart rate, other physiological measurements, and CHA remained projected outcomes. 
Candidate plans were further screened against participant-specific behavioral limits and safety constraints through feasibility-constrained optimization before the highest-ranked scenarios were passed to the coaching model. 
This design separates actions that a participant can take from physiological responses predicted by SensorWM. 


At the individual level, the projected benefits of behavioral interventions remained dependent on physiological context and participant-specific feasibility (Figure~\ref{fig:counterfactual}d). For one representative participant with greater behavioral adherence capacity, feasibility-constrained optimization selected a plan associated with a 2.3-year reduction in Health Age, approaching the maximum projected reduction of 2.9 years. In contrast, for a participant with lower adherence capacity, the optimizer selected a more conservative plan yielding a 1.3-year reduction despite an unconstrained maximum of 6.7 years. These examples illustrate how maximizing projected physiological improvement alone may favor behavioral changes that are difficult to sustain, whereas incorporating individualized feasibility constraints yields more achievable and personalized intervention targets. The simulation and optimization framework identifies candidate behavioral pathways that balance anticipated health benefits against the effort required to achieve them. These ranked trajectories form the physiological basis of the Counterfactual Coaching system. We next evaluated whether providing these projected futures to an LLM-based coach improved the quality, feasibility, and clinical appropriateness of its recommendations relative to coaching based on observed wearable history alone.

\subsection{Blinded human evaluation of counterfactual coaching by lifestyle coaches and clinicians}
\label{sec:expert_evaluation}

\begin{figure*}[t!] 
  \centering 
  \includegraphics[width=\linewidth,height=0.86\textheight,keepaspectratio]{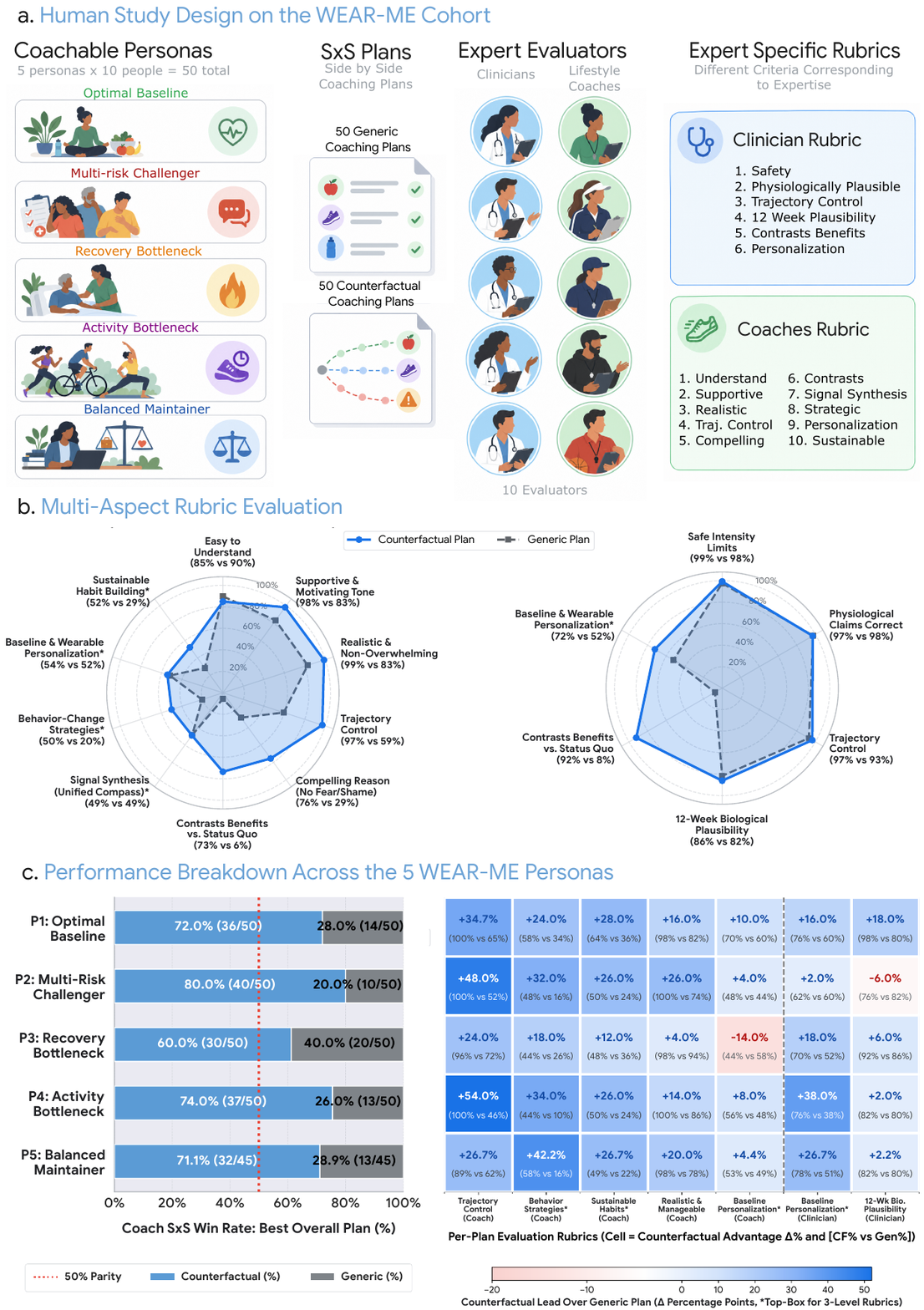} 
  \caption{\small\textbf{Human expert evaluation of Counterfactual vs. Generic coaching plans}. \textbf{(a)} Cohort stratification across five WEAR-ME personas and dual-track evaluation workflow ($5$ lifestyle coaches, $5$ clinicians). \textbf{(b)} Radar comparisons of positive endorsement rates across $10$ behavioral coaching rubrics (left) and $6$ clinical rubrics (right). \textbf{(c)} Persona-level breakdown of coach SxS win rates (left; red dotted line indicates $50\%$ parity) and multi-rubric percentage-point advantage ($\Delta = \text{CF}\% - \text{Gen}\%$, right).}
  \label{fig:human_eval}
\end{figure*}

Having established that SensorWM generates individualized, feasibility ranked behavioral trajectories, we evaluated whether incorporating these projections into an LLM coaching agent improved the resulting behavioral prescriptions. We compared \textbf{Counterfactual Coaching} which conditioned the coaching agent on each participant's baseline CHA, SensorWM-projected wearable trajectory, and feasibility ranked counterfactual pathways against a \textbf{Generic Coaching} baseline that used the identical underlying LLM, participant wearable history, plan structure, and behavior-change instructions, but without CHA or simulation-derived projections.

We conducted a blinded, randomized, dual-track evaluation across 50 cases stratified equally across five prespecified WEAR-ME behavioral personas (49 valid cases after quality control; Figure~\ref{fig:human_eval}a and Methods). Each paired plan was independently reviewed by five certified lifestyle coaches (assessing overall plan quality, anticipated long-term adherence, and 10 behavioral dimensions) and five board-certified clinicians specializing in primary care or lifestyle medicine (assessing clinical safety gates, physiological accuracy, and 6 clinical dimensions), yielding 245 paired evaluations per track across the 49 valid cases (Figures \ref{fig:main1}e and \ref{fig:human_eval}; Supplementary Information ~\ref{sec:supp_expert_evaluation} and Supplementary Tables~\ref{tab:supp_coach_results}, \ref{tab:supp_clinician_results}).

In blinded head-to-head comparisons, lifestyle coaches selected Counterfactual Coaching as the best overall plan in 71.4\% of evaluations (175/245; 95\% CI, 65.5–76.7\%; P = $1.49 \times 10^{-11}$; Figure \ref{fig:human_eval}b), compared with 28.6\% for Generic Coaching (70/245; two-sided exact binomial \(P=1.49\times10^{-11}\)). Counterfactual Coaching was preferred for anticipated long-term adherence in 71.0\% of evaluations (174/245; 95\% CI, 65.0–76.3\%; P = $3.72 \times 10^{-11}$; Figure \ref{fig:main1}e), compared with 29.0\% for Generic Coaching (71/245; \(P=3.72\times10^{-11}\); Figure \ref{fig:main1}e). Preference direction was uniform across all five lifestyle coaches, with majority rater consensus favoring Counterfactual Coaching in $87.8\%$ ($43/49$) of cases for overall quality and $89.8\%$ ($44/49$) for adherence (Supplementary Table~\ref{tab:supp_coach_results}). Crucially, these gains did not come at the expense of clinical safety. Clinicians judged $99.2\%$ ($243/245$) of Counterfactual plans and $97.6\%$ ($239/245$) of Generic plans to be within safe exercise intensity limits ($P = 0.289$), with comparably high physiological claim accuracy ($97.1\%$ versus $97.6\%$; Figure~\ref{fig:main1}e and Supplementary Table~\ref{tab:supp_clinician_results}).

Multi-aspect rubric evaluations revealed that overall preference was driven by dimensions directly enabled by counterfactual trajectory modeling (Figure~\ref{fig:human_eval}b and Supplementary Tables~\ref{tab:supp_coach_results}, \ref{tab:supp_clinician_results}). Among lifestyle coaches, Counterfactual plans substantially outperformed Generic plans in contrasting the benefits of behavior change against the status-quo trajectory ($73.5\%$ versus $5.7\%$; $\Delta = +67.8$ percentage points, $P = 3.15 \times 10^{-45}$), providing a compelling reason to change without fear- or shame-based framing ($75.5\%$ versus $28.6\%$; $\Delta = +46.9$ points), and conveying actionable control over the future trajectory ($97.1\%$ versus $59.4\%$; $\Delta = +37.7$ points). Counterfactual plans also achieved higher top-box ratings for behavior-change strategies ($50.2\%$ versus $20.4\%$) and sustainable habit building ($52.2\%$ versus $28.6\%$), and were more frequently rated realistic and non-overwhelming ($98.8\%$ versus $82.9\%$) and supportive in tone ($98.4\%$ versus $83.3\%$), whereas general comprehensibility ($84.9\%$ versus $89.8\%$, $P = 0.111$) and multi-signal synthesis ($49.0\%$ versus $48.6\%$) were comparable between arms. Clinician evaluations mirrored these findings: clinicians rated Counterfactual plans markedly higher for contrasting intervention benefits against the status quo ($92.2\%$ versus $7.8\%$; $\Delta = +84.5$ points, $P = 3.49 \times 10^{-55}$), baseline physiological personalization ($72.2\%$ versus $52.2\%$; $\Delta = +20.0$ points, $P = 5.57 \times 10^{-6}$), and 12-week biological plausibility ($86.1\%$ versus $81.6\%$, $P = 0.043$). Complementary perspectives between coaches and clinicians, including how clinicians prioritized physiological risk tailoring while coaches emphasized behavioral agency and external lifestyle constraints, are detailed in 
Supplementary Table~\ref{tab:supp_concordance}.

Stratified analyses confirmed that the preference for Counterfactual Coaching generalized across all five WEAR-ME personas (Figure~\ref{fig:human_eval}c and Supplementary Table~\ref{tab:supp_persona_matrix}), led by \textit{Multi-Risk Challengers} (Persona 2; $80.0\%$ win rate, $40/50$), \textit{Activity Bottleneck} (Persona 4; $74.0\%$, $37/50$), \textit{Optimal Baseline} (Persona 1; $72.0\%$, $36/50$), and \textit{Balanced Maintainers} (Persona 5; $71.1\%$, $32/45$). Qualitative error analysis and subgroup comparisons also surfaced two interpretable failure modes
(Supplementary Table~\ref{tab:supp_persona_matrix}): first, in the \textit{Recovery Bottleneck} persona (P3; high baseline activity but insufficient sleep, $60.0\%$ win rate, $30/50$), some plans prescribed redundant step increases in an already saturated activity domain; second, in high-risk profiles (P2), clinicians noted that a subset of 12-week CHA reductions (4--5 years) appeared overly optimistic in magnitude, indicating an opportunity for tighter physiological timescale calibration. These dual-track evaluations demonstrate that grounding an LLM coach in SensorWM counterfactual projections produced more actionable and more frequently preferred behavioral plans while maintaining high expert-rated safety across diverse user profiles.

To complement the expert evaluation and assess scalability across a larger sample, we additionally performed an automated evaluation using rubric-aligned LLM autoraters across an expanded participant cohort (Extended Data Tables~\ref{tab:dim_scores}, \ref{tab:pairwise}). Consistent with the blinded lifestyle-coach and clinician assessments, automated evaluations consistently favored Counterfactual Coaching over the Generic Coaching baseline across overall plan quality, trajectory control, and physiological personalization, confirming that the benefits of simulation-grounded coaching generalize across the broader cohort.

\section{Discussion}

In this study, we introduced CHA and embedded it within an end-to-end counterfactual coaching architecture that couples an autoregressive wearable SensorWM with a feasibility constrained coaching agent. Across population cohorts, we demonstrated that CHA, which is constructed entirely from modifiable physiological and behavioral signals passively tracked by consumer wearables, captures cardiometabolic health, providing incremental prognostic value for incident cardiovascular and metabolic disease beyond age, sex, and BMI in the \textit{All of Us} cohort, and within-person behavioral plasticity over time. We further showed that SensorWM accurately forecasts multi-week physiological trajectories across coupled wearable signals, outperforming generic time series foundation models and statistical baselines while generating biologically coherent counterfactual rollouts under candidate lifestyle modifications. Crucially, in a blinded, dual-track evaluation by certified lifestyle coaches and board-certified clinicians, conditioning an LLM coaching agent on CHA and SensorWM counterfactual trajectories substantially improved overall plan quality and projected long-term adherence compared with coaching grounded in retrospective wearable data alone, while maintaining high clinical safety and physiological plausibility.

A central methodological challenge in quantifying biological aging is the absence of a single gold-standard ground truth \cite{moqri2023biomarkers}. Because true biological age is latent, we evaluated CHA through its alignment with established cardiometabolic constructs, its prospective association with incident disease onset in the \textit{All of Us} cohort, and its longitudinal sensitivity within individuals. Across academia and industry, prior efforts to quantify biological and physiological aging span four primary paradigms, each confronting a fundamental trade-off between clinical validity and everyday behavioral actionability. First, at the molecular and clinical level, epigenetic, proteomic, and blood panel clocks, evolving from first generation chronological age predictors \cite{horvath2013dna, hannum2013genome} to second generation mortality clocks (PhenoAge \cite{levine2018epigenetic}; GrimAge \cite{lu2019dna, mccrory2021grimage}), third generation longitudinal pace of aging biomarkers (DunedinPACE \cite{belsky2022dunedinpace}), and clinical blood composites,  \cite{klemera2006new, insidetracker2021innerage}capture deep cellular and systemic senescence, yet their reliance on invasive, costly, and episodic sampling limits their utility as high frequency feedback signals for daily behavior change. Second, classical epidemiological risk age models, such as Framingham Vascular/Heart Age \cite{d2008general}, Spiegelhalter's effective age formulation \cite{spiegelhalter2016how}, and commercial actuarial engines (Vitality Age), translate multivariable proportional hazards into intuitive age equivalents, but operate as static, episodic snapshots. Third, earlier wearable-based age estimators trained to regress chronological time \cite{pyrkov2018quantitative, miller2025wearable} often latch onto age correlated features that carry limited prognostic or modifiable signal, as reflected in their weak concordance with clinical reference scores in our benchmarks. Fourth, commercial consumer wearable features have largely traded multi system clinical disease grounding for convenience: Garmin Fitness Age maps cardiorespiratory fitness ($\text{VO}_2\text{max}$) and resting heart rate against population percentiles without linking to clinical disease incidence \cite{garmin2021fitnessage}; Oura Cardiovascular Age \cite{oura2024cardiovascular} and polysomnographic Sleep Age models \cite{brink2022age} infer physiological age from passive photoplethysmography (pulse wave velocity) or sleep waveforms, offering observational state readouts that cannot be causally decomposed into specific behavioral targets; and WHOOP Healthspan (WHOOP Age and Pace of Aging \cite{whoopbuck2025healthspan}) aggregates lifestyle exposures using hazard ratios parameterized on late life all cause mortality rather than earlier onset cardiometabolic morbidity.

Our framework resolves this validity--actionability tension by unifying epidemiological survival modeling, action-conditioned trajectory forecasting, and counterfactual behavioral AI coaching agent into a modular, closed-loop architecture. Rather than regressing chronological age, matching population fitness percentiles, or awaiting late life mortality endpoints, CHA anchors high frequency wearable signals directly to published epidemiological hazard ratios for incident cardiometabolic diseases, the earliest and most modifiable burden of chronic morbidity prioritizing time varying behavioral features across physical activity, sedentary patterns, sleep regularity, and cardiorespiratory health, consistent with preventive frameworks such as Life's Essential 8 \cite{lloyd2022life}. By design, this hazard ratio formulation is inherently extensible: unlike monolithic end to end machine learning clocks that require complete retraining when inputs change, new behavioral signals, wearable sensors, or clinical biomarkers (e.g., resistance training volume, dietary patterns, or continuous glucose metrics) can be seamlessly incorporated as new epidemiological evidence or sensing modalities emerge. Expressing these multi-system risks on a shared age equivalent scale yields a responsive metric through which individuals can track their physiological state over time and transparently inspect which factors drive it. Finally, instead of treating Health Age as a static assessment, we pair CHA with an action conditioned trajectory prediction model and a counterfactual coaching agent that operates directly on the user's interpretable behavioral state space, solving for feasible, personalized habit adjustments that yield the greatest marginal reduction in cardiometabolic disease hazard and actively guide the individual toward a younger physiological trajectory.

Our work also advances recent efforts applying large language models and foundation models to biological aging and personal health coaching. Recently, Li et al.\ demonstrated that LLMs can estimate system-wide and organ-specific biological age directly from routine physical examination records and clinical biomarkers across large-scale populations \cite{li2025large}. While such approaches highlight the capacity of LLMs to synthesize cross-sectional clinical records for static risk stratification, direct zero-shot LLM estimation of health age from continuous wearable streams showed substantially lower alignment with clinical cardiometabolic references in our evaluations. Similarly, recent wearable-grounded conversational agents, including PH-LLM and multi-agent personal health frameworks \cite{khasentino2025personal, merrill2026transforming, heydari2025anatomy} have established important benchmarks for interpreting retrospective sensor summaries, yet they rely on the language model's qualitative priors to prescribe behavioral changes. Our architecture bridges these paradigms by separating physiological state estimation and trajectory simulation from natural-language synthesis. Rather than asking an LLM to implicitly guess both biological age and future intervention responses, we pair an explicit, interpretable health age target (i.e., CHA) with a dedicated quantitative world model (i.e., SensorWM) and validate the resulting prescriptions through dual lifestyle-coaching and clinical expert tracks.

This architectural separation directly addresses the explainability gap in clinical artificial intelligence, where post-hoc attribution methods typically identify which inputs drove a static risk score without indicating how an individual can alter that trajectory \cite{ghassemi2021false, kumar2020problems, karimi2021algorithmic}. In our framework, component-level age attributions in CHA are directly coupled to SensorWM's action-conditioned rollouts, translating static feature importance into concrete, quantified behavioral levers such as how a specific increase in daily steps, or sleep duration is projected to shift an individual's future Health Age. The impact of this design was directly reflected in our expert evaluations: the largest advantages of Counterfactual Coaching over Generic Coaching emerged precisely in contrasting projected intervention benefits against a trajectory of continued current habits ($\Delta = +67.8$ percentage points among coaches; $\Delta = +84.5$ points among clinicians), providing a compelling rationale for change, and fostering perceived control over one's health trajectory. Explainability, when operationalized through prospective outcome projections rather than retrospective attributions, thus serves as an active mechanism for motivational framing and sustainable habit formation.

A key technical contribution of this study is demonstrating that a domain specific Sensor World Model can learn coupled, multi week physiological dynamics across behavioral inputs and physiological responses (such as resting heart rate and physical activity), outperforming general purpose time series foundation models across forecasting horizons. Simulating "what-if" behavioral scenarios over a 12-week horizon enabled the coaching agent to evaluate candidate regimens against individualized feasibility constraints before recommending a single goal to the user. 

Our dual-track human evaluation yielded important insights into both the strengths and current failure modes of simulation guided coaching, as well as how different domain experts evaluate AI-generated coaching advice. While both expert panels strongly favored Counterfactual Coaching overall and confirmed its high safety ($99.2\%$ within safe intensity bounds), lifestyle coaches and clinicians prioritized distinct dimensions of personalization. Clinicians were particularly sensitive to physiological risk tailoring and the contrast between projected intervention benefits and the continued current habits, whereas lifestyle coaches focused heavily on behavioral agency, habit progression, and real-world barriers beyond wearable sensors noting in qualitative feedback that wearable derived personalization does not yet capture work schedules, caregiving duties, or environmental access. Furthermore, persona stratified error analyses highlighted two actionable failure modes: in already highly active users with sleep deficits (\textit{Recovery Bottleneck}), the optimizer occasionally suggested redundant step increases alongside sleep goals; and in high-risk profiles (\textit{Multi-Risk Challengers}), clinicians flagged a subset of 4--5 year projected CHA reductions over 12 weeks as overly optimistic, underscoring the need to constrain not only behavioral input feasibility but also the biological rate of physiological adaptation.

Our findings point toward a scalable paradigm for preventive medicine, primary care, and population health management. Because CHA and SensorWM operate entirely on passive signals from widely adopted consumer wearables, this framework can function as a continuous digital companion between episodic clinic visits identifying subclinical cardiometabolic drift early and translating longitudinal risk into individualized, clinician auditable lifestyle prescriptions. In primary care and lifestyle medicine workflows, presenting patients with an intuitive age equivalent metric alongside personalized "what-if" trajectories could enhance shared decision making, replace generic lifestyle handouts with tailored 12-week action plans, and enable remote monitoring of physiological response to prescribed habits. Prospective randomized controlled trials will be needed to determine whether deploying counterfactual wearable coaching in real-world populations leads to sustained behavioral adherence, durable reductions in CHA, and measurable improvements in long-term cardiometabolic outcomes.

Several limitations of this study warrant consideration and inform directions for future work. First, although SensorWM captures longitudinal temporal dynamics across behavioral and physiological signals, it is trained on observational wearable data rather than randomized interventional trials; consequently, its counterfactual projections reflect conditional observational associations rather than guaranteed causal treatment effects, and unmeasured time varying factors such as dietary changes, medication adjustments, intercurrent illness, or psychosocial stress may confound observed trajectories. Second, while consumer wearables enable passive, high frequency monitoring, wearable-owning cohorts can reflect socioeconomic and health conscious selection biases, and wrist worn sensors do not directly measure nutritional intake, resistance training load, or contextual barriers to behavior change such as work schedules, caregiving responsibilities, and environmental access. Third, whereas candidate sleep targets in the counterfactual grid were capped at 9~hours, we did not impose an upper saturation cap on candidate daily step targets because additional step volume can still influence projected downstream physiology such as resting heart rate and Active Zone Minutes through SensorWM; however, in already highly active individuals with insufficient sleep (\textit{Recovery Bottleneck}), this design choice occasionally led the optimizer to propose further step increases alongside sleep goals, suggesting that future implementations should incorporate baseline-aware saturation ceilings or diminishing-return constraints in the candidate action grid. Finally, we evaluated coaching plans using blinded reviews by clinicians and lifestyle coaches, which is an essential step before deploying AI recommendations to users. A natural next step will be to follow users over time to measure how these plans affect real-world habit adherence and CHA.

\section{Methods}
\subsection{Coachable Health Age}
\label{sec:health_age}

\subsubsection{Health Age formulation}
\label{sec:health_age_formulation}

CHA was designed to summarize health on an age-equivalent scale using physiological and behavioral characteristics that can be repeatedly measured with consumer wearables and that have established associations with adverse health outcomes. In contrast to wearable age models trained to recover chronological age, CHA was not optimized to predict elapsed time. Instead, each constituent measurement was mapped to a relative-risk contribution using published epidemiological associations, and the resulting risk contributions were expressed in age-equivalent units \cite{spiegelhalter2016how}.

For participant $j$, the Health Age Delta was defined as

\begin{equation}
\Delta \mathrm{Age}_j
=
\frac{1}{G}
\sum_{i=1}^{K}
w_i \log\left(\mathrm{HR}_{ij}\right),
\label{eq:health_age_delta}
\end{equation}

where $\mathrm{HR}_{ij}$ denotes the relative hazard ratio associated with health domain $i$ for participant $j$, $w_i$ denotes the corresponding weight in log-hazard space, and $G=0.1$ is an age-rate constant approximating a 10\% exponential increase in mortality risk per year of age. CHA was then

\begin{equation}
\mathrm{HealthAge}_j
=
\mathrm{Age}^{\mathrm{chronological}}_j
+
\Delta \mathrm{Age}_j.
\label{eq:health_age}
\end{equation}

Thus, $\Delta \mathrm{Age}>0$ denotes a Health Age older than chronological age and $\Delta \mathrm{Age}<0$ denotes a Health Age younger than chronological age.

The primary wearable-derived components were resting heart rate, mean daily step count, mean sleep duration, and weekly Active Zone Minutes (AZM). Cardiorespiratory fitness ($\mathrm{VO}_2\max$), wearable-predicted insulin resistance, and sleep regularity were incorporated as optional components when available. Resting heart rate was mapped using published associations with incident heart-failure risk \cite{zhang2016resting}; daily step count using associations with incident diabetes \cite{master2022association}; sleep duration using a U-shaped association with cardiovascular disease \cite{daghlas2019sleep}; and weekly AZM using reported associations between physical-activity volume and cardiovascular risk \cite{kany2024associations,wei2026volume}. Cardiorespiratory fitness was represented relative to age-, sex-, and weight-expected $\mathrm{VO}_{2}\max$ and mapped using reported mortality associations per metabolic equivalent \cite{mandsager2018association}. Sleep regularity was quantified using the standard deviation of sleep midpoint and mapped using reported associations between sleep irregularity and cardiovascular risk \cite{huang2020sleep}. Predicted insulin resistance was obtained from a previously validated wearable and demographic prediction model \cite{metwally2026insulin}.

Because the four primary wearable-derived quantities (daily steps, weekly AZM, resting heart rate, and sleep duration) encode correlated aspects of activity, autonomic tone, and sleep, summing their unadjusted literature log-hazards with unit weights would double-count shared risk variance. To account for multicollinearity among these four core domains, we estimated the empirical correlation matrix $\mathbf{R}$ of their log-hazard contributions in the WEAR-ME cohort and solved for mutually adjusted weights $\mathbf{w} = \mathbf{R}^{-1}\mathbf{1}$ in standardized log-hazard space, yielding $w_{\mathrm{sleep}} = 0.8790$, $w_{\mathrm{RHR}} = 0.7066$, $w_{\mathrm{AZM}} = 0.6950$, and $w_{\mathrm{steps}} = 0.4773$ in Eq.~\eqref{eq:health_age_delta} (with $w_i = 1.0$ for the optional domains when available).

\subsubsection{Detailed parameterization of CHA: core wearable domains}
\label{sec:supp_health_age_parameterization}

CHA converts each constituent wearable or physiological measurement to a log-relative-risk contribution $\log(\mathrm{HR}_i)$ and aggregates these contributions using the domain weights $w_i$ according to Eq.~\eqref{eq:health_age_delta} \cite{spiegelhalter2016how}.

\paragraph{Resting heart rate.}
Resting heart rate was mapped using reported associations with incident heart-failure risk \cite{zhang2016resting}, with $\mathrm{RHR} \leq 60\text{ bpm}$ as the reference region ($\mathrm{HR} = 1.0$):
\begin{equation}
\log(\mathrm{HR}_{\mathrm{RHR}})
=
\begin{cases}
0,
& \mathrm{RHR}\leq60,\\[4pt]
\dfrac{\mathrm{RHR}-60}{10}\log(1.08),
&60<\mathrm{RHR}\leq70,\\[10pt]
\log(1.08)
+
\dfrac{\mathrm{RHR}-70}{20}
\left[\log(1.33) - \log(1.08)\right],
&\mathrm{RHR}>70.
\end{cases}
\label{eq:supp_rhr}
\end{equation}

\paragraph{Daily steps.}
Daily step count was mapped using reported associations between daily steps and incident diabetes \cite{master2022association}, with a reference of $8{,}000\text{ steps/day}$ ($\mathrm{HR} = 1.0$) and an upper plateau at $\mathrm{Steps}_{\mathrm{eff}} = \min(\mathrm{Steps}, 9000)$:
\begin{equation}
\log(\mathrm{HR}_{\mathrm{steps}})
=
\frac{\min(\mathrm{Steps}, 9000) - 8000}{1000}
\log(0.88).
\label{eq:supp_steps}
\end{equation}

\paragraph{Sleep duration.}
Sleep duration was mapped using the reported U-shaped association with incident cardiovascular disease \cite{daghlas2019sleep}, with $6.0\text{--}9.0\text{ hours/night}$ as the optimal reference interval ($\mathrm{HR} = 1.0$):
\begin{equation}
\log(\mathrm{HR}_{\mathrm{sleep}})
=
\begin{cases}
(6.0-\mathrm{Sleep})\log(1.20),
&\mathrm{Sleep}<6.0,\\[4pt]
0,
&6.0\leq\mathrm{Sleep}\leq9.0,\\[4pt]
(\mathrm{Sleep}-9.0)\log(1.34),
&\mathrm{Sleep}>9.0.
\end{cases}
\label{eq:supp_sleep}
\end{equation}

\paragraph{Active Zone Minutes.}
Weekly AZM was mapped using reported quartile associations between moderate-to-vigorous physical-activity volume and cardiovascular risk \cite{kany2024associations,wei2026volume}, with $<115\text{ minutes/week}$ as the reference baseline ($\mathrm{HR} = 1.0$) and a plateau at $\geq 500\text{ minutes/week}$:
\begin{equation}
\log(\mathrm{HR}_{\mathrm{AZM}})
=
\begin{cases}
0,
&\mathrm{AZM}<115,\\[5pt]
\dfrac{\mathrm{AZM}-115}{130}\log(0.76),
&115\leq\mathrm{AZM}<245,\\[10pt]
\log(0.76)
+
\dfrac{\mathrm{AZM}-245}{173}
\left[\log(0.71)-\log(0.76)\right],
&245\leq\mathrm{AZM}<418,\\[10pt]
\log(0.71)
+
\dfrac{\mathrm{AZM}-418}{82}
\left[\log(0.68)-\log(0.71)\right],
&418\leq\mathrm{AZM}<500,\\[10pt]
\log(0.68),
&\mathrm{AZM}\geq500.
\end{cases}
\label{eq:supp_azm}
\end{equation}

\subsubsection{Multicollinearity weights ($w_i$) calibrated in the WEAR-ME cohort}
\label{sec:cha_collinearity_weights}

To estimate the domain weights $w_i$ in Eq.~\eqref{eq:health_age_delta} without double-counting shared physiological variance across correlated wearable domains, we first evaluated the four unadjusted log-hazard contributions $\log(\mathrm{HR}_{\mathrm{sleep}})$, $\log(\mathrm{HR}_{\mathrm{RHR}})$, $\log(\mathrm{HR}_{\mathrm{steps}})$, and $\log(\mathrm{HR}_{\mathrm{AZM}})$ across participants in the WEAR-ME cohort using Eqs.~\eqref{eq:supp_rhr}--\eqref{eq:supp_azm}. Evaluating correlations in transformed log-hazard space rather than on raw wearable variables is necessary because raw sleep duration has a non-monotonic U-shaped relationship with mortality risk, and daily steps, RHR, and AZM contain saturation plateaus; in log-hazard space, all four domains are aligned on a common monotonic risk scale.

Because the published hazard ratios originate from independent studies with distinct clinical endpoints and covariate adjustments, rather than univariate regressions against a single outcome in a single cohort, applying an unstandardized covariance inversion ($\boldsymbol{\Sigma}^{-1}\boldsymbol{\sigma}^2$) would weight off-diagonal adjustments by the ratio of domain standard deviations ($\sigma_k / \sigma_i$), allowing domains with wide numerical ranges (such as daily steps) to over-subtract from narrower-range physiological domains (such as RHR). To discount each domain strictly according to its empirical informational overlap with the remaining domains, we standardized each log-hazard column to unit variance, computed the $4 \times 4$ Pearson correlation matrix $\mathbf{R}$ across the WEAR-ME cohort ($R_{ik} = \mathrm{Corr}(\log \mathrm{HR}_i, \log \mathrm{HR}_k)$), and solved the standardized linear system for the domain weights $\mathbf{w}$:
\begin{equation}
\mathbf{R}\tilde{\mathbf{w}} = \mathbf{1}
\quad \Longleftrightarrow \quad
\tilde{\mathbf{w}} = \mathbf{R}^{-1}\mathbf{1},
\qquad
w_i = \min\!\left(\max(\tilde{w}_i, 0),\, 1\right).
\label{eq:supp_collinearity_system}
\end{equation}
In Eq.~\eqref{eq:supp_collinearity_system}, each domain's unit marginal weight ($1.0$) is decomposed into its unique direct weight $w_i$ plus its correlated overlap with the other domains ($\sum_{k \neq i} R_{ik}w_k$). Estimating Eq.~\eqref{eq:supp_collinearity_system} on the WEAR-ME cohort yielded the weights:
\begin{equation}
w_{\mathrm{sleep}} = 0.8790, \qquad
w_{\mathrm{RHR}} = 0.7066, \qquad
w_{\mathrm{AZM}} = 0.6950, \qquad
w_{\mathrm{steps}} = 0.4773.
\label{eq:supp_collinearity_weights}
\end{equation}
Multiplying each core domain's log-hazard contribution by $w_i$ in Eq.~\eqref{eq:health_age_delta} scales each knot log-hazard coefficient from $\beta_i = \log(\mathrm{HR}_i)$ to $\beta^{\mathrm{adj}}_i = w_i \log(\mathrm{HR}_i) = \log(\mathrm{HR}_i^{w_i})$, which is equivalent to replacing each unadjusted knot hazard ratio $\mathrm{HR}_i$ in Eqs.~\eqref{eq:supp_rhr}--\eqref{eq:supp_azm} with its collinearity-adjusted hazard ratio:
\begin{equation}
\log\!\left(\mathrm{HR}^{\mathrm{adj}}_{i}\right)
=
w_i \log\!\left(\mathrm{HR}_{i}\right)
\quad \Longleftrightarrow \quad
\mathrm{HR}^{\mathrm{adj}}_{i} = \exp\!\left(w_i \log \mathrm{HR}_{i}\right) = \left(\mathrm{HR}_{i}\right)^{w_i}.
\label{eq:supp_hr_power_law}
\end{equation}
This formulation guarantees that all reference baselines remain invariant ($1.0^{w_i} = 1.0$, $\log \mathrm{HR}^{\mathrm{adj}}_i = 0$) and preserves exact piecewise continuity at all interior knots. Supplementary Table~\ref{tab:supp_cha_parameters} summarizes the unadjusted literature hazard ratios, the WEAR-ME weights $w_i$, and the resulting collinearity-adjusted knot hazard ratios $\mathrm{HR}^{\mathrm{adj}}_i$.

\subsubsection{Optional CHA domains (cardiorespiratory fitness, sleep regularity, and insulin resistance)}
\label{sec:cha_optional_domains}

When available for a participant, the following three domains are evaluated with weight $w_i = 1.0$ (defaulting to $w_i = 0$ when any required input is unmeasured):

\paragraph{Cardiorespiratory fitness.}
When measured run $\mathrm{VO}_2\max$ ($\text{mL}\cdot\text{kg}^{-1}\cdot\text{min}^{-1}$), biological sex ($\mathrm{Female} = 1$ for female, $0$ for male), and body weight ($\mathrm{Weight}_{\mathrm{lb}}$, in pounds) are available, expected fitness is computed via the FRIEND registry equation \cite{mandsager2018association}:
\begin{equation}
\widehat{\mathrm{VO}_2\max}
=
79.9
-
0.39\,\mathrm{Age}
-
13.7\,\mathrm{Female}
-
0.127\,\mathrm{Weight}_{\mathrm{lb}},
\label{eq:supp_vo2_expected}
\end{equation}
and mapped to a log-hazard contribution:
\begin{equation}
\log(\mathrm{HR}_{\mathrm{VO_2}})
=
\left(
\frac{\mathrm{VO}_{2,\mathrm{measured}} - \widehat{\mathrm{VO}_2\max}}{3.5}
\right)
\log(0.88).
\label{eq:supp_vo2}
\end{equation}

\paragraph{Sleep regularity.}
When the standard deviation of sleep midpoint ($\mathrm{SD}_{\mathrm{midpoint}}$, in minutes) is available, it is capped at $\mathrm{SD}_{\mathrm{eff}} = \min(\mathrm{SD}_{\mathrm{midpoint}}, 120)$ and mapped \cite{huang2020sleep} with $\leq 30\text{ minutes}$ as the reference region:
\begin{equation}
\log(\mathrm{HR}_{\mathrm{regularity}})
=
\begin{cases}
0,
&\mathrm{SD}_{\mathrm{eff}}\leq30,\\[5pt]
\dfrac{\mathrm{SD}_{\mathrm{eff}}-30}{30}\log(1.16),
&30<\mathrm{SD}_{\mathrm{eff}}\leq60,\\[10pt]
\log(1.16) + \dfrac{\mathrm{SD}_{\mathrm{eff}}-60}{30}\left[\log(1.52)-\log(1.16)\right],
&60<\mathrm{SD}_{\mathrm{eff}}\leq90,\\[10pt]
\log(1.52) + \dfrac{\mathrm{SD}_{\mathrm{eff}}-90}{30}\left[\log(2.11)-\log(1.52)\right],
&\mathrm{SD}_{\mathrm{eff}}>90.
\end{cases}
\label{eq:supp_regularity}
\end{equation}

\paragraph{Insulin resistance.}
When wearable-predicted insulin-resistance status \cite{metwally2026insulin} is available, it contributes
\begin{equation}
\log(\mathrm{HR}_{\mathrm{IR}})
=
\begin{cases}
\log(2.00), & \mathrm{IR}=\mathrm{True},\\
0, & \mathrm{IR}=\mathrm{False}.
\end{cases}
\label{eq:supp_ir}
\end{equation}

\subsubsection{Mathematical saturation and CHA plateau}
\label{sec:cha_saturation_plateau}

Because each core physiological metric and health behavior within our framework incorporates discrete plateaus, right-censored saturation thresholds, or bounded reference ranges, the resulting CHA models naturally exhibit structural ceilings on maximum age reduction. In the unadjusted 4-metric CHA model, an individual can only achieve the maximum possible age reduction---capped at $5.13\text{ years}$ younger than chronological age ($\Delta\text{Age} = -5.135\text{ years}$)---if they simultaneously optimize all four digital inputs. This requires maintaining a resting heart rate (RHR) $\le 60\text{ bpm}$ (reference risk level, $\text{HR} = 1.00$, $0.00\text{ years}$), reaching or exceeding $9{,}000\text{ daily steps}$ (capping the step $\text{HR}$ at $0.88$, or $-1.28\text{ years}$ relative to the $8{,}000$-step baseline), sleeping between $6.0$ and $9.0\text{ hours}$ per night (avoiding U-shaped short- and long-sleep penalties, $\text{HR} = 1.00$, $0.00\text{ years}$), and logging $\ge 500\text{ weekly Active Zone Minutes}$ (AZM; capping the AZM $\text{HR}$ at $0.68$, or $-3.86\text{ years}$). Because further behavioral extensions (e.g., $20{,}000\text{ steps/day}$ or $1{,}000\text{ AZM/week}$) encounter explicit log-hazard plateaus rooted in epidemiological saturation points, the formulation mathematically prevents an individual from bypassing this lifestyle-driven rejuvenation floor through extreme volume in a single behavior.

In the collinearity-corrected CHA model, attenuating the marginal hazard ratios to account for physiological and behavioral overlap across the four core wearable metrics tightens this structural rejuvenation cap from $-5.13\text{ years}$ to $-3.29\text{ years}$ ($\Delta\text{Age} = -3.292\text{ years}$). Under collinearity correction, RHR ($\le 60\text{ bpm}$) and sleep duration ($6.0\text{--}9.0\text{ hours}$) remain neutral reference anchors ($0.00\text{ years}$), while the maximum achievable reductions from daily steps ($\ge 9{,}000\text{ steps/day}$, $\text{HR}_{\text{corr}} = 0.9408$) and weekly AZM ($\ge 500\text{ mins/week}$, $\text{HR}_{\text{corr}} = 0.7648$) attenuate to $-0.61\text{ years}$ and $-2.68\text{ years}$, respectively. When the extended optional metrics are incorporated into the collinearity-corrected model, sleep midpoint consistency ($\le 30\text{ min SD}$) and insulin resistance status (absent) operate strictly as unidirectional risk penalties ($0.00\text{ years}$ when optimal, adding up to $+7.47\text{ years}$ at $\ge 120\text{ min SD}$ and $+6.93\text{ years}$ when insulin resistant) and therefore do not expand the maximum age reduction. Only the optional cardiorespiratory fitness term ($\text{VO}_2\text{ max}$) is modeled continuously without an upper saturation plateau, conferring an additional $-1.28\text{ years}$ of age reduction ($\text{HR} = 0.88$) for each $1.0\text{ MET}$ ($3.5\text{ mL}\cdot\text{kg}^{-1}\cdot\text{min}^{-1}$) achieved above age-, sex-, and weight-predicted norms, subject only to a global lower bound of $18.0\text{ years}$ on final Health Age.

\subsubsection{Standard health and risk scores}
\label{sec:health_age_reference_measures}

We evaluated whether CHA captured clinically meaningful variation using two complementary reference measures of cardiovascular and metabolic health: a modified Life's Essential 8 score and the Metabolic Syndrome Severity Score. Neither reference measure was used to fit CHA.

\paragraph{Modified Life's Essential 8.}
We calculated a modified version of the American Heart Association Life's Essential 8 cardiovascular-health score \cite{lloyd2022life}. Standard Life's Essential 8 integrates diet, physical activity, nicotine exposure, sleep, body mass index, blood lipids, blood glucose, and blood pressure. Because standardized dietary-intake information was unavailable, we omitted the diet component and calculated a seven-component score, hereafter referred to as mod-LE8.

Physical activity was quantified from wearable-derived AZM; sleep health from wearable-measured sleep duration; nicotine exposure from self-report; BMI from clinical measurements; blood pressure from clinical systolic and diastolic measurements together with treatment status; blood lipids from non-HDL cholesterol and lipid-lowering treatment status; and blood glucose from HbA1c and diabetes history. Each component was mapped to a 0--100 cardiovascular-health score according to the corresponding Life's Essential 8 framework \cite{lloyd2022life}, and mod-LE8 was calculated as the unweighted arithmetic mean of the seven available components. Higher values indicate more favorable cardiovascular health. Additional scoring details are provided in Section~\ref{sec:supp_clinical_references}.

\paragraph{Metabolic Syndrome Severity Score.}
We quantified metabolic risk using the continuous Metabolic Syndrome Severity Score (MSSS), following the continuous latent-variable framework established using confirmatory factor analysis \cite{gurka2014examination}. MSSS combines body mass index, systolic blood pressure, fasting plasma glucose, triglycerides, and high-density lipoprotein cholesterol. Scores were standardized as sex- and race/ethnicity-specific $z$-scores, with higher values indicating greater metabolic-syndrome severity. Participants missing one or more of the five required clinical measurements were excluded from the MSSS analysis.

\subsubsection{Construct alignment, comparator benchmarks, and longitudinal modifiability}
\label{sec:health_age_validation}

To evaluate whether CHA reflects current cardiometabolic health more faithfully than alternative age formulations, we benchmarked CHA against chronological age, blood-based Phenotypic Age (PhenoAge), a wearable chronological-age prediction model, and direct LLM-derived health-age estimates. PhenoAge was calculated according to the mortality-risk formulation of Levine et al.\ \cite{levine2018epigenetic}. The wearable chronological-age comparator used embeddings from a pretrained Sensor Foundation Model (SensorFM) \cite{narayanswamy2026towards}, followed by chronological-age regression and correction for the age-dependent regression-to-the-mean bias characteristic of age-prediction models \cite{smith2019estimation}. Direct LLM health-age comparators were generated by prompting the base language model with either the curated CHA feature set or the full set of wearable features. The comparator implementations are described in Section~\ref{sec:supp_age_comparators}.

Construct alignment with established cardiometabolic reference scores was assessed by calculating Pearson correlation coefficients between each age measure and modified Life's Essential 8 (mod-LE8) and Metabolic Syndrome Severity Score (MSSS) in the WEAR-ME cohort. Because higher mod-LE8 scores indicate more favorable cardiovascular health whereas higher MSSS scores indicate greater metabolic risk, absolute correlation magnitudes ($|r|$) were used to compare alignment across age formulations (Supplementary Figure~\ref{fig:le8}). We additionally evaluated whether $\Delta\text{CHA}$ (Health Age minus chronological age) varied systematically across participant-reported overall health categories (from excellent to poor) by computing category-specific means and 95\% confidence intervals, comparing separation against age deltas from the wearable chronological-age model.

To characterize within-person longitudinal modifiability, CHA was computed across eight consecutive three-month windows (Q1--Q8; 24 months total) among $N = 1{,}464$ participants in the longitudinal Fitbit cohort (Figure~\ref{fig:main2}c). For each participant $j$ and quarterly window $t \in \{1, \dots, 8\}$, we computed the health-age delta as
\begin{equation}
\Delta \mathrm{CHA}_{j,t}
=
\mathrm{CHA}_{j,t}
-
\mathrm{Age}^{\mathrm{chronological}}_{j,t},
\label{eq:longitudinal_health_age_delta}
\end{equation}
and quantified longitudinal shifts relative to baseline ($\Delta \mathrm{CHA}_{j,t} - \Delta \mathrm{CHA}_{j,1}$), stratifying participants into improved ($\Delta \le -1\text{ year}$), stable ($|\Delta| < 1\text{ year}$), and deteriorated ($\Delta \ge +1\text{ year}$) trajectory groups. These observational analyses characterize natural within-participant plasticity in CHA over coaching-relevant timescales and are not interpreted as causal effects of isolated behavioral interventions.

\subsubsection{Prospective validation in the All of Us Research Program}
\label{sec:aou_health_age}

We independently evaluated the prospective association and prognostic utility of CHA for incident cardiometabolic disease using the \textit{All of Us} Research Program. Participants were required to have at least 730 days of Fitbit tracking span in the daily activity-summary table and at least one valid 90-day wearable window, yielding a total wearable cohort of $N = 22{,}748$ participants ($276{,}378$ three-month windows; $68{,}101.4$ person-years of active wearable tracking; Table~\ref{tab:cohort_characteristics_aou}).
Participant sex and year of birth were obtained from the OMOP \texttt{person} table. Baseline BMI was obtained from the OMOP \texttt{measurement} table (concept ID 3038553, restricted to physiologically plausible values of $12\text{--}80\text{ kg/m}^2$) by selecting the measurement closest to the start of wearable tracking, prioritizing measurements recorded on or before the end of the 90-day baseline window.
Cardiometabolic outcomes were identified from OMOP \texttt{condition\_occurrence} records using the OMOP \texttt{concept\_ancestor} hierarchy. Hypertension was defined using ancestor concept ID 316866, type 2 diabetes using ancestor concept ID 201820, and cardiovascular disease using ancestor concept IDs 316139, 4329847, 4182210, and 381591. The composite cardiometabolic outcome was defined as the earliest recorded diagnosis of any of these three conditions. Participants with a recorded diagnosis of hypertension, type 2 diabetes, or cardiovascular disease on or prior to the end of their baseline wearable window were excluded from incident-disease survival analyses, leaving $18{,}423$ disease-free participants in the primary incident-risk cohort.

\paragraph{Wearable aggregation and data-quality criteria.}
Daily step count, fairly active minutes, and very active minutes were obtained from Fitbit activity-summary records. Main-sleep duration was derived from daily sleep-summary records restricted to the primary sleep episode (\texttt{is\_main\_sleep = TRUE}). Minute-level heart-rate records were aggregated to derive daily minimum heart rate (used as the resting heart rate proxy) and the 90th percentile of daily heart rate.
A seven-day period was considered valid when at least two days contained complete activity, sleep, and heart-rate measurements. Consecutive 90-day (one-quarter) analysis windows were retained only when they contained at least five distinct valid weeks. Within each retained 90-day window, we calculated mean daily steps, mean main-sleep duration, mean daily minimum heart rate, and mean weekly active zone minutes (defined as fairly active minutes plus twice the very active minutes). CHA and Health Age Delta ($\Delta\mathrm{Age} = \mathrm{Health\ Age} - \mathrm{Chronological\ Age}$) were calculated for each 90-day window.

\subsubsection{Baseline Health Age by incident disease status and disease-free survival}
\label{sec:health_age_km}

We first compared baseline Health Age Delta (mean $\pm$ 95\% CI) between participants who remained free of all three conditions throughout follow-up (healthy controls) and those who subsequently developed incident hypertension, incident cardiovascular disease, or incident type 2 diabetes after baseline.
Next, we tested whether baseline Health Age Delta stratified longitudinal cardiometabolic disease-free survival across the primary incident-risk cohort ($N = 18{,}423$). Participants were stratified at baseline into Slower/Normal Agers ($\Delta\mathrm{Age}\leq 0$; $n=10{,}697$) and Accelerated Agers ($\Delta\mathrm{Age}>0$; $n=7{,}726$). Kaplan--Meier estimators were used to estimate the probability of remaining free of the composite cardiometabolic outcome from the end of the baseline window until the earliest recorded diagnosis of hypertension, type 2 diabetes, or cardiovascular disease, or censoring at the end of available wearable follow-up. Disease-free survival probabilities and 95\% confidence intervals were evaluated at 4, 8, 16, 24, and 32 quarters (corresponding to 1, 2, 4, 6, and 8 years of follow-up; Extended Data Table~\ref{tab:km_quarters}).

\subsubsection{Time-varying Cox proportional hazards analysis}
\label{sec:health_age_cox}

To evaluate whether longitudinal Health Age Delta remained independently associated with incident cardiometabolic disease after accounting for demographic and clinical confounders, we fitted a time-varying Cox proportional hazards model. To preserve temporal ordering and mitigate reverse causation from subclinical disease during the same interval, Health Age Delta was lagged by one 90-day window so that $\Delta\mathrm{Age}_{i,t-1}$ predicted incident diagnosis in interval $t$:
\begin{equation}
h_i(t)
=
h_0(t)
\exp
\left(
\beta_{\Delta}\Delta\mathrm{Age}_{i,t-1}
+
\beta_A \mathrm{Age}_{i,0}
+
\beta_B \mathrm{BMI}_{i,0}
+
\boldsymbol{\beta}_{S}^{\top}\mathbf{S}_i
\right),
\label{eq:health_age_cox}
\end{equation}
where $\mathrm{Age}_{i,0}$ denotes chronological age at baseline, $\mathrm{BMI}_{i,0}$ denotes baseline BMI, and $\mathbf{S}_i$ denotes indicator variables for sex. Participant trajectories were represented as start--stop intervals in days since baseline, with the stop time of event windows truncated at the exact diagnosis date. An $L_2$ ridge penalty of $10^{-4}$ was applied for numerical stability. The lagged time-varying analysis included $13{,}842$ participants ($164{,}336$ intervals; $906$ events); full model estimates are reported in Extended Data Table~\ref{tab:cox_models}.

\subsubsection{Incremental prognostic discrimination (Harrell's C-index)}
\label{sec:health_age_cindex}

To assess whether CHA adds genuine prognostic discrimination beyond routine clinical chart variables, we compared nested baseline Cox proportional hazards models in the disease-free cohort with complete baseline covariates ($N = 14{,}663$ participants; $1{,}402$ incident composite diagnoses): (1) a \textbf{Baseline Clinical Model} comprising chronological age, sex, and baseline BMI alone, and (2) an extended \textbf{Baseline + CHA} model incorporating baseline Health Age Delta.
Model discrimination was quantified using Harrell's concordance index (C-index). We evaluated the incremental gain in concordance ($\Delta C = C_{\mathrm{full}} - C_{\mathrm{base}}$) using 500-iteration paired bootstrap resampling to derive 95\% confidence intervals and two-sided $p$-values, alongside a nested likelihood ratio test ($\chi^2$ with 1 degree of freedom). To verify out-of-sample generalizability and rule out overfitting, we computed out-of-sample risk predictions using 5-fold event-stratified cross-validation. Finally, because chronological age alone strongly separates risk across wide age spans, we performed pre-specified age-stratified subgroup evaluations in younger and middle-aged adults ($<50$ years; $N = 7{,}399$, $549$ events) and older adults ($\ge 50$ years; $N = 7{,}264$, $853$ events) to test whether CHA improves risk ranking among peers of similar chronological age (Extended Data Table~\ref{tab:c_index_comparison}).

\subsubsection{Detailed clinical reference measures}
\label{sec:supp_clinical_references}

\paragraph{Modified Life's Essential 8.}
\label{sec:supp_le7}

The modified Life's Essential 8 measure was based on the American Heart Association Life's Essential 8 cardiovascular-health framework \cite{lloyd2022life}. Because standardized dietary-intake data were unavailable, the dietary domain was omitted and the remaining seven cardiovascular-health domains were retained. Each domain was scored from 0 to 100 according to the corresponding Life's Essential 8 criteria, and the final mod-LE8 score was calculated as

\begin{equation}
\mathrm{mod\mbox{-}LE8}
=
\frac{1}{7}
\sum_{k=1}^{7}
S_k,
\label{eq:supp_modle7}
\end{equation}

where $S_k$ denotes the score for each retained cardiovascular-health domain.

Physical activity was calculated from wearable-derived weekly AZM. Participants accumulating at least 150 active minutes per week received 100 points, with lower activity scaled toward zero.

Sleep health was calculated from mean wearable-measured main-sleep duration. Sleep durations between 7 and 9 hours received 100 points. Very short sleep ($<4$ hours) or very long sleep ($>10$ hours) received zero points, with intermediate durations scored according to the corresponding cardiovascular-health scoring scheme.

Nicotine exposure was derived from self-report. Current nicotine use or regular environmental exposure received zero points, whereas participants without reported exposure received 100 points.

BMI was calculated from clinical height and weight measurements. BMI below 25~kg\,m$^{-2}$ received 100 points, with progressively lower scores at higher BMI and zero points at BMI $\geq40$~kg\,m$^{-2}$.

Blood pressure scoring used clinical systolic and diastolic blood-pressure measurements and antihypertensive-treatment status. Untreated blood pressure below 120/80~mmHg received 100 points. Blood pressure controlled through antihypertensive treatment was capped at 80 points.

Blood-lipid scoring used non-HDL cholesterol and lipid-lowering-treatment status. Untreated non-HDL cholesterol below 130~mg/dL received 100 points, while scores for participants receiving lipid-lowering therapy were capped at 80 points.

Blood-glucose scoring was based on HbA1c together with diabetes history. Among participants without diabetes, HbA1c below 5.7\% received 100 points. Among participants with diabetes, scoring reflected degree of glycaemic control, with well-controlled diabetes (HbA1c $<7.0\%$) receiving at most 40 points and poorly controlled diabetes (HbA1c $>9.0\%$) receiving 10 points.

\paragraph{Metabolic Syndrome Severity Score.}
MSSS was calculated according to the continuous confirmatory-factor-analysis framework developed for Metabolic Syndrome Severity \cite{gurka2014examination}. The score incorporates BMI, systolic blood pressure, fasting plasma glucose, triglycerides, and HDL cholesterol. Scores were standardized as $z$-scores within sex and race/ethnicity strata, including Non-Hispanic Black, Hispanic, and Non-Hispanic White/general strata. Positive scores represent metabolic severity above the corresponding U.S. adult reference distribution. Only participants with complete measurements for all five components were included.

\subsubsection{Comparator age formulations}
\label{sec:supp_age_comparators}

\paragraph{Phenotypic Age.}
Phenotypic Age (PhenoAge) was calculated according to the multi-system mortality-risk formulation developed by Levine et al.\ \cite{levine2018epigenetic}. The formulation uses chronological age and nine clinical biomarkers: albumin, creatinine, glucose, C-reactive protein, lymphocyte percentage, mean cell volume, red-cell distribution width, alkaline phosphatase, and white-blood-cell count. Participants missing any required variable were excluded.

The linear predictor was

\begin{align}
xb =\;& -19.907
-0.0336\,\mathrm{Albumin}
+0.0095\,\mathrm{Creatinine}
+0.1953\,\mathrm{Glucose} \nonumber\\
&+0.0954\log(\mathrm{CRP})
-0.0120\,\mathrm{Lymphocyte}
+0.0268\,\mathrm{MCV}
+0.3306\,\mathrm{RDW} \nonumber\\
&+0.0019\,\mathrm{ALP}
+0.0554\,\mathrm{WBC}
+0.0804\,\mathrm{Age}.
\label{eq:supp_pheno_xb}
\end{align}

The corresponding cumulative hazard was

\begin{equation}
H(t)
=
\frac{
\exp(xb)\left[\exp(\gamma t)-1\right]
}{
\gamma
},
\qquad
t=120,
\qquad
\gamma=0.0076927,
\label{eq:supp_pheno_hazard}
\end{equation}

with 10-year mortality probability

\begin{equation}
M=1-\exp[-H(t)].
\label{eq:supp_pheno_mortality}
\end{equation}

This probability was transformed to a clinical Phenotypic Age,

\begin{equation}
\mathrm{ClinicalPhenoAge}
=
141.50225
+
\frac{
\log[-0.00553\log(1-M)]
}{
0.090165
},
\label{eq:supp_clinical_phenoage}
\end{equation}

followed by the scaling transformation used in the analysis,

\begin{equation}
\mathrm{PhenoAge}
=
\frac{
\mathrm{ClinicalPhenoAge}
}{
1
+
1.28047
\exp[
0.0344329(\mathrm{ClinicalPhenoAge}-182.344)
]
}.
\label{eq:supp_phenoage}
\end{equation}

\paragraph{Wearable chronological-age model.}
As a wearable-based chronological-age comparator, daily multimodal wearable measurements were encoded using the pretrained Sensor Foundation Model (SensorFM) \cite{narayanswamy2026towards}. Participants were required to have at least 14 valid wearable days. Daily embeddings were median-pooled across valid days to obtain one participant-level embedding.

Embeddings were standardized to zero mean and unit variance and reduced using principal component analysis. The minimum number of principal components explaining at least 95\% of cumulative variance was retained. An XGBoost regressor was then trained to predict chronological age using five-fold cross-validation. Hyperparameters, including L1 and L2 regularization, row and column subsampling, and tree depth, were selected within the training folds. Out-of-fold predictions were retained for evaluation.

Chronological-age prediction from physiological measurements can exhibit age-dependent estimation bias resulting from regression toward the mean \cite{smith2019estimation}. We therefore fitted a linear correction between chronological age and the raw age residual. Let

\begin{equation}
\Delta_{\mathrm{raw}}
=
\mathrm{Age}_{\mathrm{chrono}}
-
\mathrm{Age}_{\mathrm{pred,raw}}.
\label{eq:supp_raw_age_delta}
\end{equation}

We fitted

\begin{equation}
\mathrm{Bias}(\mathrm{Age}_{\mathrm{chrono}})
=
\alpha\mathrm{Age}_{\mathrm{chrono}}+\beta,
\label{eq:supp_age_bias}
\end{equation}

and calculated

\begin{equation}
\Delta_{\mathrm{unbiased}}
=
\Delta_{\mathrm{raw}}
-
\mathrm{Bias}(\mathrm{Age}_{\mathrm{chrono}}).
\label{eq:supp_unbiased_age_delta}
\end{equation}

The bias-adjusted wearable age was then

\begin{equation}
\mathrm{Age}_{\mathrm{wearable}}
=
\mathrm{Age}_{\mathrm{chrono}}
-
\Delta_{\mathrm{unbiased}}.
\label{eq:supp_wearable_age}
\end{equation}

\paragraph{Exploratory LLM-derived Health Age.}
We additionally evaluated an exploratory LLM-derived Health Age estimate using \texttt{gemini-3.1-pro-preview}. Two wearable-input configurations were considered. The first contained multi-day WEAR-ME measurements including blood-oxygen saturation, sleep score, AZM, sleep duration, steps, and heart-rate variability. The second contained the same curated measurements used by CHA. The model was instructed to return a single integer representing estimated health age.

Generation was evaluated at temperatures $T\in\{0.1,0.5,1.0\}$, with three independent generations per participant and temperature. Server-side caching was disabled. Within-temperature reliability was quantified using the average participant-level standard deviation, ICC(3,1), and ICC(3,$k$) with $k=3$.

Across-temperature sensitivity was assessed among participants with complete estimates at all three temperatures. Mean estimates across the three replicates were compared using the Friedman test. When the omnibus test was significant, pairwise comparisons were conducted with Wilcoxon signed-rank tests and Bonferroni correction across the three temperature comparisons.

\subsection{Trajectory prediction with SensorWM}
\label{sec:sensorwm_methods}

\subsubsection{Pretraining data and longitudinal wearable representation}
\label{sec:sensorwm_data}

World models learn transition functions that describe how an observed system evolves over time and can be recursively applied to generate future state trajectories \cite{ha2018world,lecun2022path,ding2025understanding}. We adapted this formulation to longitudinal wearable data by representing each participant as a sequence of physiological states and observable behavioral actions and learning how physiological state evolves as a function of both recent trajectory and behavioral context.

SensorWM was developed using a large longitudinal dataset of naturalistic wearable measurements collected during routine Fitbit use. The pretraining cohort contained 96,957 participants (2.28 million total person-weeks; 182,797 valid 12-week windows). Participants contributing to the pretraining dataset provided informed consent authorizing use of de-identified data for research and health-and-wellness product development. Data collection and analysis were conducted under centralized institutional review board protocol WCG~20253840. Demographic and longitudinal characteristics of the pretraining population are reported in  Table~\ref{tab:cohort_characteristics_aou}.

Wearable measurements spanned cardiovascular, cardiopulmonary, sleep, skin-temperature, physical-activity, and physiological-variability domains. We partitioned these measurements into physiological \textit{state variables} and behavioral \textit{action variables}. State variables represented the participant's observed physiological condition and included resting heart rate, heart-rate variability, respiratory rate, blood-oxygen saturation, sleep architecture, skin temperature, heart-rate-zone measurements, and longitudinal variability features. Demographic characteristics were included as static state information. Three directly observable behaviors were designated as actions: daily step count, main-sleep duration, and cardio-intensity minutes. The complete feature inventory and state--action assignments are reported in Supplementary Table~\ref{tab:state_action_features}.

Wearable measurements were harmonized to a weekly temporal resolution. Weekly representations provide a compromise between short-timescale sensor variability and the loss of temporal information associated with much coarser aggregation, and are consistent with previous approaches for representing longitudinal event streams and repeated health measurements \cite{mcdermott2023event,lee2025clinical,mcdermott2026meds}. This timescale also corresponds to the timescale over which many sleep and physical-activity exposures are commonly summarized \cite{atkinson2007relationships,buysse2010can,powell2011physical}.

Continuous physiological measurements, behavioral and count-based measurements were aggregated over available observation days and normalized by the number of valid days. For feature $x$ observed on the set of valid days $\mathcal{D}_t$ during week $t$, the weekly aggregate was

\begin{equation}
x_t^{\mathrm{week}}
=
\frac{
\sum_{d\in\mathcal{D}_t}x_d
}{
|\mathcal{D}_t|
}.
\label{eq:weekly_aggregation}
\end{equation}

This normalization prevented incomplete device wear from being interpreted mechanically as reduced activity or physiological function. Continuous features were standardized using the corresponding mean and standard deviation estimated from the pretraining data,

\begin{equation}
\widetilde{x}
=
\frac{x-\mu_{\mathrm{train}}}
{\sigma_{\mathrm{train}}}.
\label{eq:sensorwm_standardization}
\end{equation}

Each participant was subsequently represented as a longitudinal state--action sequence

\begin{equation}
(s_1,a_1,s_2,a_2,\ldots,s_T,a_T),
\qquad T\leq24,
\label{eq:state_action_sequence}
\end{equation}

where $s_t$ denotes the physiological state during week $t$ and $a_t$ denotes the corresponding behavioral action vector.

\subsubsection{WEAR-ME evaluation cohort}
\label{sec:sensorwm_eval_cohort}

SensorWM was evaluated in an independent subset of 1,654 participants from the WEAR-ME study who contained classification labels \cite{metwally2026insulin}; this subset was defined by the availability of longitudinal wearable measurements alongside clinical, behavioral, medication, or laboratory-derived labels, which were not present for all 2,780 cohort participants. Evaluation participants were held out from world-model pretraining. Participants contributed an average of $19.6 \pm 4.2$ weeks ($0.38 \pm 0.08$ years; up to 24 weeks) of wearable observations, providing longitudinal context for evaluating whether SensorWM preserved physiologically meaningful temporal information.

The evaluation benchmark included diabetes, hyperlipidemia, hypertension, kidney disease, respiratory disease, HOMA-IR defined insulin resistance, smoking status, menstrual regularity, and cardiovascular, diabetes, thyroid, mental-health, and general medication-use outcomes. Labels originated from participant questionnaires, medication records, or laboratory measurements depending on the task. All training, validation, and test partitions were defined at the participant level to prevent observations from the same individual appearing across data splits. Complete task definitions, sample sizes, outcome prevalence, demographic characteristics, longitudinal coverage, and task-specific split statistics are reported in Supplementary Tables~\ref{tab:downstream_task_summary}, \ref{tab:wearme_demographics}, and \ref{tab:downstream_task_splits}.

\subsubsection*{State--action feature specification}
\label{sec:supp_sensorwm_features}

The complete set of state and action variables supplied to SensorWM is reported in Supplementary Table~\ref{tab:state_action_features}. The feature representation combined measurements across demographic, cardiovascular, cardiopulmonary, sleep, skin-surface, motion, and regularity or variability domains. Longitudinal wearable and event-stream representations spanning multiple physiological domains have similarly been used to capture temporally evolving health information in prior work \cite{mcdermott2023event,lee2025clinical,mcdermott2026meds}.

Only step count, main-sleep asleep minutes, and cardio-intensity minutes were designated as behavioral actions. All remaining wearable measurements were treated as state variables, including physiological responses associated with behaviour, such as heart-rate-zone occupancy. Demographic variables were treated as static contextual state information.

This operational distinction was maintained during both pretraining and alternative-behaviour rollout. Action variables could be externally replaced by prespecified future sequences, whereas state variables were generated through the learned physiological dynamics. The state--action terminology denotes the role of a variable in SensorWM and does not imply that observational associations between actions and subsequent states identify causal intervention effects.

\subsubsection*{Temporal aggregation and preprocessing}
\label{sec:supp_sensorwm_preprocessing}

The underlying Fitbit measurements were available at heterogeneous temporal resolutions ranging from minute-level physiological streams to daily summary measurements. Directly modeling transitions at the highest available sampling frequency would emphasize substantial short-timescale variation and increase sequence length, whereas very coarse aggregation would reduce the number of available longitudinal transitions. We therefore represented the wearable data on a common weekly grid. Similar temporal harmonization is commonly used for longitudinal health-event representations \cite{mcdermott2023event,lee2025clinical,mcdermott2026meds}, and weekly summaries correspond naturally to established reporting intervals for sleep and physical activity \cite{atkinson2007relationships,buysse2010can,powell2011physical}.

Continuous physiological variables, including heart-rate variability, respiratory rate, blood-oxygen saturation, and temperature-related measurements, were summarized within each weekly interval. Behavioral and count-based variables were aggregated over valid observation days and normalized by the number of observed days according to Eq.~\ref{eq:weekly_aggregation}. Continuous variables were subsequently standardized using pretraining-set statistics according to Eq.~\ref{eq:sensorwm_standardization}. The same stored normalization parameters were used during downstream inference and to invert decoded predictions to their original physiological units.

\subsubsection{Action-conditioned latent dynamics}
\label{sec:sensorwm_architecture}

SensorWM was adapted from joint-embedding predictive approaches in which an encoder and latent transition model are optimized jointly without requiring reconstruction of the original observations \cite{lecun2022path,maes2026leworldmodel}. Let

\begin{equation}
s_t\in\mathbb{R}^{n_s},
\qquad
a_t\in\mathbb{R}^{n_a},
\end{equation}

denote the physiological state and behavioral action vector, respectively, during week $t$. A state encoder $E$ maps the physiological state to a $d=128$ dimensional latent representation,

\begin{equation}
z_t=E(s_t),
\qquad
z_t\in\mathbb{R}^{128}.
\label{eq:sensorwm_encoder}
\end{equation}

SensorWM models second-order dynamics in latent physiological space,

\begin{equation}
p(z_{t+1}\mid z_t,z_{t-1},a_t).
\label{eq:sensorwm_second_order}
\end{equation}

Equivalently, the process can be represented as first-order Markov in an augmented state containing the current latent representation and its recent displacement,

\begin{equation}
\Delta z_t=z_t-z_{t-1}.
\label{eq:sensorwm_displacement}
\end{equation}

The inclusion of $\Delta z_t$ allows the model to distinguish participants occupying similar current physiological states but moving through latent space in different directions.

The encoder was implemented as a multilayer perceptron with hidden dimension 256 and latent dimension 128, using batch normalization, ReLU nonlinearities, and dropout. This latent dimensionality provides sufficient capacity for multivariate temporal representations while limiting the dimensionality of the learned transition space \cite{wang2026temporal}.

The latent transition model was parameterized by a predictor network $g$. Given the current latent state and behavioral action, the predictor generated a state-dependent linear transition operator,

\begin{equation}
F_t
=
g(z_t,a_t),
\qquad
F_t\in\mathbb{R}^{128\times128}.
\label{eq:sensorwm_transition_operator}
\end{equation}

Thus, SensorWM does not learn a single global linear dynamical system. Instead, the transition operator varies as a function of the participant's current physiological representation and behavioral context.

Behavioral information was incorporated into the predictor using adaptive layer normalization and a multiplicative action gate following the action-conditioning formulation used in LeWorldModel \cite{maes2026leworldmodel}. The action-conditioning scale and shift parameters were initialized at zero, such that optimization began near an unconditioned transition model and action dependence was learned progressively. Spectral normalization was applied to linear layers of the predictor to constrain the magnitude of the transition mapping during recursive rollout.

Rather than predicting each future latent state independently, SensorWM transforms the participant's preceding latent displacement. The predicted displacement is

\begin{equation}
\widehat{\Delta z}_{t+1}
=
F_t\Delta z_t,
\label{eq:sensorwm_predicted_displacement}
\end{equation}

and the subsequent latent state is obtained by discrete integration,

\begin{equation}
\widehat{z}_{t+1}
=
z_t+\widehat{\Delta z}_{t+1}.
\label{eq:sensorwm_latent_update}
\end{equation}

Combining Eqs.~\ref{eq:sensorwm_displacement}--\ref{eq:sensorwm_latent_update},

\begin{equation}
\widehat{z}_{t+1}
=
z_t
+
g(z_t,a_t)
\left(z_t-z_{t-1}\right).
\label{eq:sensorwm_full_transition}
\end{equation}

This parameterization represents physiological evolution as an action-dependent transformation of the participant's recent direction of movement through latent space. The learned operator can preserve, attenuate, amplify, or rotate the previous displacement according to the current physiological state and supplied behavior.

\subsubsection*{Sketched Isotropic Gaussian Regularization}
\label{sec:supp_sigreg}

SensorWM used Sketched Isotropic Gaussian Regularization (SIGReg), following the representation regularization used in LeWorldModel \cite{maes2026leworldmodel}, to discourage collapse of the jointly learned latent representation. The target aggregate latent distribution was an isotropic standard Gaussian,

\begin{equation}
z\sim\mathcal{N}(0,I).
\end{equation}

Rather than estimating the full high-dimensional distribution directly, SIGReg projected minibatch latent representations onto randomly sampled one-dimensional directions and quantified deviation from a standard Gaussian using an Epps--Pulley characteristic-function criterion. Each regularization step used 256 random projections. 

SIGReg was applied jointly with the rollout and trajectory-path objectives described in the main Methods. It therefore provided an anti-collapse constraint without requiring an input-reconstruction objective, explicit contrastive negative examples, a momentum-updated target encoder, or stop-gradient target representations.

\subsubsection{Multi-step trajectory training}
\label{sec:sensorwm_training}

SensorWM was trained using 12-week longitudinal rollout windows. Each trajectory was initialized from the first two observed physiological states,

\begin{equation}
\widehat{z}_0=z_0^{*}=E(s_0),
\qquad
\widehat{z}_1=z_1^{*}=E(s_1),
\end{equation}

with initial latent displacement

\begin{equation}
\widehat{\Delta z}_1
=
z_1^{*}-z_0^{*}.
\end{equation}

Subsequent states were generated recursively. At rollout step $t$,

\begin{align}
\widehat{F}_t
&=
g(\widehat{z}_t,a_t),\\
\widehat{\Delta z}_{t+1}
&=
\widehat{F}_t\widehat{\Delta z}_t,\\
\widehat{z}_{t+1}
&=
\widehat{z}_t+\widehat{\Delta z}_{t+1}.
\label{eq:sensorwm_recursive_training}
\end{align}

After initialization, predicted states rather than future observed states were supplied to subsequent transitions. Prediction errors at earlier rollout steps could therefore propagate to later steps during training, exposing the model to the same recursive dynamics subsequently used during inference.

For each future week, the latent target was produced by applying the jointly trained encoder to the observed physiological state,

\begin{equation}
z_{t+1}^{*}=E(s_{t+1}).
\end{equation}

The primary rollout objective was

\begin{equation}
\mathcal{L}_{\mathrm{roll}}
=
\frac{1}{|\mathcal{M}|}
\sum_{t\in\mathcal{M}}
\left\|
\widehat{z}_{t+1}
-
z_{t+1}^{*}
\right\|_2^2,
\label{eq:sensorwm_rollout_loss}
\end{equation}

where $\mathcal{M}$ denotes the set of valid transitions in the training window.

We additionally constrained the local direction of the learned trajectory. The empirical displacement between consecutively encoded observed states was

\begin{equation}
\Delta z_{t+1}^{*}
=
z_{t+1}^{*}-z_t^{*},
\end{equation}

and the trajectory-path objective was

\begin{equation}
\mathcal{L}_{\mathrm{path}}
=
\frac{1}{|\mathcal{M}|}
\sum_{t\in\mathcal{M}}
\left\|
\widehat{\Delta z}_{t+1}
-
\Delta z_{t+1}^{*}
\right\|_2^2.
\label{eq:sensorwm_path_loss}
\end{equation}

The rollout and path objectives therefore constrain complementary properties of the learned dynamics: $\mathcal{L}_{\mathrm{roll}}$ penalizes errors in the predicted latent location, whereas $\mathcal{L}_{\mathrm{path}}$ penalizes errors in the direction and magnitude of local movement.
The complete SensorWM training objective was

\begin{equation}
\mathcal{L}_{\mathrm{SensorWM}}
=
\lambda_{\mathrm{MSE}}
\mathcal{L}_{\mathrm{roll}}
+
\lambda_{\mathrm{path}}
\mathcal{L}_{\mathrm{path}}
+
\lambda_{\mathrm{SIG}}
\mathcal{L}_{\mathrm{SIG}},
\label{eq:sensorwm_objective}
\end{equation}

with $\lambda_{\mathrm{MSE}}=1$. During model development,
$\lambda_{\mathrm{path}}$ was evaluated over the range 1.5--3.0 dynamically adjusted during pre-training  and
$\lambda_{\mathrm{SIG}}=0.1$. The implementation of SIGReg is described in  Methods~\ref{sec:supp_sigreg}.

The encoder and transition predictor were optimized jointly using AdamW with an initial learning rate of $3\times10^{-5}$ and weight decay of $10^{-3}$. Global gradient norm was clipped at 1.0.

\subsubsection{Diagnostic physiological decoder and recursive forecasting}
\label{sec:sensorwm_forecasting}

SensorWM pretraining was performed entirely in latent space and did not require reconstruction of the original wearable measurements, consistent with the joint-embedding predictive formulation on which the architecture was based \cite{lecun2022path,maes2026leworldmodel}. To evaluate predicted trajectories in physiological units, we subsequently trained a post hoc linear decoder $D$,

\begin{equation}
\widehat{s}_t=D(z_t),
\label{eq:sensorwm_decoder}
\end{equation}

while keeping the pretrained encoder and transition predictor fixed. The decoder was optimized using masked mean-squared error such that only observed physiological measurements contributed to the loss. AdamW was used with learning rate $10^{-3}$ and weight decay $10^{-4}$. Fifteen percent of decoder-training observations were reserved for internal validation. Training used early stopping, and the parameters corresponding to the lowest validation loss were restored before evaluation. The decoder did not contribute gradients to the encoder or transition predictor and therefore did not alter the pretrained SensorWM representation.

For action-conditioned forecasting, each participant was initialized from the two most recent observed physiological states,

\begin{equation}
z_0=E(s_0),
\qquad
z_1=E(s_1),
\qquad
\Delta z_1=z_1-z_0.
\end{equation}

SensorWM was then propagated autoregressively using the participant's recorded behavioral action sequence over the forecasting horizon. At each prediction step, the corresponding behavioral action vector was supplied as an exogenous input to condition the physiological transition. Future physiological states were predicted recursively without access to future observed physiological measurements.

\begin{align}
F_t &= g(z_t,a_t),\\
\Delta z_{t+1} &= F_t\Delta z_t,\\
z_{t+1} &= z_t+\Delta z_{t+1}.
\label{eq:sensorwm_recursive_inference}
\end{align}

No future physiological measurements were provided after rollout initialization. Subsequent physiological states were predicted autoregressively using the action-conditioned latent transition dynamics defined in Eq.~\ref{eq:sensorwm_recursive_inference}.

Feature-level forecasting performance was quantified using the coefficient of determination ($R^2$) and Pearson correlation coefficient over the population. SensorWM was compared with the mean of the preceding one, two, and four weeks, last observation carried forward (LOCF), and TimesFM \cite{das2024decoderonlyfoundationmodeltimeseries} as a general-purpose time-series forecasting model.

Complete feature-level $R^2$ and correlation results are reported in Supplementary Tables~\ref{tab:fitbit_worldmodel_forecasting_r2}--\ref{tab:fitbit_worldmodel_forecasting_pearson}.

\subsubsection{Trajectory and downstream information fidelity}
\label{sec:sensorwm_fidelity}

We evaluated whether SensorWM-generated trajectories preserved both the geometric structure and clinically relevant information contained in observed longitudinal wearable measurements.

To visualize trajectory geometry, observed wearable sequences were passed through the frozen SensorWM encoder to obtain 128-dimensional latent representations at each weekly time point and compared against the generated latent features. Principal-component analysis was fitted to the combined set of observed and generated latent states, and the same transformation was applied to both. The first two principal components were used to visualize participant-specific trajectories, enabling direct comparison of their direction, dispersion, and temporal evolution in a common coordinate system.

We next evaluated whether generated trajectories retained information relevant to downstream health prediction. For each task, the observed longitudinal Fitbit sequence and corresponding SensorWM-generated trajectory were evaluated using matched gated recurrent unit (GRU) sequence classifiers \cite{chung2014empirical}. The classifier architecture, optimization procedure, train--validation--test partition, and outcome definition were held fixed between observed and generated inputs so that performance differences reflected information retained by the generated trajectory rather than differences in downstream model capacity.

We additionally compared the trajectory-based representations with two non-trajectory baselines derived from the same wearable data: features obtained from the first week of wearable observations and embeddings generated by the pretrained SensorFM \cite{narayanswamy2026towards}. Representations were mapped to a common 128-dimensional space for controlled comparison. The observed and SensorWM-generated longitudinal sequences were evaluated with the same GRU-based sequence-classification framework \cite{chung2014empirical}, whereas the comparison representations supplied static participant-level information.

Classification performance was quantified using area under the receiver operating characteristic curve (AUROC). For the primary comparison between observed and SensorWM-generated trajectories, differences in paired AUROCs were assessed using DeLong's test \cite{delong1988comparing} for correlated receiver operating characteristic curves.

Failure to reject the null hypothesis was interpreted as an absence of a detectable difference rather than as formal statistical equivalence. Complete classification results and additional robustness analyses are reported in Supplementary Tables~\ref{tab:fitbit_worldmodel_lsm_firstweek_classification}--\ref{tab:fitbit_worldmodel_classification_sex_2} and Supplementary Results \ref{supp:results}.

\subsubsection{Alternative-behavior rollouts}
\label{sec:sensorwm_alternative_rollouts}

The same frozen transition model was subsequently evaluated under prespecified alternative behavioral sequences. For each participant, the physiological initialization was held fixed while the recorded future action sequence
$\{a_t\}_{t=1}^{K}$
was replaced with an alternative sequence
$\{\widetilde{a}_t\}_{t=1}^{K}$.
SensorWM was then propagated autoregressively using the behavioral action vector at each forecast step. Specifically, the transition operator was conditioned on the current predicted latent physiological state and the corresponding action vector, and was applied to the preceding latent state difference to predict the subsequent physiological state.

\begin{equation}
\widetilde{z}_{t+1}
=
\widetilde{z}_t
+
g(\widetilde{z}_t,\widetilde{a}_t)
\left(
\widetilde{z}_t-\widetilde{z}_{t-1}
\right),
\label{eq:sensorwm_alternative_rollout}
\end{equation}

and decoded to physiological space as

\begin{equation}
\widetilde{s}_t
=
D(\widetilde{z}_t).
\end{equation}

Observed-action and alternative-action trajectories therefore shared the same initial physiological history and differed in the behavioral sequence supplied to the learned transition model. Their divergence represents a model-implied conditional difference under the specified action sequences. Because SensorWM was trained using observational wearable data, these alternative trajectories were interpreted as conditional model projections rather than identified causal effects of behavioral intervention.

\subsection{Agentic Counterfactual Coaching}
\subsubsection{WEAR-ME cohort and feature preprocessing}
\label{sec:m-data}
Participants for the coaching evaluations were drawn from the WEAR-ME study ($N = 2{,}780$ participants with complete demographic linkage and at least three valid weeks of longitudinal wearable data). Inputs to plan generation comprised wearable-derived baseline metrics, chronological age, CHA, resting heart rate (RHR), daily steps, device-measured sleep duration, Active Zone Minutes (AZM), and cardiorespiratory fitness, together with each participant’s habitual distributions (up to 24 weeks; mean $19.6 \pm 4.2$ weeks, corresponding to $0.38 \pm 0.08$ years and 3,289 complete 12-week windows in Table \ref{tab:cohort_characteristics_aou}), historical peaks, and model-projected trajectories. All data were de-identified and accessed under the study's institutional governance and privacy controls.

Per-domain daily wearable records were aggregated to ISO calendar weeks by computing the weekly mean of daily values, yielding one weekly feature vector per participant. Static features were harmonized to standard units (weight in kilograms, height in meters, body mass index [BMI] in $\text{kg/m}^2$), and chronological age was derived from birth year.  Each feature was standardized to zero mean and unit variance using pre-trained cohort-level statistics.  Features were partitioned into participant-controllable \emph{actions} (volitional behaviors: daily steps, main-sleep duration in minutes, and cardiorespiratory exercise minutes) and observed physiological \emph{states} (RHR, AZM, and remaining physiological markers). Sequences were padded to a uniform length with an accompanying validity mask.

\subsubsection{Counterfactual simulation and feasibility-constrained optimization}
\label{sec:m-sim}
\label{sec:m-opt}
For each participant, we simulated a set of feasible 12-week behavior-change scenarios and their projected physiological consequences, casting plan selection as a constrained optimization problem that trades projected CHA improvement against individualized behavioral effort.

\paragraph*{Candidate space and counterfactual rollouts.}
By construction, candidate~0 (all targets equal to baseline) represents the ``do-nothing'' scenario, in which observed behavioral levels are held constant over the 12-week forecasting horizon. Its projected endpoint defines the \emph{status-quo} anchor, $\mathrm{CHA}_{\text{sq}}$. For each intervention candidate $c$, we defined the projected counterfactual benefit as $\Delta\mathrm{CHA}(c) = \mathrm{CHA}_{\text{sq}} - \mathrm{CHA}_c$, where $\mathrm{CHA}_c$ denotes the corresponding 12-week projected CHA. This future-versus-future comparison measures the incremental benefit of behavioral modification relative to continued habitual behavior, rather than the absolute change from the participant's observed baseline CHA, thereby accounting for physiological drift predicted under unchanged actions.

\paragraph*{Personalized effort and optimization objective.}
To reflect individual behavioral flexibility rather than population-average difficulty, we estimated per-action variability $\sigma^{\text{user}}_k$ as the within-participant standard deviation of action $k$ over the participant's historical wearable record, bounded below by safety floors (500 steps, 15 sleep minutes, and 10 cardiorespiratory exercise minutes) so that highly consistent participants were not assigned near-infinite effort penalties. Personalized effort was defined as the sum of squared, self-standardized deviations from baseline:
\begin{equation}
E(c) \;=\; \sum_{k\in\{\text{step},\,\text{sleep},\,\text{cardio}\}}
\left(\frac{c_k - b_k}{\sigma^{\text{user}}_k}\right)^{\!2}.
\label{eq:effort}
\end{equation}
Each candidate trajectory was scored by its projected CHA reduction per unit of personalized effort,

\begin{equation}
\mathrm{score}(c) \;=\;
\frac{\Delta\mathrm{CHA}(c)}{1 + E(c)}
\;=\;
\frac{\mathrm{CHA}_{\text{sq}} - \mathrm{CHA}_c}{1 + E(c)},
\label{eq:objective}
\end{equation}

where the additive constant regularizes the ratio at $E(c)=0$. 
We discarded candidates whose projected reduction did not exceed a minimal clinically meaningful threshold ($\mathrm{CHA}_{\text{sq}} - \mathrm{CHA}_c > 0.1\text{ years}$) and ranked the remaining feasible trajectories by Eq.~\eqref{eq:objective}. The ranked candidates were subsequently organized into three behavioral pathways corresponding to sleep-focused, movement-focused, and combined interventions. Up to three pathways were provided to the coaching agent as Option 1, Option 2, and Option 3, together with the status-quo projection $\mathrm{CHA}_{\text{sq}}$, participant-specific historical wearable ranges, and recorded behavioral peaks. Each option described the proposed behavioral targets and their projected 12-week changes in CHA and resting heart rate, allowing the agent to synthesize a personalized coaching plan from distinct behavioral alternatives.
Sensitivity analyses across objective hyperparameters are reported in the Supplementary Information.
\paragraph*{Action--outcome firewall.}
The optimization exposes only volitional behavioral actions (steps, sleep duration, cardiorespiratory exercise) as recommendable targets; projected physiological states (RHR, AZM, and CHA) are treated strictly as downstream model outcomes and are never prescribed as direct behavioral goals. This separation is enforced in the generation prompt and audited in evaluation, preventing the agent from instructing a user to ``lower your resting heart rate'' rather than prescribing the behavioral modifications projected to achieve that outcome.
\subsubsection{Paired coaching-plan generation and two-arm study design}
\label{sec:m-design}
\label{sec:m-arms}
To isolate the contribution of CHA and SensorWM counterfactual projections, we used a two-arm, within-participant paired design. For each participant, two 12-week personalized lifestyle coaching plans were generated from identical baseline demographic and longitudinal wearable histories (up to 24 weeks; recent daily averages, habitual means, standard deviations, minimum--maximum ranges, and historical peaks) using the same underlying language model (\texttt{gemini-3.5-flash}, temperature $0.2$, served via the OneTwo interface):
\begin{itemize}
    \item \textbf{Generic Coaching Arm (Baseline Control):} Conditioned on the participant's demographics, baseline wearable metrics, 24-week habitual distributions, and historical peaks, but \textbf{without} access to CHA or SensorWM counterfactual simulations. The model was instructed to synthesize the wearable history and construct a progressive 12-week behavior-change plan across three 4-week phases (\textit{Weeks 1--4: Foundation}, \textit{Weeks 5--8: Build}, and \textit{Weeks 9--12: Sustain}).
    \item \textbf{Counterfactual Coaching Arm (Full System):} Conditioned on the identical demographic and longitudinal wearable context \textbf{plus} the participant's baseline CHA, the 12-week status-quo projection (\textit{Future A: If No Habit Changes}, $\mathrm{CHA}_{\text{sq}}$), and the top feasibility-ranked counterfactual pathways (\textit{Future B: Your Counterfactual Options}), quantifying projected changes in CHA ($\Delta\text{CHA}$), RHR ($\Delta\text{RHR}$), and AZM alongside historical-best self-efficacy anchors.
\end{itemize}
To prevent confounding from stylistic or structural formatting differences, both arms were governed by strict \textbf{Universal Tone and Formatting Parity Guardrails} (Supplementary Information~\ref{c3}), enforcing an identical 500--700-word target length, an identical four-part layout (\texttt{Your Health Snapshot}, \texttt{12-Week Progressive Plan}, \texttt{Evidence-Based Habit Formation}, and \texttt{Summary}), explicit Self-Determination Theory (SDT) motivational interviewing principles (implementation intentions, coping planning, self-monitoring, and autonomy support), and a prohibition against clinical jargon or fear/shame-based framing. All plan generation and evaluations were conducted offline using retrospective participant profiles.

\subsubsection*{Experimental Prompt Architectures and Universal Tone Parity Guardrails}
To ensure a rigorous scientific comparison between the \textbf{Generic Coaching Arm} and the \textbf{Counterfactual Coaching Arm}, both prompts were engineered under strict \textbf{Tone, Structure, and Length Parity Guardrails}. In preliminary pilot testing, unaligned baseline prompts tended to produce dry, bulleted clinical summaries, whereas counterfactual prompts produced warm, narrative coaching, confounding the informational value of counterfactual simulations with stylistic warmth. We eliminated this confounder by enforcing identical persona instructions, identical Self-Determination Theory (SDT) motivational interviewing principles, identical formatting constraints, and an identical four-section output schema across both arms:
\begin{enumerate}
    \item \textbf{Tone \& Empathy:} Warm, supportive, non-judgmental, collaborative, and empowering; strict prohibition against fear-based, alarmist, or shame-inducing language.
    \item \textbf{No Clinical or Statistical Jargon:} Explicit ban on raw statistical or variable tokens (such as \texttt{std}, \texttt{min/max}, \texttt{wearable\_rhr}, \texttt{counterfactual}, \texttt{baseline delta}, or \texttt{intervention vector}). All physiological concepts (such as Resting Heart Rate and Active Zone Minutes) must be explained in accessible, everyday language.
    \item \textbf{Target Length Parity:} Strictly bounded to \textbf{500--700 words} across both arms so raters could not distinguish arms by verbosity.
    \item \textbf{Four-Section Structural Parity:}
    \begin{itemize}
        \item \textbf{Section 1:} \texttt{YOUR HEALTH SNAPSHOT} (Generic) / \texttt{YOUR HEALTH SNAPSHOT \& HEALTH AGE} (Counterfactual).
        \item \textbf{Section 2:} \texttt{YOUR 12-WEEK PROGRESSIVE PLAN} (Generic) / \texttt{YOUR COUNTERFACTUAL PATHS \& 12-WEEK PLAN} (Counterfactual), structured identically into \textit{Phase 1 (Weeks 1--4: Foundation)}, \textit{Phase 2 (Weeks 5--8: Build)}, and \textit{Phase 3 (Weeks 9--12: Sustain)}.
        \item \textbf{Section 3:} \texttt{EVIDENCE-BASED HABIT FORMATION} (incorporating habit stacking, implementation intentions, self-monitoring, and barrier troubleshooting).
        \item \textbf{Section 4:} \texttt{SUMMARY \& NEXT STEPS} (closing with an empowering one-sentence takeaway).
    \end{itemize}
\end{enumerate}

\subsubsection{Dual-track human expert evaluation and cohort stratification}
\paragraph*{Persona stratification and sampling.}
To evaluate clinical safety, behavioral quality, and anticipated adherence across diverse user profiles without over-representing median wearable users, we partitioned the WEAR-ME cohort into five prespecified Coachable Personas defined by baseline daily steps, sleep duration, BMI, and $\Delta\text{CHA}$ ($\text{CHA} - \text{chronological age}${\textbf{, where $\text{CHA}$ denotes the formulation unadjusted for multicollinearity}}):
\begin{enumerate}
    \item \textbf{Persona 1 --- Optimal Baseline:} High physical activity ($\ge 10{,}000\text{ steps/day}$; sub-cohort mean $13{,}775\text{ steps/day}$), adequate sleep ($7\text{--}9\text{ h/night}$; mean $7.42\text{ h}$), healthy BMI ($18.5\text{--}24.9\text{ kg/m}^2$; mean $22.9\text{ kg/m}^2$), and favorable $\Delta\text{CHA} < -1.0\text{ year}$ (mean $-4.99\text{ years}$; $n = 78$, $2.81\%$ of cohort), testing maintenance coaching and ceiling effects.
    \item \textbf{Persona 2 --- Multi-Risk Challenger:} Sedentary activity ($< 5{,}000\text{ steps/day}$; mean $3{,}341\text{ steps/day}$), insufficient sleep ($< 7.0\text{ h/night}$; mean $6.03\text{ h}$), elevated BMI ($\ge 25.0\text{ kg/m}^2$; mean $36.3\text{ kg/m}^2$), and accelerated $\Delta\text{CHA} \ge +1.0\text{ year}$ (mean $+3.06\text{ years}$; $n = 277$, $9.96\%$ of cohort), testing multi-lever prioritization and psychological feasibility.
    \item \textbf{Persona 3 --- Recovery Bottleneck:} High physical activity ($\ge 10{,}000\text{ steps/day}$; mean $12{,}738\text{ steps/day}$) paired with chronic sleep restriction ($< 7.0\text{ h/night}$; mean $6.23\text{ h}$; $n = 354$, $12.73\%$ of cohort), testing whether the system recognizes activity saturation and isolates sleep and recovery.
    \item \textbf{Persona 4 --- Activity Bottleneck:} Isolated inactivity ($< 5{,}000\text{ steps/day}$; mean $3{,}190\text{ steps/day}$) with preserved sleep duration ($7\text{--}9\text{ h/night}$; mean $7.67\text{ h}$) and accelerated $\Delta\text{CHA} \ge +1.0\text{ year}$ (mean $+2.78\text{ years}$; $n = 227$, $8.17\%$ of cohort), testing targeted physical activity escalation.
    \item \textbf{Persona 5 --- Balanced Maintainer:} Moderate daily movement ($5{,}000\text{--}10{,}000\text{ steps/day}$; mean $5{,}539\text{ steps/day}$), adequate sleep ($7\text{--}9\text{ h/night}$; mean $7.48\text{ h}$), and age-aligned $\Delta\text{CHA} \in [-1.0, +1.0]\text{ year}$ (mean $-0.31\text{ years}$; $n = 143$, $5.14\%$ of cohort), testing progressive habit building in typical users.
\end{enumerate}
Ten participants were randomly sampled from each persona ($N = 50$ cases total). During pre-lock quality control, one case in Persona 5 (Participant ID \texttt{57217}) was identified as having received identical plan text in both blinded columns due to a serialization collision and was excluded \textit{a priori}, leaving $N = 49$ valid participant cases ($10$ each in Personas 1--4 and $9$ in Persona 5; mean chronological age $49.5 \pm 6.6\text{ years}$, $57.1\%$ female, mean BMI $31.2 \pm 7.2\text{ kg/m}^2$, mean baseline CHA $53.6 \pm 6.1\text{ years}$, $\Delta\text{CHA} = +4.04 \pm 4.52\text{ years}$). For each case, the Generic and Counterfactual plans were randomly assigned to \texttt{Plan 1} and \texttt{Plan 2} via a Bernoulli coin flip ($49.0\%$ versus $51.0\%$), with all raters blinded to study hypotheses and arm assignments.
\paragraph*{Dual-track expert panels and evaluation rubrics.}
We convened two complementary panels of domain experts ($N = 10$ raters total):
\begin{enumerate}
    \item \textbf{Certified Lifestyle \& Health Coaches ($n = 5$):} Board-certified health, wellness, and exercise coaches evaluated each plan across \textbf{10 behavioral rubric dimensions}—comprising 6 binary (\texttt{Yes}/\texttt{No}) dimensions (\textit{Easy to Understand}, \textit{Supportive \& Motivating Tone}, \textit{Trajectory Control}, \textit{Compelling Reason [No Fear/Shame]}, \textit{Contrasts Benefits vs.\ Status Quo}, and \textit{Realistic \& Non-Overwhelming}) and 4 three-level ordinal dimensions (\textit{Signal Synthesis}, \textit{Baseline \& Wearable Personalization}, \textit{Behavior-Change Strategies}, and \textit{Sustainable Habit Building})—followed by \textbf{2 forced-choice side-by-side (SxS) primary endpoints}: \textbf{Best Overall Coaching Plan} and \textbf{Highest Likelihood of Long-Term Adherence}, with mandatory free-text rationale.
    \item \textbf{Board-Certified Clinicians ($n = 5$):} Physicians in internal medicine, preventive medicine, and lifestyle/cardiovascular medicine independently audited each plan across \textbf{6 clinical dimensions}—comprising 5 binary clinical safety and plausibility gates (\textit{Safe Intensity Limits}, \textit{Physiological Claims Correct}, \textit{12-Week Biological Plausibility}, \textit{Trajectory Control}, and \textit{Contrasts Benefits vs.\ Status Quo}) and 1 three-level ordinal clinical personalization rubric (\textit{Baseline \& Wearable Personalization}).
\end{enumerate}
Three shared dimensions, Contrasts Benefits vs. Status Quo, Trajectory Control, and Baseline \& Wearable Personalization, used identical rubric definitions in both expert panels to enable cross-disciplinary concordance analysis. Item-specific denominators are reported in Supplementary Tables \ref{tab:supp_coach_results} and \ref{tab:supp_clinician_results} with cross-disciplinary comparisons summarized in Table \ref{tab:supp_concordance}.

\subsubsection{Full-cohort automated evaluation with blinded LLM autoraters}
\label{sec:m-rubric}
\label{sec:m-rater}
To complement the 49-case human expert study and evaluate plan quality across the full eligible WEAR-ME cohort ($N = 2{,}780$; $n = 2,613$ evaluable paired plans after quality control), we conducted a blinded automated evaluation using a pre-specified, Kirkpatrick-Model inspired four-level rubric \cite{kirkpatrick1959techniques} (Extended Data Table~\ref{tab:rubric}). Each of the 12 dimensions across Level 1 (\textit{Reaction}: Clarity \& readability, Tone \& motivation, Personal relevance), Level 2 (\textit{Learning}: Data comprehension, Causal understanding, Insight/stakes), Level 3 (\textit{Behavior}: Actionability \& specificity, Feasibility \& progression, Behavior-change technique), and Level 4 (\textit{Results}: Predicted health impact, Measurability, Evidence grounding) was scored on an anchored 1--5 scale. We computed both an \emph{equal-weighted} overall score ($25\%$ per level) and a \emph{priority-weighted} overall score (Reaction $0.15$, Learning $0.30$, Behavior $0.35$, Results $0.20$) \cite{zhang2026rubricstree},
\begin{equation}
\text{overall} \;=\; \sum_{\ell=1}^{4} w_\ell\,L_\ell,
\qquad
\textstyle\sum_{\ell=1}^{4} w_\ell = 1,
\end{equation}
alongside a four-item \emph{safety and accuracy gate} rating \textbf{unsafe advice}, \textbf{fabricated numbers}, \textbf{missing medical disclaimer}, and \textbf{factual/physiological error} as \emph{none}, \emph{minor}, or \emph{major} (with any \emph{major} flag capping overall scores at $2.0$, alongside an uncapped quality index reported separately).

For each participant, the paired Generic and Counterfactual plans were randomly shuffled and relabeled as ``Plan 1'' and ``Plan 2'' prior to judging. The autorater received a ground-truth block of the participant's observed metrics, historical peaks, and model projections, with explicit instructions to evaluate substance rather than length or vocabulary and to treat clearly framed forward-looking projections as legitimate rather than fabricated numbers. Each paired plan set was evaluated three times independently ($N=3$ passes; primary judge \texttt{gemini-3.5-flash}, temperature $0.5$), aggregating dimension scores by median, safety flags by worst-case severity, and head-to-head preference by majority vote across passes. 

\subsubsection{Statistical analysis}
\label{sec:m-stats}
All primary comparisons in both the human expert study and the full-cohort automated evaluation used within-participant paired contrasts between the Counterfactual and Generic coaching arms. In the human expert evaluation ($N = 49$ cases; 245 paired evaluations per track, 245 coach evaluations with complete head-to-head responses in Figures \ref{fig:main1}e and \ref{fig:human_eval}), forced-choice head-to-head endpoints (\textit{Best Overall Plan} and \textit{Highest Likelihood of Long-Term Adherence}) were tested against the null hypothesis of equal preference ($p_0 = 0.50$) using two-sided exact binomial tests with Wilson $95\%$ confidence intervals, supplemented by case-level majority consensus ($\ge 3$ of $5$ raters). For per-plan binary and 3-level ordinal rubrics, we prespecified top-box achievement (\texttt{Yes} for binary items; highest anchor for ordinal items) and tested paired percentage-point differences ($\Delta = \text{Counterfactual}\% - \text{Generic}\%$) using two-sided exact McNemar's tests on discordant pairs.
In the full-cohort automated evaluation, paired continuous rubric contrasts were summarized by the mean within-participant difference with a $10{,}000$-sample percentile bootstrap $95\%$ confidence interval, a two-sided paired Wilcoxon signed-rank test (with Holm--Bonferroni correction across rubric families), and three paired effect sizes (Cohen's $d_z$, matched-pairs rank-biserial correlation, and Cliff's $\delta$). Binary preference rates were summarized with Wilson $95\%$ confidence intervals, and paired safety flag rates were compared using McNemar's test. Autorater self-consistency across repeated passes was quantified using Fleiss' $\kappa$ and Krippendorff's $\alpha$, and cross-model inter-judge agreement (\texttt{gemini-3.5-flash} versus \texttt{gemini-3.1-pro-preview}) was assessed via intra-class correlation $\mathrm{ICC}(2,1)$, Spearman's $\rho$, quadratic-weighted $\kappa$, and Cohen's $\kappa$ on binary preferences. Statistical significance was denoted as $^*P < 0.05$, $^{**}P < 0.01$, and $^{***}P < 0.001$.

\clearpage
\small\subsection*{Author Contributions}

SAL, ZE, AAH, AAM contributed to the conception and design of the work; SAL, ZE, MZ, MG, SS, AS, PR, DM, AAH, AAM contributed to the data acquisition and curation; SAL, ZE, MZ, MG, PR, MZP, MAX, DM, AZF, AAH, AAM contributed to the technical implementation; HP, MZP, DM, DA, YY, JNC, AZF, LS, MM, SP, AAH, AAM
provided technical and infrastructure guidance; SS, AZF, LS provided clinical inputs to the study;
SAL, ZE, AAH contributed to the supplementary data analysis; SAL, ZE, MZ, MG, SS, AS, HP, PR, MZP, MAX, DM, DA, YY, JNC, AZF, LS, MM, SP, AAH, AAM contributed to the drafting and revising of the manuscript.

\subsection*{Data availability}
\textit{All of Us} data are accessible to registered researchers through the \textit{All of Us} Researcher Workbench (\url{https://www.researchallofus.org/}). The WEAR-ME dataset is available to approved researchers for reproducibility purposes (access instructions at \url{https://github.com/Google-Health/consumer-health-research/tree/main/insulin_resistance_prediction}). De-identified longitudinal Fitbit records cannot be publicly shared due to participant consent and privacy restrictions.

\subsection*{Code availability}
Code for Coachable Health Age, SensorWM, and all statistical analyses will be released upon publication at \url{https://github.com/Google-Health/consumer-health-research}.

\small\subsection*{Acknowledgements}
We would like to thank our co-authors and collaborators from Google Research and academia for their contributions to this work. We would also like to thank the users who consented their data to make this research possible. We additionally would like to thank the clinicians and lifestyle coaches who provided their expertise to make this work possible.


\small\subsection*{Correspondence}
Correspondence should be addressed to \{aliheydari, aametwally\}@google.com

\clearpage
\bibliographystyle{unsrt}
\bibliography{main}

\include{extended_data_figures}

\include{supplemental}

\end{document}

%% file: extended_data_figures.tex
\setcounter{figure}{0}
\renewcommand{\figurename}{Extended Data Fig.}
\renewcommand{\thefigure}{\arabic{figure}}
\renewcommand{\theHfigure}{extendedfigure.\arabic{figure}}

\setcounter{table}{0}
\renewcommand{\tablename}{Extended Data Table}
\renewcommand{\thetable}{\arabic{table}}
\renewcommand{\theHtable}{extendedtable.\arabic{table}}

\section*{Extended Data}

\subsection*{Prospective cardiometabolic associations of Coachable Health Age}

\begin{table*}[htbp]
\centering
\caption{\textbf{Kaplan--Meier cardiometabolic disease-free survival probabilities by baseline Coachable Health Age status.} Disease-free survival probabilities (95\% CI) at 4, 8, 16, 24, and 32 quarters (1, 2, 4, 6, and 8 years) of follow-up comparing Slower/Normal Agers ($\Delta\text{Age} \le 0$) and Accelerated Agers ($\Delta\text{Age} > 0$).}
\label{tab:km_quarters}

\footnotesize
\setlength{\tabcolsep}{5pt}
\begin{adjustbox}{max width=\textwidth,center}
\begin{tabular}{@{}lccc@{}}
\toprule
\textbf{Follow-up Horizon} & \textbf{Slower/Normal Agers ($\Delta\text{Age} \le 0$)} & \textbf{Accelerated Agers ($\Delta\text{Age} > 0$)} & \textbf{Absolute Risk Difference} \\
 & \textbf{Disease-Free Survival (95\% CI)} & \textbf{Disease-Free Survival (95\% CI)} & \textbf{($\Delta$ Survival, \%)} \\
\midrule
4 Quarters (1 Year) & 98.31\% (98.03--98.54\%) & 97.70\% (97.31--98.03\%) & +0.61\% \\
8 Quarters (2 Years) & 96.72\% (96.34--97.06\%) & 95.51\% (94.95--96.00\%) & +1.21\% \\
16 Quarters (4 Years) & 91.91\% (91.24--92.53\%) & 90.35\% (89.41--91.22\%) & +1.55\% \\
24 Quarters (6 Years) & 86.99\% (86.00--87.91\%) & 83.72\% (82.18--85.15\%) & +3.26\% \\
32 Quarters (8 Years) & 82.41\% (80.70--83.99\%) & 79.28\% (77.03--81.34\%) & +3.13\% \\
\bottomrule
\end{tabular}
\end{adjustbox}

\vspace{2pt}
\begin{flushleft}
\scriptsize
Log-rank test comparing Slower/Normal Agers ($n = 10,697$) vs.\ Accelerated Agers ($n = 7,726$): $\chi^2 = 20.17$, $p = 7.09 \times 10^{-6}$.
\end{flushleft}
\end{table*}

\begin{table*}[htbp]
\centering
\caption{\textbf{Multivariable Cox proportional hazards regression models for incident cardiometabolic disease.} Log-hazard ratios ($\hat{\beta} \pm \text{SE}$), hazard ratios (HR) with 95\% confidence intervals (CI), $z$-statistics, and two-sided $p$-values for (A) the lagged time-varying Cox model ($N = 13{,}842$ participants; $164{,}336$ three-month intervals; $906$ events) where $\Delta\text{Age}_{t-1}$ predicts incident diagnosis in window $t$, and (B) the baseline landmark Cox model ($N = 14{,}663$ participants; $1{,}402$ events). Both models are adjusted for chronological age at baseline, baseline BMI, and sex.}
\label{tab:cox_models}
\footnotesize
\setlength{\tabcolsep}{3pt}
\begin{tabular*}{\textwidth}{@{\extracolsep{\fill}}lcccc@{}}
\toprule
\textbf{Model \& Covariate} & \textbf{Log-HR ($\hat{\beta} \pm \text{SE}$)} & \textbf{HR (95\% CI)} & \textbf{$z$-stat} & \textbf{$p$-value} \\
\midrule
\multicolumn{5}{@{}l}{\textit{\textbf{A. Lagged Time-Varying Cox Model ($N = 13{,}842$; $164{,}336$ intervals; $906$ events)$^{\text{a}}$}}} \\
\quad Lagged Health Age Delta ($\Delta\text{Age}_{t-1}$, +1 yr) & $0.04 \pm 0.01$ & \textbf{1.04} (1.02--1.05) & $4.99$ & $6.0 \times 10^{-7}$ \\
\quad Baseline chronological age (+1 yr) & $0.02 \pm 0.002$ & \textbf{1.03} (1.02--1.03) & $10.56$ & $< 10^{-16}$ \\
\quad Baseline BMI (+1 $\text{kg/m}^2$) & $0.03 \pm 0.003$ & \textbf{1.03} (1.03--1.04) & $11.18$ & $< 10^{-16}$ \\
\quad Sex: Male (vs.\ Female) & $0.09 \pm 0.07$ & 1.09 (0.94--1.27) & $1.20$ & $0.23$ \\
\quad Sex: Other / Prefer not to answer / Skipped & $-0.04 \pm 0.25$ & 0.96 (0.59--1.58) & $-0.15$ & $0.88$ \\
\addlinespace[4pt]
\multicolumn{5}{@{}l}{\textit{\textbf{B. Baseline Landmark Cox Model ($N = 14{,}663$; $1{,}402$ events)$^{\text{b}}$}}} \\
\quad Baseline Health Age Delta ($\Delta\text{Age}_{0}$, +1 yr) & $0.02 \pm 0.01$ & \textbf{1.02} (1.01--1.04) & $3.63$ & $2.8 \times 10^{-4}$ \\
\quad Baseline chronological age (+1 yr) & $0.02 \pm 0.002$ & \textbf{1.02} (1.02--1.03) & $11.72$ & $< 10^{-16}$ \\
\quad Baseline BMI (+1 $\text{kg/m}^2$) & $0.05 \pm 0.004$ & \textbf{1.05} (1.04--1.05) & $12.20$ & $< 10^{-16}$ \\
\quad Sex: Male (vs.\ Female) & $0.02 \pm 0.06$ & 1.02 (0.91--1.15) & $0.37$ & $0.71$ \\
\quad Sex: Other / Prefer not to answer / Skipped & $-0.27 \pm 0.22$ & 0.76 (0.49--1.17) & $-1.24$ & $0.21$ \\
\bottomrule
\end{tabular*}
\vspace{2pt}
\begin{flushleft}
\scriptsize
$^{\text{a}}$Partial AIC $= 15{,}657.39$; log-likelihood ratio test $= 231.45$ on 5 df ($p < 10^{-16}$). $^{\text{b}}$Partial AIC $= 24{,}648.38$; log-likelihood ratio test $= 317.80$ on 5 df ($p < 10^{-16}$).
\end{flushleft}
\end{table*}

\begin{table*}[htbp]
\centering
\caption{\textbf{Incremental prognostic discrimination of Coachable Health Age beyond routine clinical demographics.}
Comparison of nested Cox proportional hazards models predicting incident cardiometabolic disease, with and without Coachable Health Age, in the disease-free baseline cohort. Discrimination was quantified using Harrell's C-index, with 95\% confidence intervals. Differences in C-index ($\Delta C$) were evaluated using paired bootstrap resampling.}
\label{tab:c_index_comparison}
\scriptsize
\setlength{\tabcolsep}{2pt}
\begin{tabular*}{\textwidth}{@{\extracolsep{\fill}}lcccc@{}}
\toprule
\textbf{Cohort ($N$; Events)} &
\textbf{Baseline Clinical} &
\textbf{+ Coachable} &
\textbf{$\Delta$ C-index [95\% CI]} &
\textbf{$p$-value} \\
& \textbf{(Age + Sex + BMI)}
& \textbf{Health Age} & & \\
\midrule
Full Cohort ($14{,}663$; $1{,}402$)
& 0.6485 [0.6340--0.6632]
& \textbf{0.6541} [0.6390--0.6685]
& \textbf{+0.0056} [+0.0018, +0.0096]
& $0.005$ \\

5-Fold CV ($14{,}663$; $1{,}402$)
& 0.6475 [0.6330--0.6615]
& \textbf{0.6529} [0.6390--0.6668]
& \textbf{+0.0053} [+0.0012, +0.0098]
& $0.016$ \\

Adults $< 50$ yr ($7{,}399$; $549$)
& 0.6648 [0.6420--0.6910]
& \textbf{0.6762} [0.6525--0.7010]
& \textbf{+0.0114} [+0.0048, +0.0181]
& $< 0.001$ \\

\bottomrule
\end{tabular*}
\vspace{2pt}
\end{table*}

\begin{table}[htbp]
\centering
\caption{\textbf{Kirkpatrick-Inspired Evaluation Rubric.}
The automated evaluation framework comprises four levels and twelve
dimensions, each assessed using a five-point anchored rating scale.}
\label{tab:rubric}

\small
\renewcommand{\arraystretch}{1.25}
\setlength{\tabcolsep}{10pt}

\begin{tabularx}{\textwidth}{@{}p{2.5cm}p{3.6cm}X@{}}
\toprule
\textbf{Evaluation Level} &
\textbf{Guiding Question} &
\textbf{Assessment Dimensions} \\
\midrule

\multirow[t]{3}{*}{\textbf{L1. Reaction}} &
\multirow[t]{3}{3.6cm}{\emph{Will the user engage?}} &
\textbf{1a.} Clarity and readability \\
& & \textbf{1b.} Tone and motivation \\
& & \textbf{1c.} Personal relevance \\

\addlinespace[4pt]
\midrule

\multirow[t]{3}{*}{\textbf{L2. Learning}} &
\multirow[t]{3}{3.6cm}{\emph{Does the user understand?}} &
\textbf{2a.} Data comprehension \\
& & \textbf{2b.} Causal understanding \\
& & \textbf{2c.} Insight and stakes (``why now'') \\

\addlinespace[4pt]
\midrule

\multirow[t]{3}{*}{\textbf{L3. Behavior}} &
\multirow[t]{3}{3.6cm}{\emph{Will the user follow it?}} &
\textbf{3a.} Actionability and specificity \\
& & \textbf{3b.} Feasibility and progression \\
& & \textbf{3c.} Behavior-change technique \\

\addlinespace[4pt]
\midrule

\multirow[t]{3}{*}{\textbf{L4. Results}} &
\multirow[t]{3}{3.6cm}{\emph{Is improvement likely?}} &
\textbf{4a.} Predicted health impact \\
& & \textbf{4b.} Measurability \\
& & \textbf{4c.} Evidence grounding \\

\bottomrule
\end{tabularx}
\end{table}

\clearpage

\begin{table}[ht]
\centering
\caption{\textbf{Comparison of counterfactual and generic recommendations across evaluation dimensions.}
Mean scores for each of the 12 rubric dimensions, evaluated on a five-point scale (1 = lowest, 5 = highest) across 2,613 paired observations. Of 2,780 eligible participants with at least 3 weeks of wearable data, 155 were excluded prior to simulation due to missing (zero-imputed) baseline sleep duration in the final observation week and 12 were excluded due to unparseable judge outputs, yielding 2,613 paired evaluations. Scores are reported before application of the safety gate. Boldface highlights dimensions exhibiting particularly large differences between the counterfactual and generic approaches. CF, Counterfactual Coaching.}
\label{tab:dim_scores}
\small
\begin{tabular}{@{}clcc@{}}
\toprule
\textbf{Dim} & \textbf{Rubric dimension} & \textbf{CF} & \textbf{Generic} \\
\midrule
1a & Clarity \& readability           & 5.00 & 5.00 \\
1b & Tone \& motivation               & 5.00 & 4.96 \\
1c & Personal relevance               & 4.99 & 4.40 \\
2a & Data comprehension               & 4.99 & 4.22 \\
2b & Causal understanding             & 4.98 & 3.76 \\
\textbf{2c} & \textbf{Insight / stakes (``why now'')} & \textbf{4.98} & \textbf{2.86} \\
3a & Actionability \& specificity     & 4.92 & 4.65 \\
3b & Feasibility \& progression       & 4.95 & 4.27 \\
3c & Behavior-change technique        & 4.98 & 4.75 \\
4a & Predicted health impact          & 4.94 & 4.17 \\
4b & Measurability                    & 4.89 & 4.62 \\
\textbf{4c} & \textbf{Evidence grounding} & \textbf{4.92} & \textbf{4.15} \\
\bottomrule
\end{tabular}
\end{table}

\begin{table}[ht]
\centering
\caption{\textbf{Within-participant comparison of counterfactual and generic recommendations.}
The pre-specified contrast is evaluated using weighted overall scores from paired observations ($n = 2,613$). $\Delta$ denotes the mean paired difference, with positive values indicating higher scores for counterfactual recommendations; $d_T$ is Cohen's effect size for paired observations. Statistical significance is assessed using a Wilcoxon signed-rank test. Win, tie and loss counts indicate the number of paired comparisons in which counterfactual recommendations scored higher, equally or lower, respectively. Preference denotes the proportion of pooled votes favoring counterfactual recommendations in blinded head-to-head evaluations.}
\label{tab:pairwise}
\small
\begin{tabular}{@{}lccccr@{}}
\toprule
\textbf{Contrast} & $\boldsymbol{\Delta}$ & $\boldsymbol{d_T}$ & $\boldsymbol{p}$
  & \textbf{W / T / L} & \textbf{Preferred} \\
\midrule
Counterfactual $\to$ Generic & $+0.70$ & 1.86 & $< 10^{-300}$ & 2,570 / 3 / 40 & 98.2\% \\
\bottomrule
\end{tabular}
\end{table}

%% file: supplemental.tex
\appendix
\setcounter{figure}{0}
\renewcommand{\figurename}{Supplementary Figure}
\renewcommand{\thefigure}{\arabic{figure}}
\renewcommand{\theHfigure}{suppfigure.\arabic{figure}}

\setcounter{table}{0}
\renewcommand{\tablename}{Supplementary Table}
\renewcommand{\thetable}{\arabic{table}}
\renewcommand{\theHtable}{supptable.\arabic{table}}

\clearpage

\startcontents[appendices]

\begin{center}
  \Large \textbf{Appendix Table of Contents}
\end{center}

\printcontents[appendices]{l}{1}{\setcounter{tocdepth}{3}}

\clearpage
\section{Supplementary Information}
\label{supp:results}
\subsection{Coachable Health Age}

\subsubsection*{LLM Health Age \& Consistency}
\label{llm}
\textbf{Within-Temperature Consistency}
The stochastic stability of the LLM's estimations was evaluated by comparing the consistency of health age predictions across three runs ($k=3$) under different temperatures ($T \in \{0.1, 0.5, 1.0\}$) for two data input configurations: (1) Raw Signals (incorporating all raw wearable sensor streams) and (2) Identical metrics that coachable Health Age algorithm receives as input. In order to be able to compare across temperatures, in cases where LLM provided a range as health age (e.g., 45-48 years), the average of the interval was used in the analysis. 
For the Raw Signals configuration, the model exhibited greater stability across temperatures (Supplementary Table~\ref{tab:consistency_results}). At $T=0.1$, the average standard deviation (SD) between runs was low ($1.35$~years). Single-run consistency, measured by ICC(3,1), was moderate ($0.54$), and averaging across the three runs improved consistency to $\text{ICC}(3,3) = 0.78$. As temperature increased, consistency declined gradually: at $T=0.5$, average SD rose to $2.06$~years ($\text{ICC}(3,1) = 0.30$, $\text{ICC}(3,3) = 0.57$), and at $T=1.0$, SD reached $3.37$~years while retaining higher reliability than the curated baseline ($\text{ICC}(3,1) = 0.20$, $\text{ICC}(3,3) = 0.42$).

The Identical CHA configuration showed a sharp drop in consistency as temperature increased (Supplementary Table~\ref{tab:consistency_results}). While reliability at $T=0.1$ matched the all-signals configuration in ICC ($\text{ICC}(3,1) = 0.54$, $\text{ICC}(3,k) = 0.78$) albeit with higher run-to-run spread (SD = $2.72$~years), reliability collapsed at $T=0.5$ (SD = $3.54$~years), with single-run consistency dropping to $\text{ICC}(3,1) = 0.09$ and three-run average consistency only reaching $\text{ICC}(3,k) = 0.23$. A similar collapse was observed at $T=1.0$ (SD = $3.20$~years, $\text{ICC}(3,1) = 0.11$, $\text{ICC}(3,k) = 0.28$).

\textbf{Across-Temperature Sensitivity}
For the \textbf{Raw Signals} configuration, the Friedman test indicated a statistically significant overall difference in health age estimations across the three temperatures ($\chi^2(2) = 6.47, p = 0.039$). However, post-hoc Wilcoxon signed-rank tests with Bonferroni correction did not detect any statistically significant differences between specific temperature pairs ($T_{0.1}$ vs.\ $T_{0.5}$: adj.\ $p = 0.075$; $T_{0.5}$ vs.\ $T_{1.0}$: adj.\ $p = 0.0689$; $T_{0.1}$ vs.\ $T_{1.0}$: adj.\ $p = 1.00$).

For the \textbf{Curated Metrics} configuration, the sensitivity to temperature was highly pronounced. The Friedman test revealed an extremely significant difference across temperatures ($\chi^2(2) = 39.71, p = 2.39 \times 10^{-9}$). Post-hoc Wilcoxon signed-rank tests with Bonferroni correction confirmed that estimated health ages systematically shifted across all temperature pairs ($T_{0.1}$ vs.\ $T_{0.5}$: adj.\ $p = 0.0039$; $T_{0.5}$ vs.\ $T_{1.0}$: adj.\ $p = 0.0024$; $T_{0.1}$ vs.\ $T_{1.0}$: adj.\ $p = 7.71 \times 10^{-10}$; Supplementary Table~\ref{tab:across_temp}).

\textbf{Association with standard Metrics (Modified LE8 and MSSS): }To evaluate the clinical alignment of the generated health ages, we calculated the Pearson correlation coefficient ($r$) between the delta health age (estimated health age minus actual age) and two established health indicators: the modified Life's Essential 8 (mod-LE8) score and the Metabolic Syndrome Severity (MSSS) score. 
Across both data configurations, delta health age correlated negatively with the modified LE8 score and positively with the MSSS score (Supplementary Table~\ref{tab:correlations}).

\subsubsection*{Coachable Health Age Correlates with AHA Life's Essentials 8 \& Metabolic Syndrome Score}

We further evaluated the clinical alignment of Coachable Health Age by examining its associations with established measures of cardiovascular and metabolic health. The American Heart Association's Life's Essential 8 (LE8) framework \cite{lloyd2022life} provides a comprehensive assessment of cardiovascular health across eight domains, including diet, physical activity, nicotine exposure, sleep health, body mass index, blood lipids, blood glucose, and blood pressure. We considered a modified LE8 score (mod-LE8) excluding diet, alongside the Metabolic Syndrome Severity Score (MSSS) \cite{wiley2016metabolic}, a continuous measure of metabolic dysfunction incorporating BMI, blood pressure, triglycerides, HDL cholesterol, and fasting glucose. CHA demonstrated the strongest absolute correlations with both mod-LE8 ($|r| = 0.512$) and MSSS ($|r| = 0.355$), outperforming chronological age and alternative biological age estimates. These findings suggest that CHA captures clinically meaningful variation in cardiometabolic health beyond that reflected by conventional age measures.

We additionally examined the relationship between wearable-derived Delta Health Age and self-reported overall health status. Consistent with the expected relationship between perceived health and physiological aging, Delta Health Age increased monotonically with worsening self-reported health, from approximately zero years among individuals reporting excellent health to over 11 years among those reporting poor health. This graded association provides complementary evidence of construct validity, suggesting that wearable-derived health age estimates reflect meaningful differences in overall health status that are consistent with individuals' subjective assessments of their health.

\subsection{Trajectory Prediction with Sensor World Model}


\subsubsection*{Choosing the SensorWM}

We compared two world model training formulations, Trajectory Transformer \cite{janner2021offline}, which autoregressively predicts future wearable representations using a transformer backbone \cite{vaswani2017attention}, and LeWorldModel \cite{maes2026leworldmodel}, which learns action-conditioned latent dynamics through joint-embedding prediction. To determine which formulation better preserved clinically relevant information, we evaluated downstream classification performance using generated trajectories from both models against observed Fitbit data. LeWorldModel outperformed Trajectory Transformer on nine of 12 tasks, with notable improvements in insulin resistance (AUROC 0.760 vs.\ 0.719), smoking status (0.774 vs.\ 0.698), and hyperlipidemia (0.692 vs.\ 0.645). Based on its stronger preservation of clinically informative representations, we adopted the LeWorldModel-inspired training formulation for SensorWM (Supplementary Table~\ref{traj-wm}).

\subsubsection*{Scaling SensorWM}

We investigated the effect of pre-training data scale on SensorWM's ability to generate clinically informative physiological trajectories, comparing models trained on 1k, 10k, and 100k participants. We evaluated representation quality indirectly through downstream classification performance using model-generated trajectories. As shown in Supplementary Table~\ref{scaling}, increasing pre-training data generally improved downstream discrimination, with the 100k model achieving the highest AUROC on 8 of 12 tasks, including insulin resistance (0.760), respiratory conditions (0.660), and smoking status (0.774). Although improvements were not strictly monotonic across all tasks, these results suggest that scaling pre-training data enhances the preservation of clinically relevant information in generated physiological trajectories, motivating our use of the 100k-participant model for SensorWM.

\subsubsection*{Trajectory versus Static Classification}

While world models are trained to forecast future physiological states rather than optimize directly for discriminative downstream classification, an important question is whether the longitudinal dynamics captured by the model encode information that is absent from static snapshots of wearable data. To investigate this, we compare downstream prediction performance using four different feature sources: (1) the true observed longitudinal Fitbit trajectories, (2) trajectories generated by the sensor world model, (3) representations obtained from the state of the art SensorFM \cite{narayanswamy2026towards}, and (4) features derived solely from the first week of wearable observations.

This comparison isolates the value of longitudinal physiological information. Both the SensorFM and first-week baselines operate on a limited temporal view of an individual's health state and therefore primarily capture cross-sectional physiological characteristics at a single point in time. In contrast, the observed Fitbit trajectories and the world-model-generated trajectories incorporate patterns that emerge over extended periods, including behavioral consistency, physiological adaptation, recovery dynamics, and temporal variability. These longitudinal signatures may encode information about an individual's health that cannot be recovered from a single observation window alone helping motivate the need for having a world model capable of learning the physiological trajectories of human health. All models utilize the same Gated Recurrent Unit (GRU) style classifier \cite{dey2017gate} to ensure fairness and Longitudinal Models (True Fitbit Sequence \& Sensor World Model Fitbit Sequences) and SensorFM \cite{narayanswamy2026towards} latent space dimensions are fixed at $D=128$ to ensure fair comparisons. 

We therefore evaluate whether access to a longitudinal trajectory provides measurable benefits for downstream health prediction. If longitudinal physiological dynamics contain information beyond static physiological snapshots, models trained on observed or generated trajectories should outperform approaches that rely solely on single-time-point measurements. Furthermore, comparable performance between observed and generated trajectories would suggest that the sensor world model successfully preserves the clinically relevant information contained within real wearable trajectories while enabling future-state simulation and forecasting. This analysis therefore serves both as an assessment of the utility of longitudinal wearable signals and as a validation of the information fidelity of the learned world model representations.


\subsubsection*{Classification Subgroup Analysis}

To evaluate the fidelity of the generated physiological trajectories, we perform a series of subgroup analyses constructed from our demographic labels stratifying by age, body mass index (BMI), and sex. The objective of these experiments is not to assess demographic fairness, but rather to determine whether SensorWM preserves clinically relevant physiological information across populations that exhibit substantially different baseline physiology, behavioral patterns, and disease prevalence. If the generated trajectories faithfully capture the underlying structure of the observed wearable signals, downstream predictive performance should remain similar when using world-model-generated features instead of the true wearable measurements.

For each demographic subgroup, we repeat the downstream prediction experiments using both observed Fitbit features and world-model-generated features. Performance is quantified using AUROC, and differences between the two feature sources are evaluated using DeLong's test. By comparing prediction performance within individual demographic strata, we can directly assess whether the forecasting process preserves the predictive signal contained in the original wearable measurements. Large performance degradations would suggest that information relevant to a particular population is lost during forecasting, whereas comparable performance indicates that the generated trajectories maintain the physiological characteristics necessary for downstream inference.

This stratified evaluation provides a particularly stringent assessment of feature fidelity because the world model must accurately reproduce physiological dynamics across populations with distinct health profiles and behavioral patterns. Consistent agreement between models trained on observed and generated features would indicate that the learned latent dynamics capture physiologically meaningful structure and that forecasted trajectories retain much of the information present in the original wearable observations. Such preservation is essential if generated physiological states are to serve as effective surrogates for future wearable measurements in downstream health applications.

\subsubsection*{Sensitivity Analysis of Error Propagation}

A fundamental challenge in world model development is understanding how forecasting errors propagate into downstream clinical predictions. Although trajectory forecasting models are typically optimized using reconstruction objectives, improvements in reconstruction accuracy do not necessarily translate into improvements in clinically relevant decision-making. We therefore performed a sensitivity analysis to quantify the relationship between trajectory forecasting error and downstream prediction consistency across all evaluation tasks.

For each participant, we first generated a future wearable trajectory using the sensor world model and computed the trajectory reconstruction error relative to the observed wearable measurements. We then applied identical downstream classifiers to both the observed and world-model-generated trajectories and calculated the absolute difference in predicted probabilities between the two predictions. This analysis provides a direct measure of error propagation, allowing us to assess whether inaccuracies in trajectory forecasting lead to meaningful changes in downstream clinical inference. Participants were further categorized according to whether the classifier predictions obtained from observed and generated trajectories were concordant or discordant.

Across nearly all prediction tasks, we observed that downstream prediction differences remained small even when trajectory reconstruction errors varied substantially (Supplementary Figure~\ref{fig:sesnitivity}). Most participants clustered near the origin of the plots, indicating that generated trajectories frequently produced predictions that closely matched those obtained from the true wearable measurements. Notably, many individuals with moderate forecasting errors still exhibited highly consistent downstream predictions, suggesting that reconstruction accuracy alone is not the primary determinant of downstream utility. Instead, the world model appears to preserve the physiologically relevant structure necessary for clinical prediction despite imperfect reconstruction of individual measurements.

Discordant predictions were relatively uncommon and were concentrated among participants exhibiting larger differences in predicted probabilities. Importantly, these identified discordance pairs were not always associated with the largest trajectory reconstruction errors. In several tasks, participants with comparatively low reconstruction error nonetheless crossed classification decision boundaries, whereas others with substantially larger forecasting errors remained correctly classified. This observation suggests that downstream sensitivity depends not only on the magnitude of forecasting error but also on where that error occurs within the physiological state space. Errors affecting features strongly associated with a particular outcome may have a disproportionately large effect on prediction, whereas reconstruction errors in less informative regions may have minimal clinical consequence.

These findings indicate that the relationship between forecasting accuracy and downstream performance is highly nonlinear. While trajectory reconstruction error provides a useful measure of forecasting quality, it does not fully characterize the utility of generated physiological trajectories for clinical prediction. From a health modeling perspective, preserving the latent physiological signals relevant to downstream outcomes may be more important than perfectly reconstructing every observed wearable measurement. The overall concentration of points near low probability differences across tasks therefore provides evidence that the sensor world model retains much of the clinically meaningful information present in real wearable trajectories, even when forecasting errors accumulate over extended prediction horizons.

\subsubsection*{Feature Fidelity: Correlations}

A central question is whether generated trajectories preserve the physiological relationships present in real-world sensor data. While trajectory reconstruction metrics evaluate the accuracy of individual feature forecasts, they do not assess whether the model maintains the higher-order dependencies that govern interactions between wearable features. Preserving these dependencies is essential if generated trajectories are to serve as realistic surrogates for future wearable observations. We therefore evaluated feature fidelity by comparing the correlation structure of observed and world-model-generated wearable trajectories.

For each cohort, we computed pairwise correlation matrices across all physiological and behavioral features using both the observed Fitbit trajectories and the corresponding trajectories generated by the Sensor World Model. Correlations were calculated at the population level as well as within behaviorally distinct subgroups consisting of sedentary and highly active participants. To quantify deviations from the true physiological structure, we additionally computed difference matrices representing the change in pairwise correlations between observed and generated trajectories. This analysis evaluates whether the world model preserves the statistical organization of the wearable feature space rather than simply reproducing individual measurements.

\begin{figure}[h!]
    \centering
    \includegraphics[width=\linewidth]{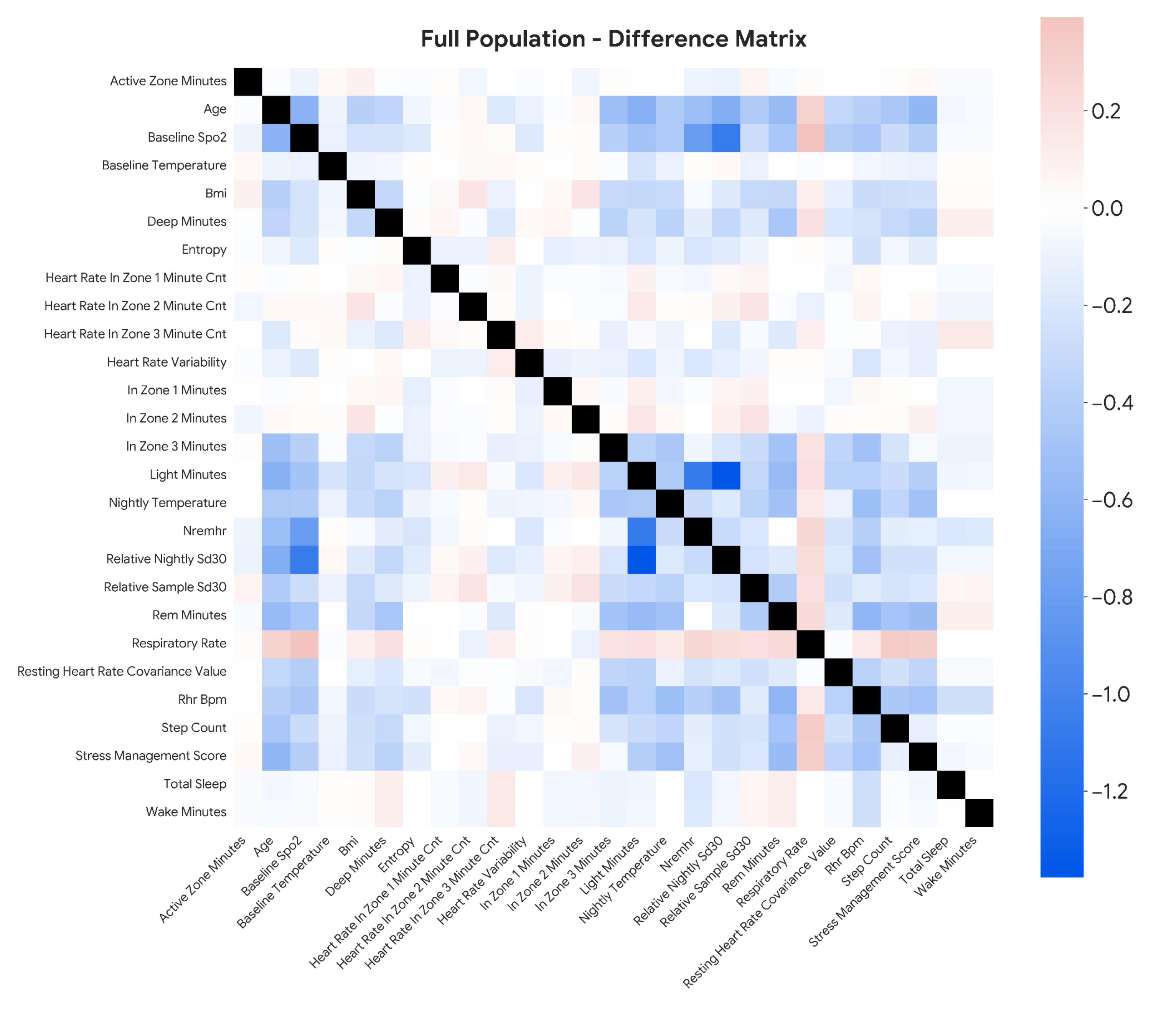}
    \caption{\textbf{Differences in pairwise correlations between observed and SensorWM-generated Fitbit features.}
The matrix shows the difference between pairwise Pearson correlations computed from observed Fitbit features and those generated by SensorWM in the full evaluation population (observed minus generated). Values near zero indicate agreement in feature correlations, whereas positive and negative values indicate stronger correlations in the observed and generated data, respectively. The distribution of differences highlights both preserved relationships and discrepancies in the correlation structure of generated physiological features.}
    \label{fig:difference}
\end{figure}

Across all cohorts, the generated trajectories closely reproduced the correlation structure observed in the original wearable data (Supplementary Figure~\ref{fig:correlation_matrices}, \ref{fig:difference}). Many of the strongest physiological relationships, including interactions between cardiovascular measurements, sleep architecture variables, activity metrics, and stress-related features, were preserved in the generated trajectories. The overall visual similarity between observed and generated correlation matrices suggests that the world model captures underlying physiological dependencies that extend beyond individual feature reconstruction which is a well known property of world models \cite{lecun2022path}. 

The difference matrices further indicate that most pairwise relationships remain stable following trajectory generation. Although localized deviations are present, particularly among a subset of activity- and sleep-related measurements, these changes are small relative to the overall correlation structure. Importantly, preservation of feature relationships was observed not only within the full population but also within sedentary and highly active cohorts, which exhibit substantially different behavioral and physiological profiles. This suggests that the world model learns physiological dynamics that generalize across heterogeneous patterns of activity rather than relying on population-average correlations alone.

These findings provide complementary evidence to the downstream prediction experiments. Whereas classification analyses evaluate whether generated trajectories preserve task-relevant predictive information, correlation analyses assess whether they preserve the broader physiological organization of the wearable feature space. The strong agreement between observed and generated correlation structures suggests that the world model retains much of the higher-order information encoded within longitudinal wearable measurements. Consequently, the generated trajectories appear to preserve both local feature behavior and global physiological structure, supporting their use as realistic simulations of future health trajectories.

\subsection{Expert Evaluation of Counterfactual Coaching}
\label{sec:supp_expert_evaluation}

The main-text expert evaluation compared Counterfactual Coaching with Generic Coaching in a blinded, randomized, paired study of 49 WEAR-ME participant cases reviewed by five certified lifestyle coaches and five board-certified clinicians. Across both expert panels, the study comprised 490 paired plan evaluations. Detailed per-rater, rubric-level, cross-disciplinary, and persona-stratified results are reported below and in Supplementary Tables~\ref{tab:supp_personas}--\ref{tab:supp_rater_breakdown}.



\subsubsection*{Head-to-head preference and anticipated adherence}

Lifestyle coaches completed head-to-head side-by-side comparisons of Counterfactual and Generic Coaching plans for each participant. In the set displayed in Fig.~\ref{fig:main1}e, 245 valid paired comparisons were available. Counterfactual Coaching was selected as the Best Overall Plan in 71.4\% of evaluations (175/245; 95\% CI, 65.5--76.7\%), compared with 28.6\% for Generic Coaching (70/245; 95\% CI, 23.3--34.5\%). This corresponded to a +42.9 percentage-point difference and a 2.50-fold preference ratio (two-sided exact binomial $P = 1.49 \times 10^{-11}$).

Lifestyle coaches also selected Counterfactual Coaching as having the Highest Likelihood of Long-Term Adherence in 71.0\% of displayed evaluations (174/245; 95\% CI, 65.0--76.3\%), compared with 29.0\% for Generic Coaching (71/245; $P = 3.72 \times 10^{-11}$).

Preference was directionally consistent across all five lifestyle coaches. Individual-rater Counterfactual win rates ranged from 53.1\% to 98.0\% for Best Overall Plan and from 44.9\% to 98.0\% for anticipated long-term adherence (Supplementary Table~\ref{tab:supp_rater_breakdown}). When evaluations were aggregated by participant using majority consensus of at least three of five coaches, Counterfactual Coaching was preferred in 43 of 49 cases (87.8\%; $P = 5.73 \times 10^{-8}$) for Best Overall Plan and in 44 of 49 cases (89.8\%; $P = 7.60 \times 10^{-9}$) for Highest Likelihood of Long-Term Adherence.

\subsubsection*{Clinical safety and physiological accuracy}

Clinical safety was independently assessed by five board-certified clinicians across 245 paired evaluations. Counterfactual plans passed the Safe Intensity Limits criterion in 99.2\% of evaluations (243/245; 95\% CI, 97.1--99.8\%), compared with 97.6\% of Generic plans (239/245; 95\% CI, 94.8--98.9\%). The difference was not statistically significant ($\Delta = +1.6$ percentage points; exact McNemar $P = 0.289$; Supplementary Table~\ref{tab:supp_clinician_results}).

Physiological claims were judged correct in 97.1\% of Counterfactual plans (238/245) and 97.6\% of Generic plans (239/245; $P = 1.00$). Clinicians also rated Trajectory Control highly in both conditions (96.7\% [237/245] for Counterfactual and 93.1\% [228/245] for Generic; $\Delta = +3.7$ percentage points; $P = 0.093$). These results indicate that the additional projected physiological information in the Counterfactual condition did not produce a detectable reduction in safety or physiological correctness.

\subsubsection*{Lifestyle coach rubric performance}

We analyzed paired top-box achievement across 10 lifestyle-coach evaluation dimensions (Supplementary Table~\ref{tab:supp_coach_results}). Counterfactual Coaching achieved statistically significant improvements across seven of the ten dimensions.

The largest difference occurred for Contrasts Benefits vs.\ Status Quo. Coaches rated 73.5\% of Counterfactual plans (180/245) as effectively contrasting the projected benefits of behavioral change with maintaining current behavior, compared with 5.7\% of Generic plans (14/245). This corresponded to a +67.8 percentage-point difference, with 170 discordant pairs favoring Counterfactual and 4 favoring Generic (exact McNemar $P = 3.15 \times 10^{-45}$).

Counterfactual Coaching also increased the proportion of plans judged to provide a Compelling Reason to Change without fear- or shame-based framing from 28.6\% (70/245) to 75.5\% (185/245), a +46.9 percentage-point difference (121 versus 6 discordant pairs; $P = 6.39 \times 10^{-29}$). Trajectory Control similarly increased from 59.4\% (145/244) for Generic Coaching to 97.1\% (238/245) for Counterfactual Coaching, with 97 versus 5 discordant pairs ($P = 3.46 \times 10^{-23}$).

On the ordinal behavior-change rubrics, Counterfactual Coaching more than doubled the frequency of Strongly Applied Behavior-Change Strategies, from 20.4\% (50/245) to 50.2\% (123/245; $\Delta = +29.8$ percentage points; 89 versus 16 discordant pairs; $P = 1.87 \times 10^{-13}$). Excellent Sustainable Habit Building increased from 28.6\% (70/245) to 52.2\% (128/245; $\Delta = +23.7$ percentage points; 75 versus 17 discordant pairs; $P = 7.29 \times 10^{-10}$). At the opposite end of the ordinal scales, ``Absent'' behavior-change strategies decreased from 4.1\% (10/245) to 0.4\% (1/245), while ``Poor'' habit building decreased from 3.3\% (8/245) to 1.2\% (3/245).

Counterfactual Coaching was also more frequently judged Realistic and Non-Overwhelming, with top-box ratings increasing from 82.9\% (203/245) to 98.8\% (242/245; $\Delta = +15.9$ percentage points; 41 versus 2 discordant pairs; $P = 2.15 \times 10^{-10}$). Supportive and Motivating Tone increased from 83.3\% (204/245) to 98.4\% (241/245; $\Delta = +15.1$ percentage points; 41 versus 4 discordant pairs; $P = 9.33 \times 10^{-9}$).

Not all dimensions differed. Top-box Signal Synthesis ratings were nearly identical between Counterfactual and Generic Coaching (49.0\% [120/245] versus 48.6\% [119/245]; $P = 1.00$). However, examination of the full ordinal distribution showed that fragmented metric reporting decreased from 19.6\% (48/245) in Generic Coaching to 7.8\% (19/245) in Counterfactual Coaching, with a corresponding increase in cohesive descriptive summaries (43.3\% versus 29.8\%).

Lifestyle-coach ratings of Baseline and Wearable Personalization showed a smaller top-box difference. In the evaluations, 54.3\% of Counterfactual plans (133/245) and 51.8\% of Generic plans (127/245) were rated Highly Tailored ($P = 0.572$). Despite similar top-box rates, bottom-box ``Generic'' ratings decreased from 4.1\% (10/245) for Generic Coaching to 2.4\% (6/245) for Counterfactual Coaching.

Easy to Understand was the only dimension with a numerically higher top-box rate for Generic Coaching: 89.8\% (220/245) compared with 84.9\% (208/245) for Counterfactual Coaching. The difference was not statistically significant ($\Delta = -4.9$ percentage points; 18 versus 30 discordant pairs; $P = 0.111$), but is consistent with a potential trade-off between the simplicity of a single recommendation pathway and the greater numerical density introduced by multiple projected options.

\subsubsection*{Clinician rubric performance}

Clinician evaluations are summarized in Supplementary Table~\ref{tab:supp_clinician_results}. The largest difference was again observed for Contrasts Benefits vs.\ Status Quo. Counterfactual Coaching achieved the top rating in 92.2\% of clinician evaluations (226/245), compared with 7.8\% for Generic Coaching (19/245). This represented a +84.5 percentage-point difference, with 213 discordant pairs favoring Counterfactual and 6 favoring Generic ($P = 3.49 \times 10^{-55}$).

Clinicians also rated Counterfactual plans as Highly Tailored to the participant's physiological profile more often than Generic plans (72.2\% [177/245] versus 52.2\% [128/245]; $\Delta = +20.0$ percentage points; 82 versus 33 discordant pairs; $P = 5.57 \times 10^{-6}$).

Twelve-week Biological Plausibility was rated positively in 86.1\% of Counterfactual plans (211/245) compared with 81.6\% of Generic plans (200/245), corresponding to a +4.5 percentage-point difference (18 versus 7 discordant pairs; $P = 0.043$). The smaller magnitude of this effect relative to the trajectory-framing and personalization measures indicates that the primary clinician-perceived contribution of the counterfactual system was not a generalized increase across all clinical dimensions, but improved linkage between participant-specific physiology, candidate actions, and projected outcomes.

\subsubsection*{Cross-disciplinary concordance and divergence}

Three evaluation dimensions were shared between lifestyle coaches and clinicians, allowing comparison of how the two expert groups interpreted the same plans.

Both groups strongly favored Counterfactual Coaching for Contrasts Benefits vs.\ Status Quo, with a pooled difference of +76.1 percentage points ($P < 10^{-98}$). This was the most consistent cross-disciplinary signal and directly reflected the additional future-versus-future information available to the Counterfactual coach.

The groups differed more substantially in their interpretation of Trajectory Control. Clinicians rated both Generic and Counterfactual plans highly on this dimension (93.1\% and 96.7\%, respectively), reflecting that both conditions provided safe, modifiable, phased recommendations. Lifestyle coaches applied a stricter behavioral interpretation: Generic plans received a top-box rating in 59.4\% of evaluations compared with 97.1\% for Counterfactual plans. Coach feedback indicated that the explicit presentation of multiple behavioral options in Counterfactual plans increased perceived autonomy and choice relative to a single prescribed pathway.

The opposite pattern was observed for Baseline and Wearable Personalization. Clinicians gave Counterfactual Coaching a +20.0 percentage-point top-box advantage (72.2\% versus 52.2\%; $P < 0.001$), whereas lifestyle-coach ratings were more similar (54.3\% versus 51.8\%; $P = 0.572$). Clinicians appeared to place greater weight on quantitative alignment between the plan and physiological drivers such as Health Age and resting heart rate. Lifestyle coaches noted in free-text responses that physiological tailoring alone did not capture important psychosocial constraints, including work schedules, caregiving responsibilities, and access to exercise resources.

\subsubsection*{Performance across WEAR-ME Coachable Personas}

To evaluate whether the observed preference was restricted to a particular participant phenotype, we stratified results across five prespecified WEAR-ME Coachable Personas (Fig.~\ref{fig:human_eval}c; Supplementary Table~\ref{tab:supp_persona_matrix}).

Among Multi-Risk Challenger participants characterized by low activity, insufficient sleep, elevated BMI, and accelerated Health Age, Counterfactual Coaching achieved its highest side-by-side preference. It was selected as the Best Overall Plan in 80.0\% of comparisons (40/50; $P = 2.39 \times 10^{-5}$) and as having the highest likelihood of long-term adherence in 82.0\% (41/50; $P = 5.61 \times 10^{-6}$). The largest rubric differences included Trajectory Control (+48.0 percentage points; 100\% versus 52\%), Behavior-Change Strategies (+32.0 points; 48\% versus 16\%), Sustainable Habits (+26.0 points; 50\% versus 24\%), and Realistic and Manageable Routine Fit (+26.0 points; 100\% versus 74\%). Counterfactual plans more often presented distinct behavioral entry points rather than requiring simultaneous changes across multiple domains.

This subgroup also revealed a calibration limitation. Clinicians rated 12-week Biological Plausibility slightly lower for Counterfactual than Generic plans (76\% versus 82\%; $\Delta = -6.0$ percentage points). Free-text comments indicated that model-projected Health Age reductions of approximately 4--5 years following modest 12-week changes in activity or sleep appeared overly optimistic in some  participant cases.

Among Activity Bottleneck participants with isolated inactivity and preserved sleep, Counterfactual Coaching was preferred in 74.0\% of comparisons (37/50; $P = 0.0009$). This subgroup showed the largest persona-specific gains in coach-rated Trajectory Control (+54.0 percentage points; 100\% versus 46\%) and Behavior-Change Strategies (+34.0 points; 44\% versus 10\%), as well as a +38.0 percentage-point gain in clinician-rated Baseline Personalization (76\% versus 38\%). Clinician feedback highlighted that Counterfactual plans more often preserved already-adequate sleep and concentrated recommendations on progressive activity increases.

Among Optimal Baseline participants with high activity, adequate sleep, and a relatively younger Health Age, Counterfactual Coaching was selected as the Best Overall Plan in 72.0\% of comparisons (36/50; $P = 0.0026$) and as having the highest likelihood of long-term adherence in 70.0\% (35/50; $P = 0.0066$). Counterfactual plans showed higher ratings for Trajectory Control (+34.7 percentage points; 100\% versus 65\%), Sustainable Habits (+28.0 points; 64\% versus 36\%), and clinician-rated 12-week Biological Plausibility (+18.0 points; 98\% versus 80\%). Coaches noted that Generic plans more often prescribed unnecessary increases in activity volume despite already-high baseline activity, whereas Counterfactual plans more often emphasized maintenance and recovery.

Among Balanced Maintainer participants with population-median behavioral profiles, Counterfactual Coaching was selected in 71.1\% of comparisons (32/45; $P = 0.0066$) for both Best Overall Plan and anticipated long-term adherence. This subgroup showed the largest persona-specific improvement in Behavior-Change Strategies (+42.2 percentage points; 58\% versus 16\%) and a +26.7 percentage-point increase in clinician-rated Baseline Personalization (78\% versus 51\%).

The smallest preference margin occurred among Recovery Bottleneck participants characterized by high activity and insufficient sleep. Counterfactual Coaching was preferred in 60.0\% of valid displayed comparisons (30/50) compared with 40.0\% for Generic Coaching. Clinicians nevertheless rated Counterfactual plans as more physiologically personalized by +18.0 percentage points (70\% versus 52\%), in part because the plans more often identified sleep extension as the principal modifiable pathway for already-active participants.

Lifestyle coaches were less favorable in this subgroup, with Baseline and Wearable Personalization ratings 14.0 percentage points lower for Counterfactual Coaching (44\% versus 58\%). Free-text feedback suggested that some Counterfactual plans continued to include additional step increases for participants already averaging approximately 10,000--14,000 steps per day because their historical peaks exceeded 16,000 steps per day. Coaches viewed these additional activity options as diluting the more appropriate recovery-focused recommendation.

Together, the persona-stratified analyses indicate that the preference for Counterfactual Coaching was not confined to participants with a single behavioral deficit or elevated Health Age. The analysis also identified two recurrent limitations: redundant recommendations in behavioral domains that were already near saturation, and potentially over-optimistic short-horizon Health Age projections in some high-risk participants.
\clearpage
\section{Supplementary Figures}
\label{supp:methods}

\subsection{Coachable Health Age}

\begin{figure}[h!]
    \centering
    \includegraphics[width=\linewidth]{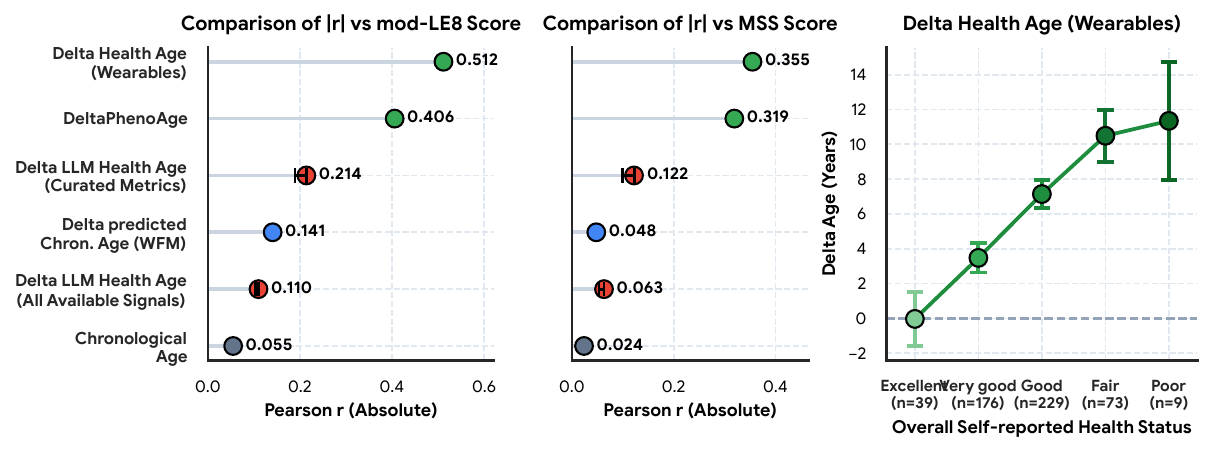}
    \caption{\textbf{Associations Between Coachable Health Age, Cardiometabolic Health, and Self-Reported Health Status.} Absolute Pearson correlation coefficients ($|r|$) comparing Coachable Health Age and alternative age estimates against the modified American Heart Association Life's Essential 8 score (mod-LE8) and Metabolic Syndrome Severity Score (MSSS), alongside wearable-derived Delta Health Age across self-reported health categories, with error bars indicating uncertainty.}
    \label{fig:le8}
\end{figure}

\begin{figure}[h!]
    \centering
    \includegraphics[width=\linewidth]{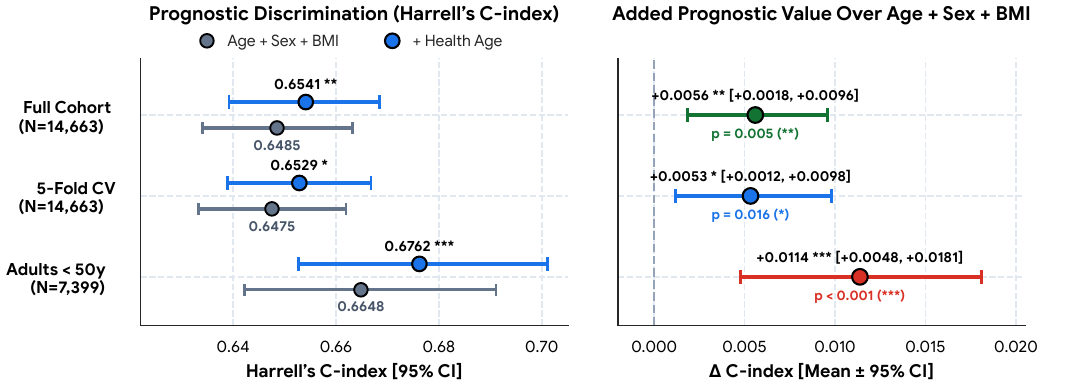}
    \caption{\textbf{Health Age provides significant prognostic discrimination beyond age, sex, and BMI in the \textit{All of Us} dataset ($N = 14{,}663$).} Harrell's C-index \textbf{(left)} and $\Delta$C-index \textbf{(right)} (with 95\% CIs) show consistent gains in the full cohort ($+0.0056^{**}$) and 5-fold cross-validation ($+0.0053^{*}$), with the strongest improvement in adults $< 50$ years ($+0.0114^{***}$, $N = 7{,}399$). $^{*}p < 0.05$, $^{**}p < 0.01$, $^{***}p \le 0.001$.}
    \label{fig:c-index}
\end{figure}

\begin{figure}[h!]
    \centering
    \includegraphics[width=0.7\linewidth]{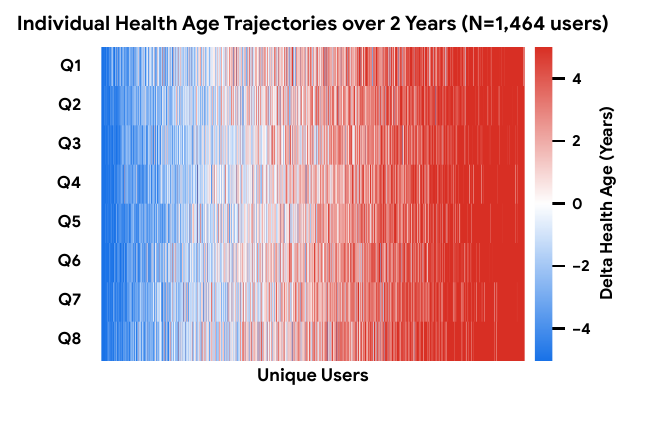}
    \caption{\textbf{Individual trajectories of Coachable Health Age over two years.} Heatmap showing longitudinal variation in Health Age Delta ($\Delta$CHA; Coachable Health Age minus chronological age) across 1,464
    participants over eight consecutive 3-month windows (Q1–Q8). Each column represents an individual participant, and each row corresponds to a 3-month observation window. Participants are ordered by baseline Health Age Delta (Q1). Colors indicate the magnitude and direction of $\Delta$CHA, with blue denoting a younger physiological health age relative to chronological age ($\Delta$CHA $<$ 0) and red denoting an older physiological health age ($\Delta$CHA $>$ 0). Temporal changes within columns illustrate individual variability in physiological and behavioral health, including trajectories of improvement, deterioration, and relative stability over the two-year observation period.}
    \label{fig:aou-longitudinal}
\end{figure}

\clearpage
\subsection{Trajectory Prediction with Sensor World Model}

\begin{figure}[h!]
    \centering
    \includegraphics[width=0.95\linewidth]{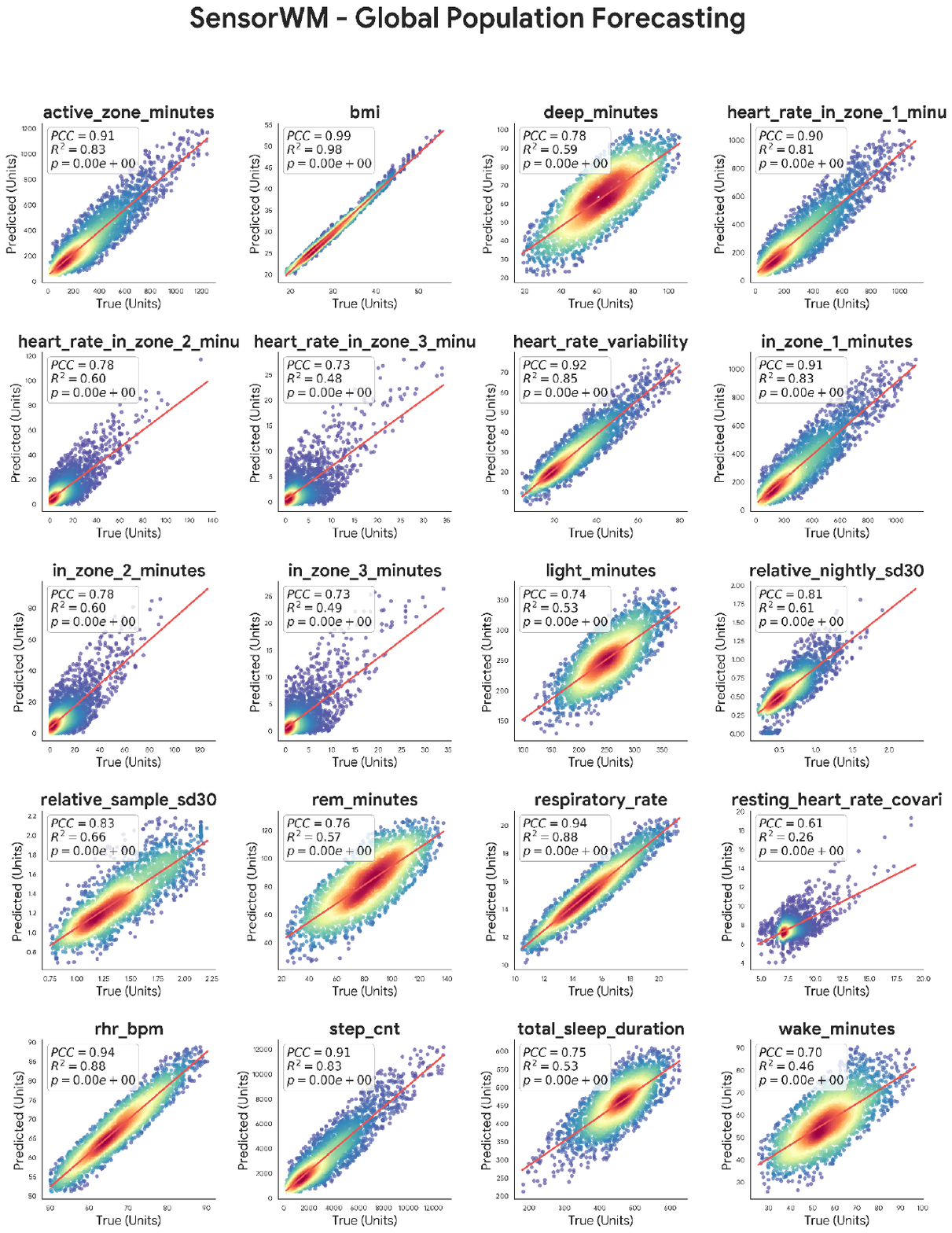}
    \caption{\textbf{Population-level forecasting performance of SensorWM.}
Comparison of observed and predicted wearable-derived physiological and behavioral measurements across the evaluation cohort. Each panel shows the relationship between ground-truth and model-predicted values for an individual feature, with point density indicated by color and the red diagonal representing perfect agreement. Pearson correlation coefficient (PCC), coefficient of determination ($R^2$), and associated $p$-values are reported for each feature.}
    \label{fig:sesnitivity}
\end{figure}

\begin{figure}[h!]
    \centering
    \includegraphics[width=0.87\linewidth]{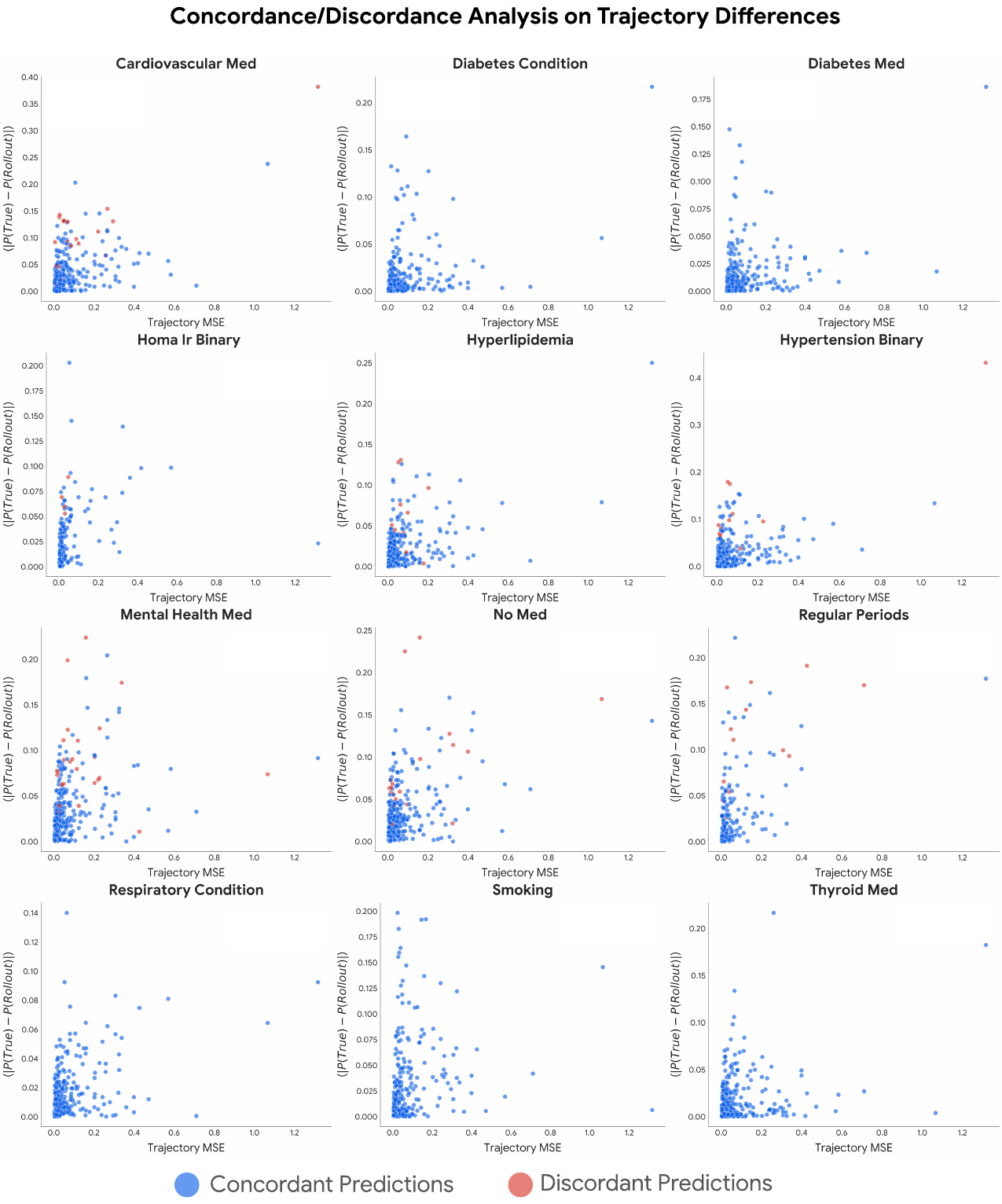}
    \caption{\textbf{Trajectory forecasting error and downstream prediction sensitivity across health outcomes.} Relationship between trajectory forecasting error and downstream prediction consistency across clinical and behavioral tasks. Each point corresponds to an individual participant and is plotted according to the trajectory reconstruction error (MSE) between the observed and world-model-generated wearable trajectories (x-axis) and the absolute difference in predicted probability between classifiers trained on true and generated trajectories (y-axis). Blue points indicate concordant classifications, whereas red points indicate discordant classifications. Across most tasks, prediction differences remain small despite variation in trajectory reconstruction error, suggesting that downstream health predictions are relatively robust to moderate forecasting inaccuracies.}
    \label{fig:sesnitivity}
\end{figure}

\begin{figure*}[t]
    \centering
    \textbf{(a) Observed Fitbit Features}

    \vspace{0.5em}
    \includegraphics[width=\textwidth]{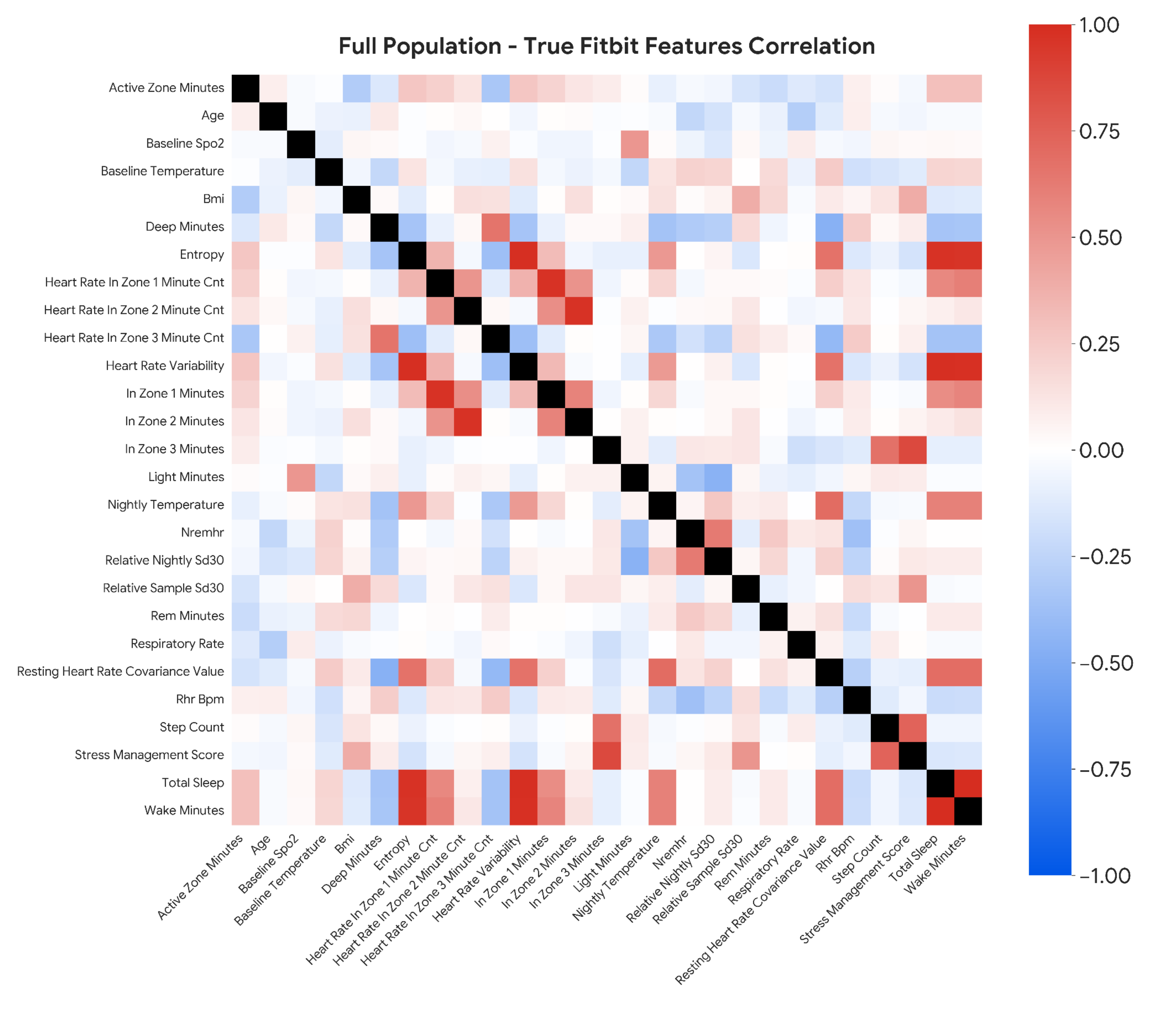}

    \caption{
    \textbf{Pairwise correlation structure of observed and SensorWM-generated wearable features.}
    \textbf{(a)} Pearson correlation matrix computed from observed Fitbit measurements across the study population. Each cell represents the correlation between two physiological or behavioral features, illustrating the underlying inter-feature dependencies in real wearable data.
    }
    \label{fig:correlation_matrices}
\end{figure*}

\begin{figure*}[t]
    \ContinuedFloat
    \centering
    \textbf{(b) SensorWM-Generated Features}

    \vspace{0.5em}
    \includegraphics[width=\textwidth]{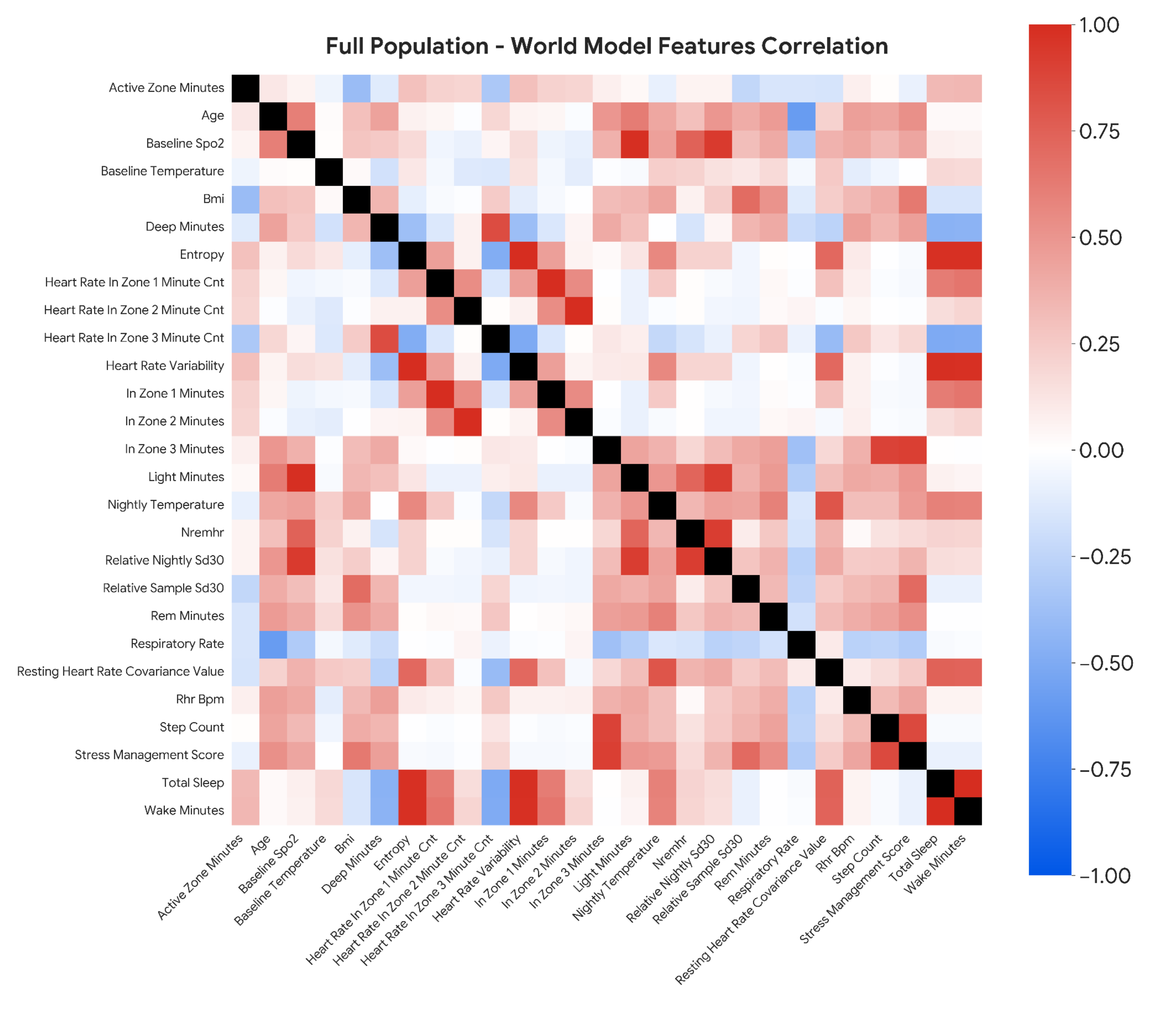}

    \caption{
    \textbf{Pairwise correlation structure of observed and SensorWM-generated wearable features (continued).}
    \textbf{(b)} Pearson correlation matrix computed from SensorWM-generated wearable measurements. Comparison with panel (a) illustrates the extent to which the model preserves positive and negative inter-feature associations observed in real Fitbit data, providing an assessment of the learned joint physiological feature distribution.
    }
    \label{fig:correlation_matrices_generated}
\end{figure*}

\clearpage
\section{Supplementary Tables}
\label{supp:data}

\subsection{Coachable Health Age}

\begin{table}[htbp]
\centering
\small
\setlength{\tabcolsep}{4pt}
\caption{\textbf{Unadjusted literature hazard ratios, WEAR-ME multicollinearity weights ($w_i$), and resulting collinearity-adjusted hazard ratios ($\mathrm{HR}^{\mathrm{adj}}_i$) for the four core Coachable Health Age domains.}}
\label{tab:supp_cha_parameters}
\begin{tabular*}{\textwidth}{@{\extracolsep{\fill}}llcccc}
\toprule
\textbf{Core} & \textbf{Knot / Slope} & \textbf{Reference} & \textbf{Unadjusted} & \textbf{WEAR-ME} & \textbf{Adjusted} \\
\textbf{Domain} & \textbf{Parameter} & \textbf{($\mathrm{HR}=1.0$)} & \textbf{$\mathrm{HR}_i$} & \textbf{Weight ($w_i$)} & \textbf{$\mathrm{HR}^{\mathrm{adj}}_i = \mathrm{HR}_i^{w_i}$} \\
\midrule
\multirow{2}{*}{Resting heart rate}
  & Knot at $70\text{ bpm}$ & $\leq 60\text{ bpm}$ & $1.08$ & $0.7066$ & $1.0559$ \\
  & Knot at $90\text{ bpm}$ & $\leq 60\text{ bpm}$ & $1.33$ & $0.7066$ & $1.2232$ \\
\midrule
Daily steps
  & Per $1,000\text{ steps}$ (cap $9\text{k}$) & $8,000\text{ steps/d}$ & $0.88$ & $0.4773$ & $0.9408$ \\
\midrule
\multirow{2}{*}{Sleep duration}
  & Per hour $<6.0\text{ h}$ & $6.0\text{--}9.0\text{ h/night}$ & $1.20$ & $0.8790$ & $1.1738$ \\
  & Per hour $>9.0\text{ h}$ & $6.0\text{--}9.0\text{ h/night}$ & $1.34$ & $0.8790$ & $1.2934$ \\
\midrule
\multirow{3}{*}{Weekly AZM}
  & Knot at $245\text{ min/wk}$ & $<115\text{ min/wk}$ & $0.76$ & $0.6950$ & $0.8263$ \\
  & Knot at $418\text{ min/wk}$ & $<115\text{ min/wk}$ & $0.71$ & $0.6950$ & $0.7881$ \\
  & Plateau $\geq 500\text{ min/wk}$ & $<115\text{ min/wk}$ & $0.68$ & $0.6950$ & $0.7648$ \\
\bottomrule
\end{tabular*}
\end{table}

\begin{table*}[h!]
\caption{
\textbf{Comparison of design characteristics across alternative age formulations.}
Approaches are summarized according to characteristics relevant to longitudinal wearable health assessment and behavioral guidance. The comparison reflects design considerations motivating Coachable Health Age rather than an empirical comparison of model performance.
}
\label{tab:modeling_comparison_5row}

\resizebox{\textwidth}{!}{%
\begin{tabular}{lccccccc}
\toprule
\textbf{Modeling Type} &
\textbf{\makecell{Predictable\\from Wearables}} &
\textbf{Explainable} &
\textbf{\makecell{Feature\\Flexibility}} &
\textbf{Engageable} &
\textbf{Coachable} &
\textbf{Movable} &
\textbf{Grounded} \\
\midrule

Chronological Age &
\xmark &
\xmark &
\xmark &
\xmark &
\xmark &
\xmark &
\xmark \\

\addlinespace

\rowcolor{highlightgray}
\makecell[l]{Predicted Chronological Age\\(SensorFM or PPG-FM)} &
\cmark &
\pmark &
\cmark &
\pmark &
\pmark &
\pmark &
\xmark \\

\addlinespace

PhenoAge &
\xmark &
\cmark &
\cmark &
\pmark &
\pmark &
\pmark &
\cmark \\

\addlinespace

\rowcolor{highlightgray}
LLM-generated Health Age &
\xmark &
\xmark &
\xmark &
\xmark &
\xmark &
\xmark &
\xmark \\

\addlinespace

\makecell[l]{Coachable Health Age\\(wearable features)} &
\cmark &
\cmark &
\cmark &
\cmark &
\cmark &
\cmark &
\cmark \\

\bottomrule
\end{tabular}%
}

\parbox{\textwidth}{
\footnotesize
\cmark~indicates that a characteristic is supported by the formulation,
\pmark~indicates limited or indirect support, and
\xmark~indicates that the characteristic is not supported as defined for the intended longitudinal coaching use case.
The categories summarize design considerations and are not empirical performance measures.
}
\end{table*}




\begin{table}[h]
\centering
\caption{\textbf{Reliability and Consistency of the Wearable Health Age Estimator Across Generation Temperatures.} Comparison of reliability and consistency metrics across different generation temperatures, evaluating the stability and reproducibility of Wearable Health Age estimates under varying sampling conditions.}
\label{tab:consistency_results}
\resizebox{\columnwidth}{!}{%
\begin{tabular}{lcccc}
\toprule
\textbf{Input Configuration} & Temperature ($T$) & Average SD (years) & Single-Run Consistency ($ICC(3,1)$) & Average Consistency ($ICC(3,k)$) \\
\midrule
\textbf{Identical Metrics as Health Age} & 0.1 & 2.72 & 0.54 & 0.78 \\
                           & 0.5 & 3.54 & 0.09 & 0.23 \\
                           & 1.0 & 3.20 & 0.11 & 0.28 \\
\midrule
\textbf{All Raw Metrics}   & 0.1 & 1.35 & 0.54 & 0.78 \\
                           & 0.5 & 2.06 & 0.30 & 0.57 \\
                           & 1.0 & 3.37 & 0.20 & 0.42 \\
\bottomrule
\end{tabular}%
}
\end{table}

\begin{table}[h]
\centering
\caption{\textbf{Bonferroni-Corrected Post-Hoc Comparisons Across Generation Temperatures.} Adjusted p-values from pairwise comparisons across generation temperatures, accounting for multiple hypothesis testing to assess statistically significant differences in estimator performance.}
\label{tab:across_temp}
\begin{tabular}{lcc}
\hline
\textbf{Comparison} & \textbf{Raw Signals (adj. $p$)} & \textbf{Curated Metrics (adj. $p$)} \\ \hline
$T_{0.1}$ vs $T_{0.5}$ & 0.0751                          & 0.0039                              \\
$T_{0.5}$ vs $T_{1.0}$ & 0.0689                          & 0.0024                              \\
$T_{0.1}$ vs $T_{1.0}$ & 1.0000                          & $7.71 \times 10^{-10}$               \\ \hline
\end{tabular}
\end{table}

\begin{table}[h]
\centering
\caption{\textbf{Correlation Between Delta LLM Health Age and Clinical Metrics.} Pearson correlation coefficients ($r$) quantifying the strength and direction of linear associations between delta LLM health age and clinical metrics, assessing the clinical alignment of wearable-derived LLM health age estimates.}
\label{tab:correlations}
\begin{tabular}{lccc}
\hline
\textbf{Input Configuration} & \textbf{Temp ($T$)} & \textbf{Modified LE8 ($r$)} & \textbf{MSSS ($r$)} \\ \hline
\textbf{All Available Signals}         & 0.1                 & -0.110                      & 0.063              \\
                                       & 0.5                 & -0.103                      & 0.061              \\
                                       & 1.0                 & -0.105                      & 0.053              \\ \hline
\textbf{Curated Metrics as Health Age} & 0.1                 & \textbf{-0.214}             & 0.122              \\
                                       & 0.5                 & -0.189                      & 0.099              \\
                                       & 1.0                 & -0.192                      & \textbf{0.123}     \\ \hline
\end{tabular}
\end{table}


\clearpage
\subsection{Trajectory Prediction with Sensor World Model}

\begin{longtable}{>{\raggedright\arraybackslash}p{0.30\textwidth}
                  >{\raggedright\arraybackslash}p{0.12\textwidth}
                  >{\small\raggedright\arraybackslash}p{0.43\textwidth}
                  c}
\caption{\textbf{State and action features grouped by physiological and behavioral domain.}
State variables represent physiological, demographic, and behavioral status variables used by the world model, while action variables represent observable behaviors that may influence future physiological states. Only \texttt{step\_cnt}, \texttt{cardio\_minutes}, and \texttt{sleep\_main\_asleep\_minute\_cnt} are treated as action variables.}
\label{tab:state_action_features}\\

\toprule
\textbf{Feature} & \textbf{Group} & \textbf{Description} & \textbf{Map} \\
\midrule
\endfirsthead

\toprule
\textbf{Feature} & \textbf{Group} & \textbf{Description} & \textbf{Map} \\
\midrule
\endhead

\midrule
\multicolumn{4}{r}{\emph{Continued on next page}} \\
\midrule
\endfoot

\bottomrule
\endlastfoot

\rowcolor{gray!20}
\multicolumn{4}{l}{\textbf{\DEMLabel\ Demographics}} \\

Age
& \DEMLabel
& Chronological age in years.
& \StateLabel \\

Gender
& \DEMLabel
& Self-reported sex or gender.
& \StateLabel \\

BMI
& \DEMLabel
& Body mass index.
& \StateLabel \\

\rowcolor{gray!20}
\multicolumn{4}{l}{\textbf{\CRDLabel\ Cardiovascular}} \\

Heart Rate Variability
& \CRDLabel
& Beat-to-beat variability in heart rhythm.
& \StateLabel \\

Resting Heart Rate BPM
& \CRDLabel
& Resting heart rate in beats per minute.
& \StateLabel \\

Resting Heart Rate Covariance
& \CRDLabel
& Covariance-based variation in rhr.
& \StateLabel \\

Heart Rate in Zone 1 Minutes
& \CRDLabel
& Heart rate recorded in zone 1.
& \StateLabel \\

Heart Rate in Zone 2 Minutes
& \CRDLabel
& Heart rate recorded in zone 2.
& \StateLabel \\

Heart Rate in Zone 3 Minutes
& \CRDLabel
& Heart rate recorded in zone 3.
& \StateLabel \\

\rowcolor{gray!20}
\multicolumn{4}{l}{\textbf{\PLMLabel\ Cardiopulmonary}} \\

Respiratory Rate
& \PLMLabel
& Average number of breaths per minute.
& \StateLabel \\

Baseline SpO$_2$
& \PLMLabel
& Baseline peripheral blood oxygen saturation.
& \StateLabel \\



\rowcolor{gray!20}
\multicolumn{4}{l}{\textbf{\SLPLabel\ Sleep}} \\

Deep Minutes
& \SLPLabel
& Number of minutes spent in deep sleep.
& \StateLabel \\

Light Minutes
& \SLPLabel
& Number of minutes spent in light sleep.
& \StateLabel \\

REM Minutes
& \SLPLabel
& Number of minutes spent in REM sleep.
& \StateLabel \\

Wake Minutes
& \SLPLabel
& Number of wake minutes during sleep.
& \StateLabel \\

Main Sleep Asleep Minute Count
& \SLPLabel
& Number of minutes asleep.
& \ActionLabel \\


\rowcolor{gray!20}
\multicolumn{4}{l}{\textbf{\MTNLabel\ Motion and Physical Activity}} \\

In Zone 1 Minutes
& \MTNLabel
& Number of activity minutes in zone 1.
& \StateLabel \\

In Zone 2 Minutes
& \MTNLabel
& Number of activity minutes in zone 2.
& \StateLabel \\

In Zone 3 Minutes
& \MTNLabel
& Number of activity minutes in zone 3.
& \StateLabel \\


Step Count
& \MTNLabel
& Step count recorded.
& \ActionLabel \\


Cardio Minutes
& \MTNLabel
& Number of minutes spent in the cardio-intensity  zone.
& \ActionLabel \\

\rowcolor{gray!20}
\multicolumn{4}{l}{\textbf{\VARLabel\ Regularity and Variability}} \\

Entropy
& \VARLabel
& Entropy-based measure of temporal regularity and predictability.
& \StateLabel \\

DeepRMSSD
& \VARLabel
& Root-mean-square successive-difference statistic computed for the corresponding deep or nightly signal.
& \StateLabel \\

Relative Nightly SD30
& \VARLabel
& Relative night-to-night variability estimated over a 30-day window.
& \StateLabel \\

Relative Sample SD30
& \VARLabel
& Relative sample-level variability estimated over a 30-day window.
& \StateLabel \\

\end{longtable}


\clearpage

\begin{table*}[h!]
\centering
\small
\setlength{\tabcolsep}{4pt}
\renewcommand{\arraystretch}{1.15}

\begin{adjustbox}{width=\textwidth}
\begin{tabular}{llcccc}
\toprule
\textbf{Prediction Task} &
\textbf{Definition} &
\textbf{Type} &
\textbf{Source} &
\textbf{Metric} &
\textbf{N (Pos\%)} \\
\midrule

\rowcolor{gray!20}
\multicolumn{6}{l}{\textbf{\textit{Health Conditions}}} \\

Diabetes Condition &
Diagnosis of diabetes condition &
\CLSLabel & \SLFLabel & \AUCLabel &
1,519 (8.6\%) \\

HOMA-IR &
Classification of HOMA-IR $\geq$ 2.9 &
\CLSLabel & \LABLabel & \AUCLabel &
775 (31.5\%) \\

Hyperlipidemia &
Diagnosis of Hyperlipidemia &
\CLSLabel & \LABLabel & \AUCLabel &
1,519 (21.7\%) \\

Hypertension Binary &
Diagnosis of hypertension &
\CLSLabel & \SLFLabel & \AUCLabel &
1,519 (23.6\%) \\


Respiratory Condition &
Diagnosis of respiratory condition &
\CLSLabel & \SLFLabel & \AUCLabel &
1,519 (17.1\%) \\

\midrule
\rowcolor{gray!20}
\multicolumn{6}{l}{\textbf{\textit{Behavioral}}} \\

Smoking &
Smoking behavior &
\CLSLabel & \SLFLabel & \AUCLabel &
1,495 (8.1\%) \\

\midrule
\rowcolor{gray!20}
\multicolumn{6}{l}{\textbf{\textit{Women's Health}}} \\

Regular Periods &
Self-reported regular menstrual cycles &
\CLSLabel & \SLFLabel & \AUCLabel &
871 (41.4\%) \\

\midrule
\rowcolor{gray!20}
\multicolumn{6}{l}{\textbf{\textit{Medication}}} \\

Cardiovascular Medication &
Use of cardiovascular medication &
\CLSLabel & \SLFLabel & \AUCLabel &
1,654 (33.0\%) \\

Diabetes Medication &
Use of diabetes medication &
\CLSLabel & \SLFLabel & \AUCLabel &
1,654 (8.7\%) \\

Mental Health Medication &
Use of depression/anxiety medication &
\CLSLabel & \SLFLabel & \AUCLabel &
1,654 (27.4\%) \\

No Medication &
No regular medication use in daily life &
\CLSLabel & \SLFLabel & \AUCLabel &
1,654 (38.8\%) \\

Thyroid Medication &
Use of thyroid medication &
\CLSLabel & \SLFLabel & \AUCLabel &
1,654 (9.9\%) \\

\bottomrule
\end{tabular}
\end{adjustbox}

\caption{
\textbf{Downstream prediction tasks grouped into categories.} Source indicates whether labels originate from self-reported questionnaires (SLF) or laboratory measurements (LAB). $N$ denotes the number of users with valid labels and Pos\% denotes positive prevalence among valid users.
}
\label{tab:downstream_task_summary}
\end{table*}

\begin{table*}[h!]
\centering

\begin{minipage}[t]{0.48\textwidth}
\centering
\caption{\textbf{Demographic characteristics of the pre-training cohort.}
Distribution of age, body mass index (BMI), and sex among participants in the Fitbit pre-training cohort.}
\label{tab:pretraining_demographics}
\small
\begin{tabular}{lcc}
\toprule
\textbf{Category} & \textbf{N} & \textbf{\%} \\
\midrule

\rowcolor{gray!15}
\multicolumn{3}{l}{\textbf{Age}} \\
<40      & 54,050 & 55.8 \\
40--59   & 27,514 & 28.3 \\
60+      & 15,393 & 15.9 \\

\rowcolor{gray!15}
\multicolumn{3}{l}{\textbf{BMI}} \\
<25      & 47,947 & 49.5 \\
25--30   & 27,481 & 28.3 \\
>30      & 21,505 & 22.2 \\
Unknown  & 24 & 0.0 \\

\rowcolor{gray!15}
\multicolumn{3}{l}{\textbf{Sex}} \\
Male     & 49,294 & 50.8 \\
Female   & 47,651 & 49.2 \\
Unknown  & 12   & 0.0 \\

\midrule
\textbf{Total Users} & \textbf{96,957} & \textbf{100.0} \\
\bottomrule
\end{tabular}
\end{minipage}
\hfill
\begin{minipage}[t]{0.48\textwidth}
\centering
\caption{\textbf{Demographic characteristics of the WEAR-ME evaluation cohort.}
Distribution of age, body mass index (BMI), and sex among participants with available downstream clinical and behavioral labels.}
\label{tab:wearme_demographics}
\small
\begin{tabular}{lcc}
\toprule
\textbf{Category} & \textbf{N} & \textbf{\%} \\
\midrule

\rowcolor{gray!15}
\multicolumn{3}{l}{\textbf{Age}} \\
<40      & 398 & 24.1 \\
40--59   & 937 & 56.7 \\
60+      & 319 & 19.3 \\

\rowcolor{gray!15}
\multicolumn{3}{l}{\textbf{BMI}} \\
<25      & 478 & 28.9 \\
25--30   & 499 & 30.2 \\
>30      & 677 & 40.9 \\

\rowcolor{gray!15}
\multicolumn{3}{l}{\textbf{Sex}} \\
Male     & 659 & 39.8 \\
Female   & 993 & 60.0 \\
Unknown  & 2   & 0.1 \\

\midrule
\textbf{Total Users} & \textbf{1,654} & \textbf{100.0} \\
\bottomrule
\end{tabular}
\end{minipage}

\end{table*}

\begin{table*}[h!]
\centering
\caption{\textbf{Downstream task splits.} Each row is a prediction task, and each split reports the number of valid samples, the number of positive samples, and the positive prevalence among valid samples.}
\small
\setlength{\tabcolsep}{4pt}
\renewcommand{\arraystretch}{1.15}

\begin{adjustbox}{width=0.7\textwidth}
\begin{tabular}{lccc ccc ccc}
\toprule
\textbf{Prediction Task} &
\multicolumn{3}{c}{\textbf{Train}} &
\multicolumn{3}{c}{\textbf{Val}} &
\multicolumn{3}{c}{\textbf{Test}} \\
\cmidrule(lr){2-4}
\cmidrule(lr){5-7}
\cmidrule(lr){8-10}
& \textbf{N} & \textbf{Pos} & \textbf{Pos\%}
& \textbf{N} & \textbf{Pos} & \textbf{Pos\%}
& \textbf{N} & \textbf{Pos} & \textbf{Pos\%} \\
\midrule

\rowcolor{gray!20}
\multicolumn{10}{l}{\textbf{\textit{Health Conditions}}} \\


Diabetes Condition &
899 & 85 & 9.5\% &
315 & 20 & 6.3\% &
305 & 26 & 8.5\% \\

HOMA-IR Binary &
470 & 150 & 31.9\% &
164 & 56 & 34.1\% &
141 & 38 & 27.0\% \\

Hyperlipidemia &
899 & 194 & 21.6\% &
315 & 76 & 24.1\% &
305 & 60 & 19.7\% \\

Hypertension Binary &
899 & 214 & 23.8\% &
315 & 82 & 26.0\% &
305 & 63 & 20.7\% \\


Respiratory Condition &
899 & 147 & 16.4\% &
315 & 56 & 17.8\% &
305 & 57 & 18.7\% \\

\midrule
\rowcolor{gray!20}
\multicolumn{10}{l}{\textbf{\textit{Behavioral}}} \\

Smoking &
884 & 69 & 7.8\% &
311 & 30 & 9.6\% &
300 & 22 & 7.3\% \\

\midrule
\rowcolor{gray!20}
\multicolumn{10}{l}{\textbf{\textit{Women's Health}}} \\


Regular Periods &
516 & 216 & 41.9\% &
188 & 79 & 42.0\% &
167 & 66 & 39.5\% \\

\midrule
\rowcolor{gray!20}
\multicolumn{10}{l}{\textbf{\textit{Medication}}} \\

Cardiovascular Medication &
992 & 326 & 32.9\% &
331 & 118 & 35.6\% &
331 & 102 & 30.8\% \\

Diabetes Medication &
992 & 85 & 8.6\% &
331 & 27 & 8.2\% &
331 & 32 & 9.7\% \\

Mental Health Medication &
992 & 257 & 25.9\% &
331 & 99 & 29.9\% &
331 & 97 & 29.3\% \\

No Medication &
992 & 388 & 39.1\% &
331 & 123 & 37.2\% &
331 & 131 & 39.6\% \\

Thyroid Medication &
992 & 92 & 9.3\% &
331 & 44 & 13.3\% &
331 & 28 & 8.5\% \\

\bottomrule
\end{tabular}
\end{adjustbox}
\label{tab:downstream_task_splits}
\end{table*}

\clearpage

\begin{table*}[h]
\centering
\caption{\textbf{Forecasting Performance of the Sensor World Model ($R^2$).} Forecasting accuracy is evaluated using the coefficient of determination ($R^2$), with results reported as mean $\pm$ standard error across evaluation runs.}
\label{tab:fitbit_worldmodel_forecasting_r2}
\begin{adjustbox}{width=\textwidth}
\begin{tabular}{lcc|cccccc}
\toprule
\textbf{Prediction Task} &
\textbf{Type} &
\textbf{Metric} &
\multicolumn{6}{c}{\textbf{Feature Source}} \\
\cmidrule(lr){4-9}
&
&
&
\textbf{1 Week Mean} &
\textbf{2 Week Mean} &
\textbf{4 Week Mean} &
\textbf{LOCF} &
\textbf{TimesFM} &
\textbf{SensorWM} \\
\midrule
\rowcolor{gray!20}
\multicolumn{9}{l}{\textbf{\textit{Forecasting}}} \\
active\_zone\_minutes &
\ForLabel &
\RSquaredLabel &
0.4937 $\pm$ 0.0025 &
0.6465 $\pm$ 0.0020 &
\underline{0.7166 $\pm$ 0.0017} &
0.7129 $\pm$ 0.0017 &
0.6355 $\pm$ 0.0047 &
\textbf{0.8260 $\pm$ 0.0012} \\
bmi &
\ForLabel &
\RSquaredLabel &
\textbf{1.0000 $\pm$ 0.0000} &
\textbf{1.0000 $\pm$ 0.0000} &
\textbf{1.0000 $\pm$ 0.0000} &
\textbf{1.0000 $\pm$ 0.0000} &
0.9352 $\pm$ 0.0010 &
\underline{0.9835 $\pm$ 0.0001} \\
deep\_minutes &
\ForLabel &
\RSquaredLabel &
-0.4998 $\pm$ 0.0000 &
0.1603 $\pm$ 0.0023 &
\underline{0.3917 $\pm$ 0.0026} &
0.3302 $\pm$ 0.0055 &
0.3421 $\pm$ 0.0062 &
\textbf{0.5929 $\pm$ 0.0023} \\
heart\_rate\_in\_zone\_1\_minute\_cnt &
\ForLabel &
\RSquaredLabel &
0.4006 $\pm$ 0.0026 &
0.5808 $\pm$ 0.0022 &
\underline{0.6699 $\pm$ 0.0019} &
0.6681 $\pm$ 0.0043 &
0.6215 $\pm$ 0.0048 &
\textbf{0.8143 $\pm$ 0.0012} \\
heart\_rate\_in\_zone\_2\_minute\_cnt &
\ForLabel &
\RSquaredLabel &
0.1016 $\pm$ 0.0020 &
0.4582 $\pm$ 0.0025 &
0.5392 $\pm$ 0.0023 &
\underline{0.5981 $\pm$ 0.0036} &
0.5468 $\pm$ 0.0054 &
\textbf{0.6015 $\pm$ 0.0023} \\
\bottomrule
\end{tabular}
\end{adjustbox}
\end{table*}


\begin{table*}[h!]
\centering
\caption{\textbf{Forecasting Performance of the Sensor World Model (r).} Forecasting accuracy is evaluated using the Pearson correlation coefficient ($r$), with results reported as mean $\pm$ standard error across evaluation runs.}
\label{tab:fitbit_worldmodel_forecasting_pearson}
\begin{adjustbox}{width=\textwidth}
\begin{tabular}{lcc|cccccc}
\toprule
\textbf{Prediction Task} &
\textbf{Type} &
\textbf{Metric} &
\multicolumn{6}{c}{\textbf{Feature Source}} \\
\cmidrule(lr){4-9}
&
&
&
\textbf{1 Week Mean} &
\textbf{2 Week Mean} &
\textbf{4 Week Mean} &
\textbf{LOCF} &
\textbf{TimesFM} &
\textbf{SensorWM} \\
\midrule
\rowcolor{gray!20}
\multicolumn{9}{l}{\textbf{\textit{Forecasting}}} \\
active\_zone\_minutes &
\ForLabel &
\PearsonLabel &
0.7791 $\pm$ 0.0022 &
0.8287 $\pm$ 0.0019 &
0.8565 $\pm$ 0.0018 &
\underline{0.8583 $\pm$ 0.0018} &
0.7985 $\pm$ 0.0049 &
\textbf{0.9098 $\pm$ 0.0015} \\
bmi &
\ForLabel &
\PearsonLabel &
\textbf{1.0000 $\pm$ 0.0000} &
\textbf{1.0000 $\pm$ 0.0000} &
\textbf{1.0000 $\pm$ 0.0000} &
\textbf{1.0000 $\pm$ 0.0000} &
0.9693 $\pm$ 0.0020 &
\underline{0.9942 $\pm$ 0.0004} \\
deep\_minutes &
\ForLabel &
\PearsonLabel &
0.5362 $\pm$ 0.0029 &
0.6379 $\pm$ 0.0027 &
\underline{0.6970 $\pm$ 0.0025} &
0.6713 $\pm$ 0.0026 &
0.5966 $\pm$ 0.0065 &
\textbf{0.7769 $\pm$ 0.0023} \\
heart\_rate\_in\_zone\_1\_minute\_cnt &
\ForLabel &
\PearsonLabel &
0.7442 $\pm$ 0.0023 &
0.7997 $\pm$ 0.0021 &
0.8332 $\pm$ 0.0019 &
\underline{0.8370 $\pm$ 0.0019} &
0.7913 $\pm$ 0.0050 &
\textbf{0.9034 $\pm$ 0.0016} \\
heart\_rate\_in\_zone\_2\_minute\_cnt &
\ForLabel &
\PearsonLabel &
0.6831 $\pm$ 0.0025 &
0.7336 $\pm$ 0.0024 &
0.7515 $\pm$ 0.0023 &
\textbf{0.7939 $\pm$ 0.0021} &
0.7543 $\pm$ 0.0053 &
\underline{0.7842 $\pm$ 0.0023} \\
\bottomrule
\end{tabular}
\end{adjustbox}
\end{table*}

\begin{table*}[h!]
\centering
\caption{
\textbf{Trajectory Based versus Static Downstream Classification Performance.} Classification performance comparing downstream prediction using true Fitbit features against Trajectory Transformer and LeWorldModel. Values are AUROC with 95\% confidence intervals.
}
\small
\setlength{\tabcolsep}{5pt}
\renewcommand{\arraystretch}{1.18}

\begin{adjustbox}{width=\textwidth}
\begin{tabular}{lccc|c|cc}
\toprule
\textbf{Prediction Task} &
\textbf{Type} &
\textbf{Metric} &
\multicolumn{4}{c}{\textbf{Feature Source}} \\
\cmidrule(lr){4-7}
&
&
&
\textbf{N (Pos\%)} &
\textbf{True Fitbit Features} &
\textbf{Trajectory Transformer} &
\textbf{LEWorldModel} \\
\midrule

\rowcolor{gray!20}
\multicolumn{7}{l}{\textbf{\textit{Health Conditions}}} \\

Diabetes Condition &
\CLSLabel &
\AUCLabel &
305 (8.5\%) &
0.723 [0.64--0.80] &
0.713 [0.60--0.85] &
\textbf{0.724 [0.63--0.83]} \\

HOMA-IR Binary &
\CLSLabel &
\AUCLabel &
141 (27.0\%) &
0.780 [0.70--0.86] &
0.719 [0.64--0.82] &
\textbf{0.760 [0.67--0.85]} \\

Hyperlipidemia &
\CLSLabel &
\AUCLabel &
305 (19.7\%) &
0.691 [0.62--0.76] &
0.645 [0.58--0.73] &
\textbf{0.692 [0.61--0.76]} \\

Hypertension Binary &
\CLSLabel &
\AUCLabel &
305 (20.7\%) &
0.796 [0.74--0.86] &
\textbf{0.793 [0.73--0.86]} &
\textbf{0.793 [0.73--0.85]} \\

Respiratory Condition &
\CLSLabel &
\AUCLabel &
305 (18.7\%) &
0.629 [0.57--0.71] &
0.634 [0.56--0.73] &
\textbf{0.660 [0.58--0.72]} \\
\bottomrule
\end{tabular}
\end{adjustbox}
\label{traj-wm}
\end{table*}

\begin{table*}[h!]
\centering
\caption{
\textbf{Trajectory Based versus Static Downstream Classification Performance across Data Scales.} Classification performance comparing downstream prediction using true Fitbit features against LEWorldModel representations fine-tuned across different scale baselines (1k, 10k, and 100k). Values are AUROC with 95\% confidence intervals.
}
\small
\setlength{\tabcolsep}{5pt}
\renewcommand{\arraystretch}{1.18}

\begin{adjustbox}{width=\textwidth}
\begin{tabular}{lccc|cccc}
\toprule
\textbf{Prediction Task} &
\textbf{Type} &
\textbf{Metric} &
\multicolumn{5}{c}{\textbf{Feature Source}} \\
\cmidrule(lr){4-8}
&
&
&
\textbf{N (Pos\%)} &
\textbf{True Fitbit Features} &
\textbf{1k} &
\textbf{10k} &
\textbf{100k} \\
\midrule

\rowcolor{gray!20}
\multicolumn{8}{l}{\textbf{\textit{Health Conditions}}} \\

Diabetes Condition &
\CLSLabel &
\AUCLabel &
305 (8.5\%) &
0.723 [0.64--0.80] &
\textbf{0.736 [0.66--0.81]} &
0.706 [0.61--0.80] &
0.724 [0.63--0.83] \\

HOMA-IR Binary &
\CLSLabel &
\AUCLabel &
141 (27.0\%) &
0.780 [0.70--0.86] &
0.699 [0.61--0.81] &
0.710 [0.61--0.80] &
\textbf{0.760 [0.67--0.85]} \\

Hyperlipidemia &
\CLSLabel &
\AUCLabel &
305 (19.7\%) &
0.691 [0.62--0.76] &
0.673 [0.61--0.75] &
0.666 [0.60--0.75] &
\textbf{0.692 [0.61--0.76]} \\

Hypertension Binary &
\CLSLabel &
\AUCLabel &
305 (20.7\%) &
0.796 [0.74--0.86] &
\textbf{0.807 [0.75--0.87]} &
0.797 [0.74--0.85] &
0.793 [0.73--0.85] \\

Respiratory Condition &
\CLSLabel &
\AUCLabel &
305 (18.7\%) &
0.629 [0.57--0.71] &
0.619 [0.56--0.70] &
0.579 [0.51--0.65] &
\textbf{0.660 [0.58--0.72]} \\
\bottomrule
\end{tabular}
\end{adjustbox}
\label{scaling}
\end{table*}

\begin{table*}[h!]
\centering
\caption{
\textbf{Trajectory Based versus Static Downstream Classification Performance.} Classification performance comparing downstream prediction using true Fitbit features, world-model-generated Fitbit features, SensorFM, and first-week-only Fitbit features. Values are AUROC with 95\% confidence intervals.
}
\small
\setlength{\tabcolsep}{5pt}
\renewcommand{\arraystretch}{1.18}

\begin{adjustbox}{width=\textwidth}
\begin{tabular}{lccccc|cc}
\toprule
\textbf{Prediction Task} &
\textbf{Type} &
\textbf{Metric} &
\multicolumn{5}{c}{\textbf{Feature Source}} \\
\cmidrule(lr){4-8}
&
&
&
\textbf{N (Pos\%)} &
\textbf{True Fitbit Features} &
\textbf{Sensor World Model Fitbit Features} &
\textbf{SensorFM} &
\textbf{First Week Fitbit Features} \\
\midrule

\rowcolor{gray!20}
\multicolumn{8}{l}{\textbf{\textit{Health Conditions}}} \\

Diabetes Condition &
\CLSLabel &
\AUCLabel &
305 (8.5\%) &
0.723 [0.64--0.80] &
0.724 [0.63--0.83] &
0.643 [0.52--0.75] &
0.655 [0.54--0.77] \\

HOMA-IR Binary &
\CLSLabel &
\AUCLabel &
141 (27.0\%) &
0.780 [0.70--0.86] &
0.760 [0.67--0.85] &
0.774 [0.69--0.85] &
0.717 [0.61--0.82] \\

Hyperlipidemia &
\CLSLabel &
\AUCLabel &
305 (19.7\%) &
0.691 [0.62--0.76] &
0.692 [0.61--0.76] &
0.614 [0.53--0.69] &
0.535 [0.45--0.63] \\

Hypertension Binary &
\CLSLabel &
\AUCLabel &
305 (20.7\%) &
0.796 [0.74--0.86] &
0.793 [0.73--0.85] &
0.753 [0.68--0.82] &
0.766 [0.70--0.83] \\

Respiratory Condition &
\CLSLabel &
\AUCLabel &
305 (18.7\%) &
0.629 [0.57--0.71] &
0.660 [0.58--0.72] &
0.658 [0.58--0.73] &
0.535 [0.45--0.62] \\
\bottomrule
\end{tabular}
\end{adjustbox}
\label{tab:fitbit_worldmodel_lsm_firstweek_classification}
\end{table*}

\begin{table*}[h!]
\centering
\caption{
\textbf{Classification Performance of Fitbit and SensorWM Features on Entire Cohort.} Classification performance comparing downstream prediction using true Fitbit features and world-model-generated Fitbit features. Values are AUROC with 95\% confidence intervals. Statistical significance between AUROC estimates was assessed using DeLong's test.
}
\small
\setlength{\tabcolsep}{5pt}
\renewcommand{\arraystretch}{1.18}
\begin{adjustbox}{width=\textwidth}
\begin{tabular}{lccccccl}
\toprule
\textbf{Prediction Task} &
\textbf{Type} &
\textbf{Metric} &
\multicolumn{4}{c}{\textbf{Feature Source}} \\
\cmidrule(lr){4-7}
&
&
&
\textbf{N (Pos\%)} &
\textbf{True Fitbit Features} &
\textbf{World Model Fitbit Features} &
\textbf{DeLong $p$-val} \\
\midrule
\rowcolor{gray!20}
\multicolumn{7}{l}{\textbf{\textit{Health Conditions}}} \\
Diabetes Condition &
\CLSLabel &
\AUCLabel &
305 (8.5\%) &
0.723 [0.64--0.80] &
0.724 [0.63--0.83] &
0.7037 \\
HOMA-IR Binary &
\CLSLabel &
\AUCLabel &
141 (27.0\%) &
0.780 [0.70--0.86] &
0.760 [0.67--0.85] &
0.2782 \\
Hyperlipidemia &
\CLSLabel &
\AUCLabel &
305 (19.7\%) &
0.691 [0.62--0.76] &
0.692 [0.61--0.76] &
0.9234 \\
Hypertension Binary &
\CLSLabel &
\AUCLabel &
305 (20.7\%) &
0.796 [0.74--0.86] &
0.793 [0.73--0.85] &
0.7784 \\
Respiratory Condition &
\CLSLabel &
\AUCLabel &
305 (18.7\%) &
0.629 [0.57--0.71] &
0.660 [0.58--0.72] &
0.6494 \\
\midrule
\rowcolor{gray!20}
\multicolumn{7}{l}{\textbf{\textit{Behavioral}}} \\
Smoking &
\CLSLabel &
\AUCLabel &
300 (7.3\%) &
0.784 [0.69--0.87] &
0.774 [0.69--0.85] &
0.2013 \\
\midrule
\rowcolor{gray!20}
\multicolumn{7}{l}{\textbf{\textit{Women's Health}}} \\
Regular Periods &
\CLSLabel &
\AUCLabel &
167 (39.5\%) &
0.852 [0.79--0.91] &
0.834 [0.77--0.90] &
0.0328 \\
\midrule
\rowcolor{gray!20}
\multicolumn{7}{l}{\textbf{\textit{Medication}}} \\
Cardiovascular Medication &
\CLSLabel &
\AUCLabel &
331 (30.8\%) &
0.785 [0.75--0.84] &
0.775 [0.73--0.83] &
0.1427 \\
Diabetes Medication &
\CLSLabel &
\AUCLabel &
331 (9.7\%) &
0.719 [0.63--0.80] &
0.730 [0.64--0.80] &
0.0639 \\
Mental Health Medication &
\CLSLabel &
\AUCLabel &
331 (29.3\%) &
0.631 [0.57--0.69] &
0.616 [0.56--0.68] &
0.2299 \\
No Medication &
\CLSLabel &
\AUCLabel &
331 (39.6\%) &
0.710 [0.64--0.76] &
0.695 [0.64--0.75] &
0.1244 \\
Thyroid Medication &
\CLSLabel &
\AUCLabel &
331 (8.5\%) &
0.750 [0.65--0.84] &
0.716 [0.62--0.80] &
0.0332 \\
\bottomrule
\end{tabular}
\end{adjustbox}
\label{tab:fitbit_worldmodel_classification}
\end{table*}

\begin{table*}[h!]
\centering
\caption{
\textbf{Classification Performance of Fitbit and SensorWM Features on Users Ages 18-39.} Classification performance comparing downstream prediction using true Fitbit features and world-model-generated Fitbit features for the age bucket 18--39. Values are AUROC with 95\% confidence intervals. Statistical significance between AUROC estimates was assessed using DeLong's test.
}
\small
\setlength{\tabcolsep}{5pt}
\renewcommand{\arraystretch}{1.18}
\begin{adjustbox}{width=\textwidth}
\begin{tabular}{lcccccc}
\toprule
\textbf{Prediction Task} &
\textbf{Type} &
\textbf{Metric} &
\multicolumn{4}{c}{\textbf{Feature Source}} \\
\cmidrule(lr){4-7}
&
&
&
\textbf{N (Pos\%)} &
\textbf{True Fitbit Features} &
\textbf{World Model Fitbit Features} &
\textbf{DeLong $p$-val} \\
\midrule
\rowcolor{gray!20}
\multicolumn{7}{l}{\textbf{\textit{Health Conditions}}} \\
Diabetes Condition &
\CLSLabel &
\AUCLabel &
74 (2.7\%) &
0.875 [0.71--1.00] &
0.792 [0.62--0.93] &
0.0006 \\
HOMA-IR Binary &
\CLSLabel &
\AUCLabel &
33 (27.3\%) &
0.764 [0.59--0.89] &
0.755 [0.57--0.89] &
0.8322 \\
Hyperlipidemia &
\CLSLabel &
\AUCLabel &
74 (8.1\%) &
0.424 [0.18--0.74] &
0.414 [0.16--0.75] &
0.8301 \\
Hypertension Binary &
\CLSLabel &
\AUCLabel &
74 (8.1\%) &
0.787 [0.51--0.98] &
0.767 [0.46--0.98] &
0.3932 \\
Respiratory Condition &
\CLSLabel &
\AUCLabel &
74 (13.5\%) &
0.608 [0.47--0.79] &
0.564 [0.41--0.77] &
0.2493 \\
\midrule
\rowcolor{gray!20}
\multicolumn{7}{l}{\textbf{\textit{Behavioral}}} \\
Smoking &
\CLSLabel &
\AUCLabel &
71 (11.3\%) &
0.837 [0.72--0.97] &
0.669 [0.51--0.83] &
0.0127 \\
\midrule
\rowcolor{gray!20}
\multicolumn{7}{l}{\textbf{\textit{Women's Health}}} \\
Regular Periods &
\CLSLabel &
\AUCLabel &
36 (69.4\%) &
0.538 [0.36--0.76] &
0.549 [0.38--0.76] &
0.7467 \\
\midrule
\rowcolor{gray!20}
\multicolumn{7}{l}{\textbf{\textit{Medication}}} \\
Cardiovascular Medication &
\CLSLabel &
\AUCLabel &
81 (13.6\%) &
0.673 [0.52--0.83] &
0.604 [0.45--0.78] &
0.0008 \\
Diabetes Medication &
\CLSLabel &
\AUCLabel &
81 (3.7\%) &
0.893 [0.81--0.96] &
0.821 [0.74--0.91] &
0.0129 \\
Mental Health Medication &
\CLSLabel &
\AUCLabel &
81 (33.3\%) &
0.664 [0.55--0.78] &
0.623 [0.52--0.73] &
0.1587 \\
No Medication &
\CLSLabel &
\AUCLabel &
81 (49.4\%) &
0.751 [0.66--0.85] &
0.746 [0.66--0.85] &
0.8439 \\
Thyroid Medication &
\CLSLabel &
\AUCLabel &
81 (2.5\%) &
0.930 [0.86--0.98] &
0.956 [0.86--1.00] &
0.4132 \\
\bottomrule
\end{tabular}
\end{adjustbox}
\label{tab:fitbit_worldmodel_classification_age_18_39}
\end{table*}

\begin{table*}[h!]
\centering
\caption{
\textbf{Classification Performance of Fitbit and SensorWM Features on Users Ages 40-59.} Classification performance comparing downstream prediction using true Fitbit features and world-model-generated Fitbit features for the age bucket 40--59. Values are AUROC with 95\% confidence intervals. Statistical significance between AUROC estimates was assessed using DeLong's test.
}
\small
\setlength{\tabcolsep}{5pt}
\renewcommand{\arraystretch}{1.18}
\begin{adjustbox}{width=\textwidth}
\begin{tabular}{lccccccc}
\toprule
\textbf{Prediction Task} &
\textbf{Type} &
\textbf{Metric} &
\multicolumn{4}{c}{\textbf{Feature Source}} \\
\cmidrule(lr){4-7}
&
&
&
\textbf{N (Pos\%)} &
\textbf{True Fitbit Features} &
\textbf{World Model Fitbit Features} &
\textbf{DeLong $p$-val} \\
\midrule
\rowcolor{gray!20}
\multicolumn{7}{l}{\textbf{\textit{Health Conditions}}} \\
Diabetes Condition &
\CLSLabel &
\AUCLabel &
161 (9.9\%) &
0.731 [0.59--0.84] &
0.736 [0.60--0.84] &
0.8045 \\
HOMA-IR Binary &
\CLSLabel &
\AUCLabel &
69 (26.1\%) &
0.845 [0.74--0.94] &
0.829 [0.71--0.93] &
0.5330 \\
Hyperlipidemia &
\CLSLabel &
\AUCLabel &
161 (18.0\%) &
0.676 [0.56--0.79] &
0.699 [0.59--0.80] &
0.2625 \\
Hypertension Binary &
\CLSLabel &
\AUCLabel &
161 (18.0\%) &
0.766 [0.66--0.83] &
0.783 [0.68--0.86] &
0.3111 \\
Respiratory Condition &
\CLSLabel &
\AUCLabel &
161 (19.9\%) &
0.622 [0.53--0.72] &
0.614 [0.52--0.72] &
0.7114 \\
\midrule
\rowcolor{gray!20}
\multicolumn{7}{l}{\textbf{\textit{Behavioral}}} \\
Smoking &
\CLSLabel &
\AUCLabel &
160 (8.1\%) &
0.707 [0.57--0.83] &
0.730 [0.59--0.86] &
0.4850 \\
\midrule
\rowcolor{gray!20}
\multicolumn{7}{l}{\textbf{\textit{Women's Health}}} \\
Regular Periods &
\CLSLabel &
\AUCLabel &
94 (41.5\%) &
0.847 [0.77--0.92] &
0.808 [0.70--0.90] &
0.0242 \\
\midrule
\rowcolor{gray!20}
\multicolumn{7}{l}{\textbf{\textit{Medication}}} \\
Cardiovascular Medication &
\CLSLabel &
\AUCLabel &
177 (24.9\%) &
0.754 [0.67--0.83] &
0.755 [0.67--0.84] &
0.9123 \\
Diabetes Medication &
\CLSLabel &
\AUCLabel &
177 (10.7\%) &
0.644 [0.52--0.77] &
0.631 [0.50--0.76] &
0.3941 \\
Mental Health Medication &
\CLSLabel &
\AUCLabel &
177 (31.1\%) &
0.597 [0.50--0.70] &
0.591 [0.51--0.70] &
0.7604 \\
No Medication &
\CLSLabel &
\AUCLabel &
177 (41.8\%) &
0.713 [0.63--0.79] &
0.682 [0.61--0.76] &
0.0301 \\
Thyroid Medication &
\CLSLabel &
\AUCLabel &
177 (9.0\%) &
0.692 [0.53--0.83] &
0.641 [0.50--0.78] &
0.0382 \\
\bottomrule
\end{tabular}
\end{adjustbox}
\label{tab:fitbit_worldmodel_classification_age_40_59}
\end{table*}

\begin{table*}[h!]
\centering
\caption{
\textbf{Classification Performance of Fitbit and SensorWM Features on Users Ages 60-79.} Classification performance comparing downstream prediction using true Fitbit features and world-model-generated Fitbit features for the age bucket 60--79. Values are AUROC with 95\% confidence intervals. Statistical significance between AUROC estimates was assessed using DeLong's test.
}
\small
\setlength{\tabcolsep}{5pt}
\renewcommand{\arraystretch}{1.18}
\begin{adjustbox}{width=\textwidth}
\begin{tabular}{lccccccc}
\toprule
\textbf{Prediction Task} &
\textbf{Type} &
\textbf{Metric} &
\multicolumn{4}{c}{\textbf{Feature Source}} \\
\cmidrule(lr){4-7}
&
&
&
\textbf{N (Pos\%)} &
\textbf{True Fitbit Features} &
\textbf{World Model Fitbit Features} &
\textbf{DeLong $p$-val} \\
\midrule
\rowcolor{gray!20}
\multicolumn{7}{l}{\textbf{\textit{Health Conditions}}} \\
Diabetes Condition &
\CLSLabel &
\AUCLabel &
68 (11.8\%) &
0.519 [0.40--0.67] &
0.533 [0.41--0.68] &
0.6658 \\
HOMA-IR Binary &
\CLSLabel &
\AUCLabel &
38 (26.3\%) &
0.686 [0.51--0.88] &
0.618 [0.42--0.85] &
0.0329 \\
Hyperlipidemia &
\CLSLabel &
\AUCLabel &
68 (36.8\%) &
0.476 [0.34--0.63] &
0.463 [0.31--0.59] &
0.6502 \\
Hypertension Binary &
\CLSLabel &
\AUCLabel &
68 (39.7\%) &
0.613 [0.49--0.75] &
0.593 [0.45--0.73] &
0.5674 \\
Respiratory Condition &
\CLSLabel &
\AUCLabel &
68 (20.6\%) &
0.669 [0.48--0.84] &
0.679 [0.49--0.84] &
0.5310 \\
\midrule
\rowcolor{gray!20}
\multicolumn{7}{l}{\textbf{\textit{Behavioral}}} \\
Smoking &
\CLSLabel &
\AUCLabel &
67 (1.5\%) &
0.864 [0.79--0.93] &
0.848 [0.76--0.92] &
--- \\
\midrule
\rowcolor{gray!20}
\multicolumn{7}{l}{\textbf{\textit{Women's Health}}} \\
Regular Periods &
\CLSLabel &
\AUCLabel &
37 (5.4\%) &
0.629 [0.46--0.80] &
0.600 [0.46--0.80] &
0.5953 \\
\midrule
\rowcolor{gray!20}
\multicolumn{7}{l}{\textbf{\textit{Medication}}} \\
Cardiovascular Medication &
\CLSLabel &
\AUCLabel &
71 (63.4\%) &
0.579 [0.42--0.71] &
0.535 [0.39--0.67] &
0.1262 \\
Diabetes Medication &
\CLSLabel &
\AUCLabel &
71 (14.1\%) &
0.708 [0.56--0.85] &
0.707 [0.51--0.86] &
0.9589 \\
Mental Health Medication &
\CLSLabel &
\AUCLabel &
71 (19.7\%) &
0.669 [0.50--0.84] &
0.648 [0.46--0.83] &
0.3585 \\
No Medication &
\CLSLabel &
\AUCLabel &
71 (23.9\%) &
0.517 [0.37--0.68] &
0.486 [0.34--0.66] &
0.2923 \\
Thyroid Medication &
\CLSLabel &
\AUCLabel &
71 (14.1\%) &
0.692 [0.52--0.87] &
0.661 [0.51--0.86] &
0.3490 \\
\bottomrule
\end{tabular}
\end{adjustbox}
\label{tab:fitbit_worldmodel_classification_age_60_79}
\end{table*}

\begin{table*}[h!]
\centering
\caption{\textbf{Classification Performance of Fitbit and SensorWM Features on Users with a BMI $<$25} Classification performance comparing downstream prediction using true Fitbit features and world-model-generated Fitbit features for the BMI bucket $<25$. Values are AUROC with 95\% confidence intervals. Statistical significance between AUROC estimates was assessed using DeLong's test.
}
\small
\setlength{\tabcolsep}{5pt}
\renewcommand{\arraystretch}{1.18}
\begin{adjustbox}{width=\textwidth}
\begin{tabular}{lccccccc}
\toprule
\textbf{Prediction Task} &
\textbf{Type} &
\textbf{Metric} &
\multicolumn{4}{c}{\textbf{Feature Source}} \\
\cmidrule(lr){4-7}
&
&
&
\textbf{N (Pos\%)} &
\textbf{True Fitbit Features} &
\textbf{World Model Fitbit Features} &
\textbf{DeLong $p$-val} \\
\midrule
\rowcolor{gray!20}
\multicolumn{7}{l}{\textbf{\textit{Health Conditions}}} \\
Diabetes Condition &
\CLSLabel &
\AUCLabel &
105 (3.8\%) &
0.738 [0.60--0.91] &
0.666 [0.54--0.83] &
0.0057 \\
HOMA-IR Binary &
\CLSLabel &
\AUCLabel &
50 (10.0\%) &
0.800 [0.59--0.95] &
0.809 [0.59--0.97] &
0.9028 \\
Hyperlipidemia &
\CLSLabel &
\AUCLabel &
105 (10.5\%) &
0.617 [0.41--0.77] &
0.620 [0.42--0.78] &
0.8651 \\
Hypertension Binary &
\CLSLabel &
\AUCLabel &
105 (13.3\%) &
0.919 [0.86--0.97] &
0.908 [0.85--0.97] &
0.3072 \\
Respiratory Condition &
\CLSLabel &
\AUCLabel &
105 (9.5\%) &
0.544 [0.36--0.70] &
0.575 [0.39--0.73] &
0.2395 \\
\midrule
\rowcolor{gray!20}
\multicolumn{7}{l}{\textbf{\textit{Behavioral}}} \\
Smoking &
\CLSLabel &
\AUCLabel &
104 (9.6\%) &
0.807 [0.63--0.92] &
0.760 [0.60--0.88] &
0.2884 \\
\midrule
\rowcolor{gray!20}
\multicolumn{7}{l}{\textbf{\textit{Women's Health}}} \\
Regular Periods &
\CLSLabel &
\AUCLabel &
49 (40.8\%) &
0.855 [0.75--0.94] &
0.848 [0.74--0.94] &
0.6268 \\
\midrule
\rowcolor{gray!20}
\multicolumn{7}{l}{\textbf{\textit{Medication}}} \\
Cardiovascular Medication &
\CLSLabel &
\AUCLabel &
110 (24.5\%) &
0.816 [0.70--0.92] &
0.791 [0.67--0.89] &
0.0205 \\
Diabetes Medication &
\CLSLabel &
\AUCLabel &
110 (3.6\%) &
0.776 [0.55--1.00] &
0.705 [0.45--0.99] &
0.0375 \\
Mental Health Medication &
\CLSLabel &
\AUCLabel &
110 (17.3\%) &
0.665 [0.53--0.81] &
0.636 [0.50--0.76] &
0.2815 \\
No Medication &
\CLSLabel &
\AUCLabel &
110 (53.6\%) &
0.702 [0.62--0.79] &
0.669 [0.58--0.78] &
0.1063 \\
Thyroid Medication &
\CLSLabel &
\AUCLabel &
110 (3.6\%) &
0.825 [0.71--0.97] &
0.842 [0.72--0.99] &
0.2457 \\
\bottomrule
\end{tabular}
\end{adjustbox}
\label{tab:fitbit_worldmodel_classification_bmi_lt25}
\end{table*}

\begin{table*}[h!]
\centering
\caption{
\textbf{Classification Performance of Fitbit and SensorWM Features on Users with a BMI 25--30} Classification performance comparing downstream prediction using true Fitbit features and world-model-generated Fitbit features for the BMI bucket \textbf{25--30}. Values are AUROC with 95\% confidence intervals. Statistical significance between AUROC estimates was assessed using DeLong's test.
}
\small
\setlength{\tabcolsep}{5pt}
\renewcommand{\arraystretch}{1.18}
\begin{adjustbox}{width=\textwidth}
\begin{tabular}{lccccccc}
\toprule
\textbf{Prediction Task} &
\textbf{Type} &
\textbf{Metric} &
\multicolumn{4}{c}{\textbf{Feature Source}} \\
\cmidrule(lr){4-7}
&
&
&
\textbf{N (Pos\%)} &
\textbf{True Fitbit Features} &
\textbf{World Model Fitbit Features} &
\textbf{DeLong $p$-val} \\
\midrule
\rowcolor{gray!20}
\multicolumn{7}{l}{\textbf{\textit{Health Conditions}}} \\
Diabetes Condition &
\CLSLabel &
\AUCLabel &
101 (6.9\%) &
0.713 [0.52--0.87] &
0.679 [0.47--0.89] &
0.3086 \\
HOMA-IR Binary &
\CLSLabel &
\AUCLabel &
53 (30.2\%) &
0.694 [0.52--0.83] &
0.664 [0.51--0.81] &
0.3352 \\
Hyperlipidemia &
\CLSLabel &
\AUCLabel &
101 (25.7\%) &
0.761 [0.67--0.85] &
0.758 [0.68--0.86] &
0.8727 \\
Hypertension Binary &
\CLSLabel &
\AUCLabel &
101 (17.8\%) &
0.744 [0.62--0.89] &
0.723 [0.59--0.86] &
0.4795 \\
Respiratory Condition &
\CLSLabel &
\AUCLabel &
101 (21.8\%) &
0.627 [0.48--0.77] &
0.622 [0.46--0.77] &
0.8209 \\
\midrule
\rowcolor{gray!20}
\multicolumn{7}{l}{\textbf{\textit{Behavioral}}} \\
Smoking &
\CLSLabel &
\AUCLabel &
99 (2.0\%) &
0.732 [0.51--0.93] &
0.716 [0.50--0.90] &
0.5847 \\
\midrule
\rowcolor{gray!20}
\multicolumn{7}{l}{\textbf{\textit{Women's Health}}} \\
Regular Periods &
\CLSLabel &
\AUCLabel &
47 (40.4\%) &
0.838 [0.71--0.94] &
0.846 [0.74--0.95] &
0.7195 \\
\midrule
\rowcolor{gray!20}
\multicolumn{7}{l}{\textbf{\textit{Medication}}} \\
Cardiovascular Medication &
\CLSLabel &
\AUCLabel &
108 (33.3\%) &
0.795 [0.69--0.90] &
0.785 [0.69--0.88] &
0.4623 \\
Diabetes Medication &
\CLSLabel &
\AUCLabel &
108 (8.3\%) &
0.707 [0.54--0.86] &
0.681 [0.48--0.85] &
0.2734 \\
Mental Health Medication &
\CLSLabel &
\AUCLabel &
108 (38.0\%) &
0.664 [0.58--0.78] &
0.641 [0.55--0.76] &
0.2887 \\
No Medication &
\CLSLabel &
\AUCLabel &
108 (35.2\%) &
0.788 [0.67--0.87] &
0.768 [0.65--0.86] &
0.2004 \\
Thyroid Medication &
\CLSLabel &
\AUCLabel &
108 (11.1\%) &
0.748 [0.59--0.88] &
0.726 [0.55--0.87] &
0.2418 \\
\bottomrule
\end{tabular}
\end{adjustbox}
\label{tab:fitbit_worldmodel_classification_bmi_25_30}
\end{table*}

\begin{table*}[h!]
\centering
\caption{
\textbf{Classification Performance of Fitbit and SensorWM Features on Users with a BMI $>$30} Classification performance comparing downstream prediction using true Fitbit features and world-model-generated Fitbit features for the BMI bucket $>30$. Values are AUROC with 95\% confidence intervals. Statistical significance between AUROC estimates was assessed using DeLong's test.
}
\small
\setlength{\tabcolsep}{5pt}
\renewcommand{\arraystretch}{1.18}
\begin{adjustbox}{width=\textwidth}
\begin{tabular}{lccccccc}
\toprule
\textbf{Prediction Task} &
\textbf{Type} &
\textbf{Metric} &
\multicolumn{4}{c}{\textbf{Feature Source}} \\
\cmidrule(lr){4-7}
&
&
&
\textbf{N (Pos\%)} &
\textbf{True Fitbit Features} &
\textbf{World Model Fitbit Features} &
\textbf{DeLong $p$-val} \\
\midrule
\rowcolor{gray!20}
\multicolumn{7}{l}{\textbf{\textit{Health Conditions}}} \\
Diabetes Condition &
\CLSLabel &
\AUCLabel &
99 (15.2\%) &
0.656 [0.52--0.80] &
0.688 [0.54--0.82] &
0.2623 \\
HOMA-IR Binary &
\CLSLabel &
\AUCLabel &
38 (44.7\%) &
0.711 [0.55--0.84] &
0.675 [0.51--0.84] &
0.4551 \\
Hyperlipidemia &
\CLSLabel &
\AUCLabel &
99 (23.2\%) &
0.662 [0.57--0.79] &
0.668 [0.58--0.80] &
0.7222 \\
Hypertension Binary &
\CLSLabel &
\AUCLabel &
99 (31.3\%) &
0.685 [0.57--0.78] &
0.701 [0.60--0.80] &
0.4107 \\
Respiratory Condition &
\CLSLabel &
\AUCLabel &
99 (25.3\%) &
0.562 [0.42--0.69] &
0.542 [0.42--0.68] &
0.5405 \\
\midrule
\rowcolor{gray!20}
\multicolumn{7}{l}{\textbf{\textit{Behavioral}}} \\
Smoking &
\CLSLabel &
\AUCLabel &
97 (10.3\%) &
0.786 [0.64--0.90] &
0.746 [0.59--0.87] &
0.4425 \\
\midrule
\rowcolor{gray!20}
\multicolumn{7}{l}{\textbf{\textit{Women's Health}}} \\
Regular Periods &
\CLSLabel &
\AUCLabel &
71 (38.0\%) &
0.850 [0.74--0.94] &
0.826 [0.71--0.91] &
0.1579 \\
\midrule
\rowcolor{gray!20}
\multicolumn{7}{l}{\textbf{\textit{Medication}}} \\
Cardiovascular Medication &
\CLSLabel &
\AUCLabel &
113 (34.5\%) &
0.731 [0.63--0.81] &
0.731 [0.63--0.82] &
0.9766 \\
Diabetes Medication &
\CLSLabel &
\AUCLabel &
113 (16.8\%) &
0.631 [0.49--0.78] &
0.622 [0.48--0.77] &
0.6933 \\
Mental Health Medication &
\CLSLabel &
\AUCLabel &
113 (32.7\%) &
0.542 [0.42--0.67] &
0.544 [0.44--0.65] &
0.9287 \\
No Medication &
\CLSLabel &
\AUCLabel &
113 (30.1\%) &
0.569 [0.44--0.70] &
0.579 [0.44--0.71] &
0.6592 \\
Thyroid Medication &
\CLSLabel &
\AUCLabel &
113 (10.6\%) &
0.702 [0.55--0.83] &
0.634 [0.47--0.77] &
0.0353 \\
\bottomrule
\end{tabular}
\end{adjustbox}
\label{tab:fitbit_worldmodel_classification_bmi_gt30}
\end{table*}

\begin{table*}[h!]
\centering
\caption{
\textbf{Classification Performance of Fitbit and SensorWM Features on Male Users} Classification performance comparing downstream prediction using true Fitbit features and world-model-generated Fitbit features for male users. Values are AUROC with 95\% confidence intervals. Statistical significance between AUROC estimates was assessed using DeLong's test.
}
\small
\setlength{\tabcolsep}{5pt}
\renewcommand{\arraystretch}{1.18}
\begin{adjustbox}{width=\textwidth}
\begin{tabular}{lccccccc}
\toprule
\textbf{Prediction Task} &
\textbf{Type} &
\textbf{Metric} &
\multicolumn{4}{c}{\textbf{Feature Source}} \\
\cmidrule(lr){4-7}
&
&
&
\textbf{N (Pos\%)} &
\textbf{True Fitbit Features} &
\textbf{World Model Fitbit Features} &
\textbf{DeLong $p$-val} \\
\midrule
\rowcolor{gray!20}
\multicolumn{7}{l}{\textbf{\textit{Health Conditions}}} \\
Diabetes Condition &
\CLSLabel &
\AUCLabel &
131 (11.5\%) &
0.763 [0.64--0.86] &
0.739 [0.61--0.83] &
0.2654 \\
HOMA-IR Binary &
\CLSLabel &
\AUCLabel &
67 (29.9\%) &
0.815 [0.71--0.90] &
0.797 [0.70--0.88] &
0.4511 \\
Hyperlipidemia &
\CLSLabel &
\AUCLabel &
131 (23.7\%) &
0.664 [0.58--0.75] &
0.675 [0.57--0.76] &
0.3843 \\
Hypertension Binary &
\CLSLabel &
\AUCLabel &
131 (23.7\%) &
0.843 [0.75--0.91] &
0.848 [0.76--0.92] &
0.5870 \\
Respiratory Condition &
\CLSLabel &
\AUCLabel &
131 (10.7\%) &
0.747 [0.63--0.87] &
0.725 [0.61--0.85] &
0.1468 \\
\midrule
\rowcolor{gray!20}
\multicolumn{7}{l}{\textbf{\textit{Behavioral}}} \\
Smoking &
\CLSLabel &
\AUCLabel &
129 (5.4\%) &
0.905 [0.78--0.99] &
0.854 [0.74--0.97] &
0.2366 \\
\midrule
\rowcolor{gray!20}
\multicolumn{7}{l}{\textbf{\textit{Medication}}} \\
Cardiovascular Medication &
\CLSLabel &
\AUCLabel &
139 (33.1\%) &
0.875 [0.80--0.93] &
0.871 [0.80--0.93] &
0.5765 \\
Diabetes Medication &
\CLSLabel &
\AUCLabel &
139 (10.1\%) &
0.762 [0.65--0.84] &
0.728 [0.60--0.82] &
0.0597 \\
Mental Health Medication &
\CLSLabel &
\AUCLabel &
139 (21.6\%) &
0.617 [0.51--0.71] &
0.594 [0.48--0.69] &
0.3424 \\
No Medication &
\CLSLabel &
\AUCLabel &
139 (48.9\%) &
0.789 [0.70--0.86] &
0.773 [0.69--0.84] &
0.2678 \\
Thyroid Medication &
\CLSLabel &
\AUCLabel &
139 (3.6\%) &
0.687 [0.51--0.92] &
0.649 [0.46--0.91] &
0.3696 \\
\bottomrule
\end{tabular}
\end{adjustbox}
\label{tab:fitbit_worldmodel_classification_sex_1}
\end{table*}

\begin{table*}[h!]
\centering
\caption{
\textbf{Classification Performance of Fitbit and SensorWM Features on Female Users} Classification performance comparing downstream prediction using true Fitbit features and world-model-generated Fitbit features for female users. Values are AUROC with 95\% confidence intervals. Statistical significance between AUROC estimates was assessed using DeLong's test.
}
\small
\setlength{\tabcolsep}{5pt}
\renewcommand{\arraystretch}{1.18}
\begin{adjustbox}{width=\textwidth}
\begin{tabular}{lccccccc}
\toprule
\textbf{Prediction Task} &
\textbf{Type} &
\textbf{Metric} &
\multicolumn{4}{c}{\textbf{Feature Source}} \\
\cmidrule(lr){4-7}
&
&
&
\textbf{N (Pos\%)} &
\textbf{True Fitbit Features} &
\textbf{World Model Fitbit Features} &
\textbf{DeLong $p$-val} \\
\midrule
\rowcolor{gray!20}
\multicolumn{7}{l}{\textbf{\textit{Health Conditions}}} \\
Diabetes Condition &
\CLSLabel &
\AUCLabel &
174 (6.3\%) &
0.738 [0.63--0.87] &
0.756 [0.64--0.87] &
0.5171 \\
HOMA-IR Binary &
\CLSLabel &
\AUCLabel &
74 (24.3\%) &
0.752 [0.62--0.88] &
0.730 [0.59--0.86] &
0.4309 \\
Hyperlipidemia &
\CLSLabel &
\AUCLabel &
174 (16.7\%) &
0.720 [0.60--0.83] &
0.714 [0.60--0.81] &
0.6873 \\
Hypertension Binary &
\CLSLabel &
\AUCLabel &
174 (18.4\%) &
0.759 [0.64--0.85] &
0.747 [0.64--0.82] &
0.5723 \\
Respiratory Condition &
\CLSLabel &
\AUCLabel &
174 (24.7\%) &
0.554 [0.46--0.63] &
0.562 [0.47--0.63] &
0.6274 \\
\midrule
\rowcolor{gray!20}
\multicolumn{7}{l}{\textbf{\textit{Behavioral}}} \\
Smoking &
\CLSLabel &
\AUCLabel &
171 (8.8\%) &
0.704 [0.56--0.81] &
0.685 [0.56--0.81] &
0.6094 \\
\midrule
\rowcolor{gray!20}
\multicolumn{7}{l}{\textbf{\textit{Women's Health}}} \\
Regular Periods &
\CLSLabel &
\AUCLabel &
167 (39.5\%) &
0.852 [0.79--0.91] &
0.834 [0.77--0.90] &
0.0328 \\
\midrule
\rowcolor{gray!20}
\multicolumn{7}{l}{\textbf{\textit{Medication}}} \\
Cardiovascular Medication &
\CLSLabel &
\AUCLabel &
192 (29.2\%) &
0.716 [0.64--0.80] &
0.697 [0.61--0.79] &
0.1015 \\
Diabetes Medication &
\CLSLabel &
\AUCLabel &
192 (9.4\%) &
0.713 [0.61--0.81] &
0.699 [0.58--0.80] &
0.4083 \\
Mental Health Medication &
\CLSLabel &
\AUCLabel &
192 (34.9\%) &
0.613 [0.53--0.71] &
0.592 [0.51--0.69] &
0.3061 \\
No Medication &
\CLSLabel &
\AUCLabel &
192 (32.8\%) &
0.615 [0.54--0.68] &
0.603 [0.53--0.68] &
0.4308 \\
Thyroid Medication &
\CLSLabel &
\AUCLabel &
192 (12.0\%) &
0.699 [0.61--0.78] &
0.652 [0.57--0.73] &
0.0638 \\
\bottomrule
\end{tabular}
\end{adjustbox}
\label{tab:fitbit_worldmodel_classification_sex_2}
\end{table*}

\clearpage
\subsection{Expert Evaluation of Counterfactual Coaching}
\label{c3}

\subsubsection*{Verbatim Prompt 1: Generic Coaching Arm (Baseline)}
\begin{tcolorbox}[
    breakable,
    enhanced,
    title=Prompt for Personalized 12-Week Lifestyle Coaching,
    title after break={Prompt for Personalized 12-Week Lifestyle Coaching (continued)},
    before skip=10pt,
    after skip=10pt
]
\small
\ttfamily

\#\#\# ROLE \& OBJECTIVE

You are an expert, certified Health and Wellness Coach grounded in Self-Determination Theory (SDT) and Motivational Interviewing. Your objective is to write a personalized, supportive, and actionable 12-week lifestyle coaching plan (500--700 words) for a participant based on their demographic profile and 24-week wearable history.

\vspace{4mm}
\#\#\# PARTICIPANT CONTEXT

* Demographics: Age: \{age\} years | Sex: \{sex\} | BMI: \{bmi\} kg/m2

* Recent Measured Baseline (Daily Averages):

  - Resting Heart Rate (RHR): \{baseline\_rhr\} bpm

  - Daily Steps: \{baseline\_steps\} steps/day

  - Nightly Sleep Duration: \{baseline\_sleep\} hours/night

  - Daily Cardio Workouts: \{baseline\_cardio\} mins/day

  - Active Zone Minutes (AZM): \{baseline\_azm\} mins/day

* 24-Week Habitual Distribution \& Historical Peaks:

  - Steps: 24-wk Average \{hab\_steps\_mean\} \\
      (Range: \{hab\_steps\_min\}-\{hab\_steps\_max\}, Std: +/-\{hab\_steps\_std\}) | Best Recorded Peak: \{peak\_steps\} steps/day

  - Sleep: 24-wk Average \{hab\_sleep\_mean\}h \\ 
      (Range: \{hab\_sleep\_min\}-\{hab\_sleep\_max\}h, Std: +/-\{hab\_sleep\_std\}h) | Best Recorded Peak: \{peak\_sleep\} hours/night

  - Cardio: 24-wk Average \{hab\_cardio\_mean\} mins \\ 
      (Range: \{hab\_cardio\_min\}-\{hab\_cardio\_max\} mins) | Best Recorded Peak: \{peak\_cardio\} mins/day

\vspace{4mm}
\#\#\# COACHING \& CLINICAL GUARDRAILS

1. Tone: Warm, empathetic, collaborative, and autonomy-supportive. Never use guilt, fear, or shame. Celebrate existing strengths before suggesting improvements.

2. Jargon Ban: Do not use clinical acronyms, variable names, or statistical terms (e.g., do not say ``std'', ``min/max'', or ``BMI class''). Translate Resting Heart Rate (RHR) and Active Zone Minutes (AZM) into clear, everyday concepts of heart efficiency and energizing movement.

3. Safe \& Realistic Calibration: Anchor all behavioral goals within the participant's current baseline and proven 24-week historical peaks. If BMI is elevated or baseline activity is sedentary, recommend joint-friendly movement options (e.g., brisk walking, swimming, cycling).

4. Length: Keep the complete response between 500 and 700 words.

\vspace{4mm}
\#\#\# REQUIRED 4-SECTION STRUCTURE

1. YOUR HEALTH SNAPSHOT

   - Warmly welcome the participant and synthesize their recent wearable metrics (Steps, Sleep, RHR, and AZM) alongside their 24-week history into a cohesive picture of where their body is thriving and where their biggest opportunity lies.

   - Reference their historical peaks as encouraging proof of what their body has already achieved.

2. YOUR 12-WEEK PROGRESSIVE PLAN

   - Design a realistic, step-by-step 12-week roadmap divided into three 4-week phases:

     * Phase 1 (Weeks 1--4: Foundation): Gentle, low-friction adjustments to build consistency around sleep and daily movement.

     * Phase 2 (Weeks 5--8: Build): Gradual progression toward their historical capacity, introducing moderate heart-pumping movement (AZM).

     * Phase 3 (Weeks 9--12: Sustain): Consolidating habits into a sustainable long-term lifestyle routine.

3. EVIDENCE-BASED HABIT FORMATION

   - Provide concrete behavioral strategies tailored to this plan, including Habit Stacking (anchoring new habits to existing daily cues), Implementation Intentions (``If-Then'' planning), Environmental Design, and compassionate troubleshooting for busy or low-energy days.

4. SUMMARY \& NEXT STEPS

   - Conclude with an encouraging summary and a memorable one-sentence takeaway that reinforces their agency and partnership.

\end{tcolorbox}

\clearpage
\subsubsection*{Verbatim Prompt 2: Counterfactual Coaching Arm}

\begin{tcolorbox}[
    breakable,
    enhanced,
    title=Prompt for Counterfactual 12-Week Lifestyle Coaching,
    title after break={Prompt for Counterfactual 12-Week Lifestyle Coaching (continued)},
    before skip=10pt,
    after skip=10pt
]
\small
\ttfamily

\#\#\# ROLE \& OBJECTIVE

You are an expert, certified Health and Wellness Coach grounded in Self-Determination Theory (SDT) and Motivational Interviewing. Your objective is to write a personalized, supportive, and actionable 12-week lifestyle coaching plan (500--700 words) for a participant using their 24-week wearable history, their Coachable Health Age, and their personalized 12-week counterfactual health simulations.

\vspace{4mm}
\#\#\# PARTICIPANT CONTEXT \& COUNTERFACTUAL SIMULATIONS

* Demographics: Age: \{age\} years | Sex: \{sex\} | BMI: \{bmi\} kg/m2

* Recent Measured Baseline (Daily Averages):

  - Chronological Age: \{age\} years

  - Current Coachable Health Age: \{baseline\_health\_age\} years

  - Resting Heart Rate (RHR): \{baseline\_rhr\} bpm

  - Daily Steps: \{baseline\_steps\} steps/day

  - Nightly Sleep Duration: \{baseline\_sleep\} hours/night

  - Daily Cardio Workouts: \{baseline\_cardio\} mins/day

  - Active Zone Minutes (AZM): \{baseline\_azm\} mins/day

* 24-Week Habitual Distribution \& Historical Peaks:

  - Steps: 24-wk Average \{hab\_steps\_mean\}\\
      (Range: \{hab\_steps\_min\}-\{hab\_steps\_max\}, Std: +/-\{hab\_steps\_std\}) | Best Recorded Peak: \{peak\_steps\} steps/day

  - Sleep: 24-wk Average \{hab\_sleep\_mean\}h \\
      (Range: \{hab\_sleep\_min\}-\{hab\_sleep\_max\}h, Std: +/-\{hab\_sleep\_std\}h) | Best Recorded Peak: \{peak\_sleep\} hours/night

  - Cardio: 24-wk Average \{hab\_cardio\_mean\} mins \\
      (Range: \{hab\_cardio\_min\}-\{hab\_cardio\_max\} mins) | Best Recorded Peak: \{peak\_cardio\} mins/day

* 12-Week Status-Quo Projection (Future A -- If Current Habits Continue Unchanged):

  - Projected Coachable Health Age: \{projected\_ha\_status\_quo\} years

  - Projected Resting Heart Rate: \{projected\_rhr\_status\_quo\} bpm

  - Projected Active Zone Minutes: \{projected\_azm\_status\_quo\} mins/day

* 12-Week Personalized Counterfactual Pathways (Future B -- Simulated Habit Options Bounded by Historical Peaks):

  - Option 1 (\{option\_1\_name\}): \{option\_1\_habit\_delta\} -> Projected Health Age: \{option\_1\_ha\} years (\{option\_1\_ha\_delta\} yrs), Projected RHR: \{option\_1\_rhr\} bpm

  - Option 2 (\{option\_2\_name\}): \{option\_2\_habit\_delta\} -> Projected Health Age: \{option\_2\_ha\} years (\{option\_2\_ha\_delta\} yrs), Projected RHR: \{option\_2\_rhr\} bpm

  - Option 3 (\{option\_3\_name\}): \{option\_3\_habit\_delta\} -> Projected Health Age: \{option\_3\_ha\} years (\{option\_3\_ha\_delta\} yrs), Projected RHR: \{option\_3\_rhr\} bpm

\vspace{4mm}
\#\#\# COACHING \& CLINICAL GUARDRAILS

1. Tone: Warm, empathetic, collaborative, and autonomy-supportive. Frame Coachable Health Age and Future A vs. Future B as an empowering compass--never use fear, guilt, or alarmism.

2. Jargon Ban: Never use the word ``counterfactual'', ``simulation'', ``model'', ``std'', or ``min/max''. Present Future A and Future B naturally as ``Where Your Current Routine Leads'' versus ``Your Personalized Pathways Forward''.

3. Autonomy \& Choice Architecture: Present the 2--3 pathways clearly so the participant feels complete ownership over choosing the path (Sleep, Movement, or Combined) that best fits their life right now.

4. Safe \& Realistic Calibration: All pathways are bounded by the participant's own 24-week historical peaks; reinforce that their body has already demonstrated it can achieve these levels. Recommend joint-friendly modalities when appropriate.

5. Length: Keep the complete response between 500 and 700 words.

\vspace{4mm}
\#\#\# REQUIRED 4-SECTION STRUCTURE

1. YOUR HEALTH SNAPSHOT \& HEALTH AGE

   - Warmly welcome the participant and synthesize their wearable metrics (Steps, Sleep, RHR, AZM) and Coachable Health Age into a unified compass.

   - Gently contrast Future A (what happens over the next 12 weeks if current habits stay the same) with Future B (how modifying specific daily levers can lower their Health Age and Resting Heart Rate), referencing their 24-week historical peaks as proof of capability.

2. YOUR PERSONALIZED PATHWAYS \& 12-WEEK PROGRESSIVE PLAN

   - Present the distinct, quantified Options (Option 1, Option 2, Option 3) with their exact 12-week Health Age and RHR payoffs, inviting the participant to choose their preferred focus.

   - Lay out a progressive 12-week roadmap divided into three 4-week phases:

     * Phase 1 (Weeks 1--4: Foundation): Starting gently on their chosen pathway.

     * Phase 2 (Weeks 5--8: Build): Stepping up toward their target habit change and heart-pumping movement (AZM).

     * Phase 3 (Weeks 9--12: Sustain): Locking in the routine to realize their 12-week Health Age and RHR improvements.

3. EVIDENCE-BASED HABIT FORMATION

   - Provide concrete behavioral strategies (Habit Stacking, ``If-Then'' Implementation Intentions, Self-Monitoring of RHR/energy, and compassionate barrier planning) to support long-term adherence.

4. SUMMARY \& NEXT STEPS

   - Conclude with an encouraging summary and a memorable one-sentence takeaway celebrating their control over their future health trajectory.

\end{tcolorbox}

\clearpage
\begin{landscape}

\small
\setlength{\tabcolsep}{3pt}
\renewcommand{\arraystretch}{1.15}

\begin{longtable}{
    L{0.135\linewidth}
    L{0.135\linewidth}
    L{0.090\linewidth}
    L{0.135\linewidth}
    L{0.130\linewidth}
    L{0.175\linewidth}
    L{0.120\linewidth}
}

\caption{\textbf{Baseline Characteristics of the Five WEAR-ME Coachable Personas.} Summary of demographic, physiological, and wearable-derived characteristics across the five personas, highlighting differences in baseline health profiles and activity patterns.}
\label{tab:supp_personas}
\\

\toprule
\textbf{Persona ID \& Clinical Archetype} &
\textbf{Selection Criteria (Steps, Sleep, BMI, $\Delta$HA)} &
\textbf{Full WEAR-ME Sub-Cohort ($n$, \% of 2,780)} &
\textbf{Sub-Cohort Means (Age, Steps, Sleep, BMI, RHR, $\Delta$HA)} &
\textbf{Evaluated Sample ($n$ Valid Cases; PIDs)} &
\textbf{Evaluated Sample Baseline (Mean $\pm$ SD)} &
\textbf{Primary Counterfactual Coaching Test} \\
\midrule
\endfirsthead

\multicolumn{7}{l}{\textit{Supplementary Table~\ref{tab:supp_personas} continued from previous page}}\\[2mm]

\toprule
\textbf{Persona ID \& Clinical Archetype} &
\textbf{Selection Criteria (Steps, Sleep, BMI, $\Delta$HA)} &
\textbf{Full WEAR-ME Sub-Cohort ($n$, \% of 2,780)} &
\textbf{Sub-Cohort Means (Age, Steps, Sleep, BMI, RHR, $\Delta$HA)} &
\textbf{Evaluated Sample ($n$ Valid Cases; PIDs)} &
\textbf{Evaluated Sample Baseline (Mean $\pm$ SD)} &
\textbf{Primary Counterfactual Coaching Test} \\
\midrule
\endhead

\midrule
\multicolumn{7}{r}{\textit{Continued on next page}}\\
\endfoot

\bottomrule
\endlastfoot


\textbf{P1: Optimal Baseline}

\textit{``The High-Vitality Longevity Pacer''}
&
Steps: $\geq 10$k/day

Sleep: 7.0--9.0 h

BMI: 18.5--24.9

$\Delta$HA: $< -1.0$ yr
&
$n = 78$

(2.81\%)
&
Age: 53.0 yrs

Steps: 13,775

Sleep: 7.42 h

BMI: 22.9 kg/m$^2$

RHR: 60.1 bpm

$\Delta$HA: $-4.99$ yrs
&
$n = 10$

\texttt{12611},
\texttt{72404},
\texttt{58662},
\texttt{55814},
\texttt{56084},
\texttt{55957},
\texttt{71670},
\texttt{72398},
\texttt{71845},
\texttt{56638}
&
Age: $57.6 \pm 7.2$ yrs

Female: 40.0\% (4/10)

BMI: $23.2 \pm 0.7$ kg/m$^2$

Steps: $13,111 \pm 2,455$

Sleep: $7.22 \pm 0.44$ h

RHR: $59.1 \pm 3.5$ bpm

HA: $56.4 \pm 7.2$ yrs

$\Delta$HA: $-1.22 \pm 0.16$ yrs
&
\textbf{Ceiling Effects \& Maintenance:}

Preventing overtraining and framing vitality preservation against age-related drift.
\\
\midrule


\textbf{P2: Multi-Risk Challenger}

\textit{``The Sedentary Fatigued Metabolic Seeker''}
&
Steps: $< 5$k/day

Sleep: $< 7.0$ h

BMI: $\geq 25.0$

$\Delta$HA: $\geq +1.0$ yr
&
$n = 277$

(9.96\%)
&
Age: 49.8 yrs

Steps: 3,341

Sleep: 6.03 h

BMI: 36.3 kg/m$^2$

RHR: 65.4 bpm

$\Delta$HA: $+3.06$ yrs
&
$n = 10$

\texttt{186449},
\texttt{56440},
\texttt{13533},
\texttt{70268},
\texttt{70494},
\texttt{56314},
\texttt{57778},
\texttt{58098},
\texttt{72185},
\texttt{56458}
&
Age: $49.4 \pm 4.6$ yrs

Female: 70.0\% (7/10)

BMI: $40.6 \pm 5.9$ kg/m$^2$

Steps: $3,415 \pm 1,862$

Sleep: $5.31 \pm 1.38$ h

RHR: $68.8 \pm 3.0$ bpm

HA: $57.4 \pm 4.9$ yrs

$\Delta$HA: $+8.04 \pm 1.09$ yrs
&
\textbf{Multi-Lever Prioritization:}

Avoiding cognitive overwhelm by offering low-friction entry pathways (Sleep vs.\ Steps) in high-risk patients.
\\
\midrule


\textbf{P3: Recovery Bottleneck}

\textit{``The Active Striver at Burnout Risk''}
&
Steps: $\geq 10$k/day

Sleep: $< 7.0$ h

BMI: Healthy/OW

$\Delta$HA: Youthful/Sync
&
$n = 354$

(12.73\%)
&
Age: 49.1 yrs

Steps: 12,738

Sleep: 6.23 h

BMI: 26.9 kg/m$^2$

RHR: 62.6 bpm

$\Delta$HA: $-4.64$ yrs
&
$n = 10$

\texttt{70621},
\texttt{71746},
\texttt{71100},
\texttt{58093},
\texttt{70552},
\texttt{13851},
\texttt{55974},
\texttt{70488},
\texttt{71651},
\texttt{58419}
&
Age: $48.6 \pm 2.5$ yrs

Female: 10.0\% (1/10)

BMI: $26.6 \pm 1.3$ kg/m$^2$

Steps: $12,404 \pm 3,439$

Sleep: $5.92 \pm 1.54$ h

RHR: $64.7 \pm 5.8$ bpm

HA: $49.2 \pm 4.1$ yrs

$\Delta$HA: $+0.65 \pm 2.84$ yrs
&
\textbf{Lever Discrimination:}

Recognizing saturated activity volume and isolating sleep extension and recovery rather than adding steps.
\\
\midrule


\textbf{P4: Activity Bottleneck}

\textit{``The Sedentary Age-Accelerated Individual''}
&
Steps: $< 5$k/day

Sleep: 7.0--9.0 h

BMI: OW/Healthy

$\Delta$HA: $\geq +1.0$ yr
&
$n = 227$

(8.17\%)
&
Age: 48.2 yrs

Steps: 3,190

Sleep: 7.67 h

BMI: 32.0 kg/m$^2$

RHR: 64.3 bpm

$\Delta$HA: $+2.78$ yrs
&
$n = 10$

\texttt{12803},
\texttt{13530},
\texttt{13656},
\texttt{56193},
\texttt{72532},
\texttt{72585},
\texttt{72223},
\texttt{71774},
\texttt{57671},
\texttt{58005}
&
Age: $46.1 \pm 4.9$ yrs

Female: 100.0\% (10/10)

BMI: $35.1 \pm 4.7$ kg/m$^2$

Steps: $2,796 \pm 788$

Sleep: $6.95 \pm 1.41$ h

RHR: $72.0 \pm 7.7$ bpm

HA: $54.3 \pm 4.6$ yrs

$\Delta$HA: $+8.17 \pm 2.65$ yrs
&
\textbf{Targeted Activity Escalation:}

Isolating progressive step accumulation while validating and preserving healthy sleep habits.
\\
\midrule


\textbf{P5: Balanced Maintainer}

\textit{``The Balanced Health Seeker''}
&
Steps: 5k--10k/day

Sleep: 7.0--9.0 h

BMI: Healthy/OW

$\Delta$HA: $-1.0$ to $+1.0$
&
$n = 143$

(5.14\%)
&
Age: 46.6 yrs

Steps: 5,539

Sleep: 7.48 h

BMI: 29.9 kg/m$^2$

RHR: 67.1 bpm

$\Delta$HA: $-0.31$ yrs
&
$n = 9$ valid$^\dagger$

\texttt{55807},
\texttt{56085},
\texttt{56144},
\texttt{58421},
\texttt{70909},
\texttt{57709},
\texttt{57065},
\texttt{58023},
\texttt{57497}
&
Age: $45.4 \pm 5.0$ yrs

Female: 66.7\% (6/9)

BMI: $30.2 \pm 3.5$ kg/m$^2$

Steps: $5,490 \pm 2,093$

Sleep: $7.12 \pm 1.38$ h

RHR: $69.7 \pm 7.7$ bpm

HA: $50.1 \pm 5.1$ yrs

$\Delta$HA: $+4.62 \pm 3.66$ yrs
&
\textbf{Fine-Tuning \& Progression:}

Nudging median compliant users from baseline maintenance into cardiovascular fitness gains.
\\
\midrule


\textbf{Total / Overall}
&
\textbf{5 Stratified Personas}
&
\textbf{1,079 / 2,780}

(38.81\%)
&
\textbf{Weighted Sub-Cohort}
&
\textbf{$N = 49$ valid}

(50 sampled)
&
Age: $49.5 \pm 6.6$ yrs

Female: 57.1\% (28/49)

BMI: $31.2 \pm 7.2$ kg/m$^2$

Steps: $7,483 \pm 5,026$

Sleep: $6.49 \pm 1.46$ h

RHR: $66.8 \pm 7.3$ bpm

HA: $53.6 \pm 6.1$ yrs

$\Delta$HA: $+4.04 \pm 4.52$ yrs
&
\textbf{Comprehensive Evaluation Across the Risk Continuum}
\\

\midrule

\multicolumn{7}{p{0.96\linewidth}}{
\footnotesize
$^\dagger$ In Persona 5, Participant ID \texttt{57217} (Row 47) was excluded \textit{a priori} because \texttt{Plan 1} and \texttt{Plan 2} contained identical text. Sub-cohort means reflect 24-week habitual means in the WEAR-ME persona registry, whereas evaluated sample means reflect the recent daily baseline snapshot presented to raters. OW, overweight.
}
\\

\end{longtable}

\end{landscape}

\begin{longtable}{
    L{0.28\textwidth}
    L{0.32\textwidth}
    L{0.32\textwidth}
}

\caption{\textbf{Comprehensive Evaluation Results of the Lifestyle Coach.} Evaluation outcomes across five coaches and 49 valid cases, including head-to-head side-by-side (SxS) preference comparisons and score distributions across a 10-aspect evaluation rubric.}
\label{tab:supp_coach_results}
\\

\toprule
\textbf{Evaluation Criterion / Statistic} &
\textbf{Counterfactual Coaching} &
\textbf{Generic Coaching} \\
\midrule
\endfirsthead

\multicolumn{3}{l}{
\textit{Supplementary Table~\ref{tab:supp_coach_results} continued from previous page}
}
\\[1.5mm]

\toprule
\textbf{Evaluation Criterion / Statistic} &
\textbf{Counterfactual Coaching} &
\textbf{Generic Coaching} \\
\midrule
\endhead

\midrule
\multicolumn{3}{r}{\textit{Continued on next page}} \\
\endfoot

\bottomrule
\endlastfoot


\multicolumn{3}{l}{
\textbf{SxS: Best Overall Plan}
}
\\
\addlinespace[1mm]

Scale &
Forced-choice win &
Forced-choice win
\\

Full export ($n=245$) &
\textbf{71.4\%} ($175/245$) &
\textbf{28.6\%} ($70/245$)
\\

95\% CI &
65.5--76.7\% &
23.3--34.5\%
\\

Net difference &
\multicolumn{2}{L{0.65\textwidth}}{
\textbf{+42.9 percentage points} favoring Counterfactual Coaching.
}
\\

Exact $P$-value &
\multicolumn{2}{L{0.65\textwidth}}{
Two-sided exact binomial $P = 1.49 \times 10^{-11}$.
}
\\

Participant-level majority consensus &
\textbf{43/49 cases (87.8\%)} &
6/49 cases (12.2\%)
\\

\midrule


\multicolumn{3}{l}{
\textbf{SxS: Highest Likelihood of Long-Term Adherence}
}
\\
\addlinespace[1mm]

Scale &
Forced-choice win &
Forced-choice win
\\

Full export ($n=245$) &
\textbf{71.0\%} ($174/245$) &
\textbf{29.0\%} ($71/245$)
\\

95\% CI &
65.0--76.3\% &
23.7--35.0\%
\\

Net difference &
\multicolumn{2}{L{0.65\textwidth}}{
\textbf{+42.0 percentage points} favoring Counterfactual Coaching.
}
\\

Exact $P$-value &
\multicolumn{2}{L{0.65\textwidth}}{
Two-sided exact binomial $P = 3.72 \times 10^{-11}$.
}
\\

Participant-level majority consensus &
\textbf{44/49 cases (89.8\%)} &
5/49 cases (10.2\%)
\\

\midrule


\multicolumn{3}{l}{
\textbf{Contrasts Benefits vs.\ Status Quo}
}
\\
\addlinespace[1mm]

Scale &
Binary (Yes/No) &
Binary (Yes/No)
\\

Full export: Yes &
\textbf{180/245 (73.5\%)} &
\textbf{14/245 (5.7\%)}
\\

95\% CI &
67.6--78.6\% &
3.4--9.4\%
\\

Full distribution: No &
65/245 (26.5\%) &
231/245 (94.3\%)
\\

Net difference &
\multicolumn{2}{L{0.65\textwidth}}{
\textbf{+67.8 percentage points} favoring Counterfactual Coaching.
}
\\

Discordant pairs &
170 &
4
\\

Exact McNemar $P$-value &
\multicolumn{2}{L{0.65\textwidth}}{
$P = 3.15 \times 10^{-45}$.
}
\\

\midrule


\multicolumn{3}{l}{
\textbf{Compelling Reason to Change (No Fear/Shame)}
}
\\
\addlinespace[1mm]

Scale &
Binary (Yes/No) &
Binary (Yes/No)
\\

Full export: Yes &
\textbf{185/245 (75.5\%)} &
\textbf{70/245 (28.6\%)}
\\

95\% CI &
69.8--80.5\% &
23.3--34.5\%
\\

Full distribution: No &
60/245 (24.5\%) &
175/245 (71.4\%)
\\

Net difference &
\multicolumn{2}{L{0.65\textwidth}}{
\textbf{+46.9 percentage points} favoring Counterfactual Coaching.
}
\\

Discordant pairs &
121 &
6
\\

Exact McNemar $P$-value &
\multicolumn{2}{L{0.65\textwidth}}{
$P = 6.39 \times 10^{-29}$.
}
\\

\midrule


\multicolumn{3}{l}{
\textbf{Trajectory Control}
}
\\
\addlinespace[1mm]

Scale &
Binary (Yes/No) &
Binary (Yes/No)
\\

Full export: Yes &
\textbf{238/245 (97.1\%)} &
\textbf{145/244 (59.4\%)}$^\ddagger$
\\

95\% CI &
94.2--98.6\% &
53.2--65.4\%
\\

Full distribution: No &
7/245 (2.9\%) &
99/244 (40.6\%)
\\

Net difference &
\multicolumn{2}{L{0.65\textwidth}}{
\textbf{+37.7 percentage points} favoring Counterfactual Coaching.
}
\\

Discordant pairs &
97 &
5
\\

Exact McNemar $P$-value &
\multicolumn{2}{L{0.65\textwidth}}{
$P = 3.46 \times 10^{-23}$.
}
\\

\midrule


\multicolumn{3}{l}{
\textbf{Behavior-Change Strategies}
}
\\
\addlinespace[1mm]

Scale &
3-level ordinal &
3-level ordinal
\\

Top-box definition &
Strongly Applied &
Strongly Applied
\\

Strongly Applied &
\textbf{123/245 (50.2\%)} &
\textbf{50/245 (20.4\%)}
\\

95\% CI for top-box &
44.0--56.4\% &
15.8--25.9\%
\\

Superficial &
121/245 (49.4\%) &
185/245 (75.5\%)
\\

Absent &
1/245 (0.4\%) &
10/245 (4.1\%)
\\

Net top-box difference &
\multicolumn{2}{L{0.65\textwidth}}{
\textbf{+29.8 percentage points} favoring Counterfactual Coaching.
}
\\

Discordant pairs &
89 &
16
\\

Exact McNemar $P$-value &
\multicolumn{2}{L{0.65\textwidth}}{
$P = 1.87 \times 10^{-13}$.
}
\\

\midrule


\multicolumn{3}{l}{
\textbf{Sustainable Habit Building}
}
\\
\addlinespace[1mm]

Scale &
3-level ordinal &
3-level ordinal
\\

Top-box definition &
Excellent &
Excellent
\\

Excellent &
\textbf{128/245 (52.2\%)} &
\textbf{70/245 (28.6\%)}
\\

95\% CI for top-box &
46.0--58.4\% &
23.3--34.5\%
\\

Satisfactory &
114/245 (46.5\%) &
167/245 (68.2\%)
\\

Poor &
3/245 (1.2\%) &
8/245 (3.3\%)
\\

Net top-box difference &
\multicolumn{2}{L{0.65\textwidth}}{
\textbf{+23.7 percentage points} favoring Counterfactual Coaching.
}
\\

Discordant pairs &
75 &
17
\\

Exact McNemar $P$-value &
\multicolumn{2}{L{0.65\textwidth}}{
$P = 7.29 \times 10^{-10}$.
}
\\

\midrule


\multicolumn{3}{l}{
\textbf{Realistic \& Non-Overwhelming Routine Fit}
}
\\
\addlinespace[1mm]

Scale &
Binary (Yes/No) &
Binary (Yes/No)
\\

Full export: Yes &
\textbf{242/245 (98.8\%)} &
\textbf{203/245 (82.9\%)}
\\

95\% CI &
96.5--99.6\% &
77.6--87.1\%
\\

Full distribution: No &
3/245 (1.2\%) &
42/245 (17.1\%)
\\

Net difference &
\multicolumn{2}{L{0.65\textwidth}}{
\textbf{+15.9 percentage points} favoring Counterfactual Coaching.
}
\\

Discordant pairs &
41 &
2
\\

Exact McNemar $P$-value &
\multicolumn{2}{L{0.65\textwidth}}{
$P = 2.15 \times 10^{-10}$.
}
\\

\midrule


\multicolumn{3}{l}{
\textbf{Supportive \& Motivating Tone}
}
\\
\addlinespace[1mm]

Scale &
Binary (Yes/No) &
Binary (Yes/No)
\\

Full export: Yes &
\textbf{241/245 (98.4\%)} &
\textbf{204/245 (83.3\%)}
\\

95\% CI &
95.9--99.4\% &
78.1--87.4\%
\\

Full distribution: No &
4/245 (1.6\%) &
41/245 (16.7\%)
\\

Net difference &
\multicolumn{2}{L{0.65\textwidth}}{
\textbf{+15.1 percentage points} favoring Counterfactual Coaching.
}
\\

Discordant pairs &
41 &
4
\\

Exact McNemar $P$-value &
\multicolumn{2}{L{0.65\textwidth}}{
$P = 9.33 \times 10^{-9}$.
}
\\

\midrule


\multicolumn{3}{l}{
\textbf{Baseline \& Wearable Personalization}
}
\\
\addlinespace[1mm]

Scale &
3-level ordinal &
3-level ordinal
\\

Top-box definition &
Highly Tailored &
Highly Tailored
\\

Highly Tailored &
\textbf{133/245 (54.3\%)} &
\textbf{127/245 (51.8\%)}
\\

95\% CI for top-box &
48.0--60.4\% &
45.6--58.0\%
\\

Partially Tailored &
106/245 (43.3\%) &
108/245 (44.1\%)
\\

Generic &
6/245 (2.4\%) &
10/245 (4.1\%)$^\S$
\\

Net top-box difference &
\multicolumn{2}{L{0.65\textwidth}}{
\textbf{+2.4 percentage points} favoring Counterfactual Coaching.
}
\\

Discordant pairs &
42 &
36
\\

Exact McNemar $P$-value &
\multicolumn{2}{L{0.65\textwidth}}{
$P = 0.572$; not statistically significant.
}
\\

\midrule


\multicolumn{3}{l}{
\textbf{Signal Synthesis (Unified Compass)}
}
\\
\addlinespace[1mm]

Scale &
3-level ordinal &
3-level ordinal
\\

Top-box definition &
Unified Compass &
Unified Compass
\\

Unified Compass &
\textbf{120/245 (49.0\%)} &
\textbf{119/245 (48.6\%)}
\\

95\% CI for top-box &
42.8--55.2\% &
42.4--54.8\%
\\

Descriptive Summary &
106/245 (43.3\%) &
73/245 (29.8\%)
\\

Fragmented &
19/245 (7.8\%)$^\S$ &
48/245 (19.6\%)
\\

Net top-box difference &
\multicolumn{2}{L{0.65\textwidth}}{
\textbf{+0.4 percentage points}.
}
\\

Discordant pairs &
47 &
46
\\

Exact McNemar $P$-value &
\multicolumn{2}{L{0.65\textwidth}}{
$P = 1.000$; not statistically significant.
}
\\

\midrule


\multicolumn{3}{l}{
\textbf{Easy to Understand}
}
\\
\addlinespace[1mm]

Scale &
Binary (Yes/No) &
Binary (Yes/No)
\\

Full export: Yes &
\textbf{208/245 (84.9\%)} &
\textbf{220/245 (89.8\%)}
\\

95\% CI &
79.9--88.8\% &
85.4--93.0\%
\\

Full distribution: No &
37/245 (15.1\%) &
25/245 (10.2\%)
\\

Net difference &
\multicolumn{2}{L{0.65\textwidth}}{
\textbf{$-4.9$ percentage points} for Counterfactual relative to Generic Coaching.
}
\\

Discordant pairs &
18 &
30
\\

Exact McNemar $P$-value &
\multicolumn{2}{L{0.65\textwidth}}{
$P = 0.111$; not statistically significant.
}
\\

\midrule


\multicolumn{3}{L{0.94\textwidth}}{
\footnotesize
$^\ddagger$ One Generic \texttt{trajectory\_control} rating was left blank by a coach, resulting in $n=244$ for this item. In 5/245 Generic \texttt{wearable\_synthesis} cells, a text comment was entered in place of a categorical rating.
}
\\[1mm]

\multicolumn{3}{L{0.94\textwidth}}{
\footnotesize
$^\S$ Although top-box \textit{Unified Compass} rates were similar (49.0\% vs.\ 48.6\%), Counterfactual Coaching reduced bottom-box \textit{Fragmented} synthesis from 19.6\% to 7.8\%. Counterfactual Coaching also reduced bottom-box \textit{Generic} personalization from 4.1\% to 2.4\%.
}
\\[1mm]

\multicolumn{3}{L{0.94\textwidth}}{
\footnotesize
CF, Counterfactual Coaching; Gen, Generic Coaching; SxS, side-by-side comparison; CI, confidence interval. Unless otherwise specified, values are reported as Counterfactual versus Generic. Paired rubric comparisons were assessed using exact McNemar tests.
}
\\

\end{longtable}

\clearpage

\begin{longtable}{
    L{0.28\textwidth}
    L{0.32\textwidth}
    L{0.32\textwidth}
}

\caption{\textbf{Comprehensive Evaluation Results of the Clinician Assessment.} Evaluation outcomes across five clinicians and 49 valid cases, comprising 245 paired evaluations across six dimensions assessing safety, accuracy, plausibility, and personalization.}
\label{tab:supp_clinician_results}
\\

\toprule
\textbf{Evaluation Criterion / Statistic} &
\textbf{Generic Coaching} &
\textbf{Counterfactual Coaching} \\
\midrule
\endfirsthead

\multicolumn{3}{l}{
\textit{Supplementary Table~\ref{tab:supp_clinician_results} continued from previous page}
}
\\[1.5mm]

\toprule
\textbf{Evaluation Criterion / Statistic} &
\textbf{Generic Coaching} &
\textbf{Counterfactual Coaching} \\
\midrule
\endhead

\midrule
\multicolumn{3}{r}{\textit{Continued on next page}} \\
\endfoot

\bottomrule
\endlastfoot


\multicolumn{3}{l}{
\textbf{Safe Intensity Limits (Primary Safety Gate)}
}
\\
\addlinespace[1mm]

Scale &
Binary (Yes/No) &
Binary (Yes/No)
\\

Fig.~\ref{fig:main1} \& \ref{fig:human_eval} bar values &
\textbf{97.6\% Yes} (panel \ref{fig:main1}e)

\textbf{98\% Yes} (panel \ref{fig:human_eval}b)
&
\textbf{99.2\% Yes} (panel \ref{fig:main1}e)

\textbf{99\% Yes} (panel \ref{fig:human_eval}b)
\\

Full export: Yes &
\textbf{239/245 (97.6\%)} &
\textbf{243/245 (99.2\%)}
\\

95\% CI &
94.8--98.9\% &
97.1--99.8\%
\\

Full distribution: No &
6/245 (2.4\%) &
2/245 (0.8\%)
\\

Net difference (CF$-$Gen) &
\multicolumn{2}{L{0.65\textwidth}}{
\textbf{+1.6 percentage points}.
}
\\

Discordant pairs ($b$ vs.\ $c$) &
6 &
2
\\

Exact McNemar $P$-value &
\multicolumn{2}{L{0.65\textwidth}}{
$P = 0.289$; not statistically significant.
}
\\

\midrule


\multicolumn{3}{l}{
\textbf{Physiological Claims Correct}
}
\\
\addlinespace[1mm]

Scale &
Binary (Yes/No) &
Binary (Yes/No)
\\

Fig.~\ref{fig:human_eval} bar values &
\textbf{98\% Yes} &
\textbf{97\% Yes}
\\

Full export: Yes &
\textbf{239/245 (97.6\%)} &
\textbf{238/245 (97.1\%)}
\\

95\% CI &
94.8--98.9\% &
94.2--98.6\%
\\

Full distribution: No &
6/245 (2.4\%) &
7/245 (2.9\%)
\\

Net difference (CF$-$Gen) &
\multicolumn{2}{L{0.65\textwidth}}{
\textbf{$-0.4$ percentage points}.
}
\\

Discordant pairs ($b$ vs.\ $c$) &
2 &
3
\\

Exact McNemar $P$-value &
\multicolumn{2}{L{0.65\textwidth}}{
$P = 1.000$; not statistically significant.
}
\\

\midrule


\multicolumn{3}{l}{
\textbf{Trajectory Control}
}
\\
\addlinespace[1mm]

Scale &
Binary (Yes/No) &
Binary (Yes/No)
\\

Fig.~\ref{fig:human_eval} bar values &
\textbf{93\% Yes} &
\textbf{97\% Yes}
\\

Full export: Yes &
\textbf{228/245 (93.1\%)} &
\textbf{237/245 (96.7\%)}
\\

95\% CI &
89.2--95.6\% &
93.7--98.3\%
\\

Full distribution: No &
17/245 (6.9\%) &
8/245 (3.3\%)
\\

Net difference (CF$-$Gen) &
\multicolumn{2}{L{0.65\textwidth}}{
\textbf{+3.7 percentage points}.
}
\\

Discordant pairs ($b$ vs.\ $c$) &
16 &
7
\\

Exact McNemar $P$-value &
\multicolumn{2}{L{0.65\textwidth}}{
$P = 0.093$; not statistically significant.
}
\\

\midrule


\multicolumn{3}{l}{
\textbf{12-Week Biological Plausibility}
}
\\
\addlinespace[1mm]

Scale &
Binary (Yes/No) &
Binary (Yes/No)
\\

Fig.~\ref{fig:human_eval} bar values &
\textbf{82\% Yes} &
\textbf{86\% Yes}
\\

Full export: Yes &
\textbf{200/245 (81.6\%)} &
\textbf{211/245 (86.1\%)}
\\

95\% CI &
76.3--86.0\% &
81.2--89.9\%
\\

Full distribution: No &
45/245 (18.4\%) &
34/245 (13.9\%)
\\

Net difference (CF$-$Gen) &
\multicolumn{2}{L{0.65\textwidth}}{
\textbf{+4.5 percentage points}.
}
\\

Discordant pairs ($b$ vs.\ $c$) &
18 &
7
\\

Exact McNemar $P$-value &
\multicolumn{2}{L{0.65\textwidth}}{
$P = 0.043$.
}
\\

\midrule


\multicolumn{3}{l}{
\textbf{Contrasts Benefits vs.\ Status Quo}
}
\\
\addlinespace[1mm]

Scale &
Binary (Yes/No) &
Binary (Yes/No)
\\

Fig.~\ref{fig:human_eval} bar values &
\textbf{8\% Yes} &
\textbf{92\% Yes}
\\

Full export: Yes &
\textbf{19/245 (7.8\%)} &
\textbf{226/245 (92.2\%)}
\\

95\% CI &
5.0--11.8\% &
88.2--95.0\%
\\

Full distribution: No &
226/245 (92.2\%) &
19/245 (7.8\%)
\\

Net difference (CF$-$Gen) &
\multicolumn{2}{L{0.65\textwidth}}{
\textbf{+84.5 percentage points}.
}
\\

Discordant pairs ($b$ vs.\ $c$) &
213 &
6
\\

Exact McNemar $P$-value &
\multicolumn{2}{L{0.65\textwidth}}{
$P = 3.49 \times 10^{-55}$.
}
\\

\midrule


\multicolumn{3}{l}{
\textbf{Baseline \& Wearable Personalization}
}
\\
\addlinespace[1mm]

Scale &
3-level ordinal &
3-level ordinal
\\

Top-box definition &
Highly Tailored &
Highly Tailored
\\

Fig.~\ref{fig:human_eval} bar values &
\textbf{52\%} &
\textbf{72\%}
\\

Highly Tailored &
\textbf{128/245 (52.2\%)} &
\textbf{177/245 (72.2\%)}
\\

95\% CI for top-box &
46.0--58.4\% &
66.3--77.5\%
\\

Partially Tailored &
109/245 (44.5\%) &
50/245 (20.4\%)
\\

Generic &
8/245 (3.3\%) &
18/245 (7.3\%)
\\

Net top-box difference (CF$-$Gen) &
\multicolumn{2}{L{0.65\textwidth}}{
\textbf{+20.0 percentage points}.
}
\\

Discordant pairs ($b$ vs.\ $c$) &
82 &
33
\\

Exact McNemar $P$-value &
\multicolumn{2}{L{0.65\textwidth}}{
$P = 5.57 \times 10^{-6}$.
}
\\

\midrule


\multicolumn{3}{L{0.94\textwidth}}{
\footnotesize
Gen, Generic Coaching; CF, Counterfactual Coaching; CI, confidence interval. Values are reported for $n=245$ paired clinician evaluations unless otherwise stated.
}
\\[1mm]

\multicolumn{3}{L{0.94\textwidth}}{
\footnotesize
Paired comparisons were assessed using two-sided exact McNemar tests. Exact $P$-values are reported directly; $P<0.05$ was considered statistically significant.
}
\\

\end{longtable}

\clearpage

\begin{longtable}{
    L{0.28\textwidth}
    L{0.32\textwidth}
    L{0.32\textwidth}
}
\caption{\textbf{Cross-Disciplinary Concordance and Divergence in Evaluation Outcomes.} Agreement and disagreement between lifestyle coaches and clinicians across shared evaluation dimensions, based on 490 pooled paired evaluations.}
\label{tab:supp_concordance}
\\

\toprule
\textbf{Shared Dimension / Statistic} &
\textbf{Lifestyle Coaches ($n=245$)} &
\textbf{Clinicians ($n=245$)} \\
\midrule
\endfirsthead

\multicolumn{3}{l}{
\textit{Supplementary Table~\ref{tab:supp_concordance} continued from previous page}
}
\\[1.5mm]

\toprule
\textbf{Shared Dimension / Statistic} &
\textbf{Lifestyle Coaches ($n=245$)} &
\textbf{Clinicians ($n=245$)} \\
\midrule
\endhead

\midrule
\multicolumn{3}{r}{\textit{Continued on next page}} \\
\endfoot

\bottomrule
\endlastfoot


\multicolumn{3}{l}{
\textbf{Contrasts Benefits vs.\ Status Quo}
}
\\
\addlinespace[1mm]

Scale &
Binary (\% Yes) &
Binary (\% Yes)
\\

Counterfactual &
\textbf{73.5\%} &
\textbf{92.2\%}
\\

Generic &
5.7\% &
7.8\%
\\

Net difference (CF$-$Gen) &
\textbf{+67.8 percentage points} &
\textbf{+84.5 percentage points}
\\

Exact $P$-value &
$3.15 \times 10^{-45}$ &
$3.49 \times 10^{-55}$
\\

Pooled Counterfactual &
\multicolumn{2}{L{0.65\textwidth}}{
\textbf{82.9\% (406/490)}
}
\\

Pooled Generic &
\multicolumn{2}{L{0.65\textwidth}}{
\textbf{6.7\% (33/490)}
}
\\

Pooled difference &
\multicolumn{2}{L{0.65\textwidth}}{
\textbf{+76.1 percentage points}; $P = 2.20 \times 10^{-99}$.
}
\\

Cross-disciplinary interpretation &
\multicolumn{2}{L{0.65\textwidth}}{
\textbf{Strong Concordance:}
Both expert groups showed the same qualitative pattern. Generic plans rarely contrasted the projected status quo with the benefits of behavioral change, whereas Counterfactual plans consistently provided this comparison.
}
\\

\midrule


\multicolumn{3}{l}{
\textbf{Trajectory Control}
}
\\
\addlinespace[1mm]

Scale &
Binary (\% Yes) &
Binary (\% Yes)
\\

Counterfactual &
\textbf{97.1\%} &
\textbf{96.7\%}
\\

Generic &
59.4\% &
93.1\%
\\

Net difference (CF$-$Gen) &
\textbf{+37.7 percentage points} &
\textbf{+3.7 percentage points}
\\

Exact $P$-value &
$3.46 \times 10^{-23}$ &
0.093
\\

Pooled Counterfactual &
\multicolumn{2}{L{0.65\textwidth}}{
\textbf{96.9\% (475/490)}
}
\\

Pooled Generic &
\multicolumn{2}{L{0.65\textwidth}}{
\textbf{76.3\% (373/489)}
}
\\

Pooled difference &
\multicolumn{2}{L{0.65\textwidth}}{
\textbf{+20.7 percentage points}; $P = 9.25 \times 10^{-22}$.
}
\\

Cross-disciplinary interpretation &
\multicolumn{2}{L{0.65\textwidth}}{
\textbf{Behavioral Stringency:}
Both groups rated Counterfactual plans near ceiling for trajectory control ($\sim97\%$). Lifestyle coaches, however, rated Generic plans substantially lower (59.4\%), consistent with a stricter emphasis on explicit choice architecture and multiple behavioral pathways. Clinicians rated Generic plans highly (93.1\%), indicating that a gradual, modifiable 12-week plan was often sufficient to satisfy their interpretation of trajectory control.
}
\\

\midrule


\multicolumn{3}{l}{
\textbf{Baseline \& Wearable Personalization}
}
\\
\addlinespace[1mm]

Scale &
3-level ordinal (\% Highly Tailored) &
3-level ordinal (\% Highly Tailored)
\\

Counterfactual &
\textbf{54.3\%} &
\textbf{72.2\%}
\\

Generic &
51.8\% &
52.2\%
\\

Net difference (CF$-$Gen) &
\textbf{+2.4 percentage points} &
\textbf{+20.0 percentage points}
\\

Exact $P$-value &
0.572 &
$5.57 \times 10^{-6}$
\\

Pooled Counterfactual &
\multicolumn{2}{L{0.65\textwidth}}{
\textbf{63.3\% (310/490)}
}
\\

Pooled Generic &
\multicolumn{2}{L{0.65\textwidth}}{
\textbf{52.0\% (255/490)}
}
\\

Pooled difference &
\multicolumn{2}{L{0.65\textwidth}}{
\textbf{+11.2 percentage points}; $P = 9.18 \times 10^{-5}$.
}
\\

Cross-disciplinary interpretation &
\multicolumn{2}{L{0.65\textwidth}}{
\textbf{Clinical vs.\ Psychosocial Lens:}
Clinicians showed a substantially larger personalization difference between Counterfactual and Generic plans, consistent with greater emphasis on physiological tailoring to Health Age and resting-heart-rate patterns. Lifestyle coaches rated the two conditions more similarly, reflecting a broader personalization standard that also considers psychosocial context such as life schedules, barriers, caregiving responsibilities, and access to exercise resources.
}
\\

\midrule


\multicolumn{3}{L{0.94\textwidth}}{
\footnotesize
CF, Counterfactual Coaching; Gen, Generic Coaching. Pooled values combine the 245 lifestyle-coach and 245 clinician evaluations. For Trajectory Control, the Generic pooled denominator is $n=489$ because one lifestyle-coach Generic rating was missing.
}
\\[1mm]

\multicolumn{3}{L{0.94\textwidth}}{
\footnotesize
The pooled statistics summarize the combined expert evaluations, whereas the discipline-specific columns show the corresponding results separately for lifestyle coaches and clinicians.
}
\\

\end{longtable}

\clearpage

\begin{longtable}{
    L{0.34\textwidth}
    L{0.26\textwidth}
    L{0.32\textwidth}
}
\caption{\textbf{Stratified Evaluation Performance Across the Five WEAR-ME Coachable Personas.} Comprehensive breakdown of evaluation outcomes stratified by coachable persona, highlighting variations in performance across distinct user profiles and corresponding to the results presented in Fig.~\ref{fig:human_eval}c.}
\label{tab:supp_persona_matrix}
\\

\toprule
\textbf{Dimension / Endpoint} &
\textbf{Coachable Persona} &
\textbf{Result} \\
\midrule
\endfirsthead

\multicolumn{3}{l}{
\textit{Supplementary Table~\ref{tab:supp_persona_matrix} continued from previous page}
}
\\[1.5mm]

\toprule
\textbf{Dimension / Endpoint} &
\textbf{Coachable Persona} &
\textbf{Result} \\
\midrule
\endhead

\midrule
\multicolumn{3}{r}{\textit{Continued on next page}} \\
\endfoot

\bottomrule
\endlastfoot


\multicolumn{3}{l}{
\textbf{Coach SxS: Best Overall Plan}
\quad
\textit{(Counterfactual vs.\ Generic; exact binomial $P$)}
}
\\
\addlinespace[1mm]

&
\textbf{P1: Optimal Baseline}

$\geq 10$k steps/day; 7--9 h sleep

$n=50$
&
\textbf{72.0\% vs.\ 28.0\%}

36/50

$P = 0.0026$
\\

&
\textbf{P2: Multi-Risk Challenger}

$<5$k steps/day; $<7$ h sleep; high BMI

$n=50$
&
\textbf{80.0\% vs.\ 20.0\%}

40/50

$P = 2.39 \times 10^{-5}$
\\

&
\textbf{P3: Recovery Bottleneck}

$\geq 10$k steps/day; $<7$ h sleep

$n=50$
&
\textbf{60.0\% vs.\ 40.0\%}

30/50

$P = 0.203$
\\

&
\textbf{P4: Activity Bottleneck}

$<5$k steps/day; 7--9 h sleep

$n=50$
&
\textbf{74.0\% vs.\ 26.0\%}

37/50

$P = 0.0009$
\\

&
\textbf{P5: Balanced Maintainer}

5--10k steps/day; 7--9 h sleep

$n=45$; 9 cases
&
\textbf{71.1\% vs.\ 28.9\%}

32/45

$P = 0.0066$
\\

\midrule


\multicolumn{3}{l}{
\textbf{Coach SxS: Long-Term Adherence}
\quad
\textit{(Counterfactual vs.\ Generic; exact binomial $P$)}
}
\\
\addlinespace[1mm]

&
\textbf{P1: Optimal Baseline}
&
\textbf{70.0\% vs.\ 30.0\%}

35/50

$P = 0.0066$
\\

&
\textbf{P2: Multi-Risk Challenger}
&
\textbf{82.0\% vs.\ 18.0\%}

41/50

$P = 5.61 \times 10^{-6}$
\\

&
\textbf{P3: Recovery Bottleneck}
&
\textbf{58.0\% vs.\ 42.0\%}

29/50

$P = 0.322$
\\

&
\textbf{P4: Activity Bottleneck}
&
\textbf{74.0\% vs.\ 26.0\%}

37/50

$P = 0.0009$
\\

&
\textbf{P5: Balanced Maintainer}
&
\textbf{71.1\% vs.\ 28.9\%}

32/45

$P = 0.0066$
\\

\midrule


\multicolumn{3}{l}{
\textbf{Trajectory Control (Lifestyle Coach)}
\quad
\textit{$\Delta$ = Counterfactual $-$ Generic}
}
\\
\addlinespace[1mm]

&
\textbf{P1: Optimal Baseline}
&
\textbf{+34.7 percentage points}

100\% vs.\ 65\%
\\

&
\textbf{P2: Multi-Risk Challenger}
&
\textbf{+48.0 percentage points}

100\% vs.\ 52\%
\\

&
\textbf{P3: Recovery Bottleneck}
&
\textbf{+24.0 percentage points}

96\% vs.\ 72\%
\\

&
\textbf{P4: Activity Bottleneck}
&
\textbf{+54.0 percentage points}

100\% vs.\ 46\%
\\

&
\textbf{P5: Balanced Maintainer}
&
\textbf{+26.7 percentage points}

89\% vs.\ 62\%
\\

\midrule


\multicolumn{3}{l}{
\textbf{Behavior-Change Strategies$^*$ (Lifestyle Coach)}
\quad
\textit{$\Delta$ = Counterfactual $-$ Generic}
}
\\
\addlinespace[1mm]

&
\textbf{P1: Optimal Baseline}
&
\textbf{+24.0 percentage points}

58\% vs.\ 34\%
\\

&
\textbf{P2: Multi-Risk Challenger}
&
\textbf{+32.0 percentage points}

48\% vs.\ 16\%
\\

&
\textbf{P3: Recovery Bottleneck}
&
\textbf{+18.0 percentage points}

44\% vs.\ 26\%
\\

&
\textbf{P4: Activity Bottleneck}
&
\textbf{+34.0 percentage points}

44\% vs.\ 10\%
\\

&
\textbf{P5: Balanced Maintainer}
&
\textbf{+42.2 percentage points}

58\% vs.\ 16\%
\\

\midrule


\multicolumn{3}{l}{
\textbf{Sustainable Habit Building$^*$ (Lifestyle Coach)}
\quad
\textit{$\Delta$ = Counterfactual $-$ Generic}
}
\\
\addlinespace[1mm]

&
\textbf{P1: Optimal Baseline}
&
\textbf{+28.0 percentage points}

64\% vs.\ 36\%
\\

&
\textbf{P2: Multi-Risk Challenger}
&
\textbf{+26.0 percentage points}

50\% vs.\ 24\%
\\

&
\textbf{P3: Recovery Bottleneck}
&
\textbf{+12.0 percentage points}

48\% vs.\ 36\%
\\

&
\textbf{P4: Activity Bottleneck}
&
\textbf{+26.0 percentage points}

50\% vs.\ 24\%
\\

&
\textbf{P5: Balanced Maintainer}
&
\textbf{+26.7 percentage points}

49\% vs.\ 22\%
\\

\midrule


\multicolumn{3}{l}{
\textbf{Realistic \& Manageable Routine Fit (Lifestyle Coach)}
\quad
\textit{$\Delta$ = Counterfactual $-$ Generic}
}
\\
\addlinespace[1mm]

&
\textbf{P1: Optimal Baseline}
&
\textbf{+16.0 percentage points}

98\% vs.\ 82\%
\\

&
\textbf{P2: Multi-Risk Challenger}
&
\textbf{+26.0 percentage points}

100\% vs.\ 74\%
\\

&
\textbf{P3: Recovery Bottleneck}
&
\textbf{+4.0 percentage points}

98\% vs.\ 94\%
\\

&
\textbf{P4: Activity Bottleneck}
&
\textbf{+14.0 percentage points}

100\% vs.\ 86\%
\\

&
\textbf{P5: Balanced Maintainer}
&
\textbf{+20.0 percentage points}

98\% vs.\ 78\%
\\

\midrule


\multicolumn{3}{l}{
\textbf{Baseline Personalization$^*$ (Lifestyle Coach)}
\quad
\textit{$\Delta$ = Counterfactual $-$ Generic}
}
\\
\addlinespace[1mm]

&
\textbf{P1: Optimal Baseline}
&
\textbf{+10.0 percentage points}

70\% vs.\ 60\%
\\

&
\textbf{P2: Multi-Risk Challenger}
&
\textbf{+4.0 percentage points}

48\% vs.\ 44\%
\\

&
\textbf{P3: Recovery Bottleneck}
&
\textbf{$-14.0$ percentage points}

44\% vs.\ 58\%
\\

&
\textbf{P4: Activity Bottleneck}
&
\textbf{+8.0 percentage points}

56\% vs.\ 48\%
\\

&
\textbf{P5: Balanced Maintainer}
&
\textbf{+4.4 percentage points}

53\% vs.\ 49\%
\\

\midrule


\multicolumn{3}{l}{
\textbf{Contrasts Benefits vs.\ Status Quo (Lifestyle Coach)}
\quad
\textit{$\Delta$ = Counterfactual $-$ Generic}
}
\\
\addlinespace[1mm]

&
\textbf{P1: Optimal Baseline}
&
\textbf{+72.0 percentage points}

78\% vs.\ 6\%
\\

&
\textbf{P2: Multi-Risk Challenger}
&
\textbf{+68.0 percentage points}

72\% vs.\ 4\%
\\

&
\textbf{P3: Recovery Bottleneck}
&
\textbf{+66.0 percentage points}

72\% vs.\ 6\%
\\

&
\textbf{P4: Activity Bottleneck}
&
\textbf{+72.0 percentage points}

74\% vs.\ 2\%
\\

&
\textbf{P5: Balanced Maintainer}
&
\textbf{+60.0 percentage points}

71\% vs.\ 11\%
\\

\midrule


\multicolumn{3}{l}{
\textbf{Compelling Reason to Change (Lifestyle Coach)}
\quad
\textit{$\Delta$ = Counterfactual $-$ Generic}
}
\\
\addlinespace[1mm]

&
\textbf{P1: Optimal Baseline}
&
\textbf{+42.0 percentage points}

78\% vs.\ 36\%
\\

&
\textbf{P2: Multi-Risk Challenger}
&
\textbf{+54.0 percentage points}

72\% vs.\ 18\%
\\

&
\textbf{P3: Recovery Bottleneck}
&
\textbf{+44.0 percentage points}

76\% vs.\ 32\%
\\

&
\textbf{P4: Activity Bottleneck}
&
\textbf{+54.0 percentage points}

78\% vs.\ 24\%
\\

&
\textbf{P5: Balanced Maintainer}
&
\textbf{+40.0 percentage points}

73\% vs.\ 33\%
\\

\midrule


\multicolumn{3}{l}{
\textbf{Supportive \& Motivating Tone (Lifestyle Coach)}
\quad
\textit{$\Delta$ = Counterfactual $-$ Generic}
}
\\
\addlinespace[1mm]

&
\textbf{P1: Optimal Baseline}
&
\textbf{+16.0 percentage points}

98\% vs.\ 82\%
\\

&
\textbf{P2: Multi-Risk Challenger}
&
\textbf{+16.0 percentage points}

98\% vs.\ 82\%
\\

&
\textbf{P3: Recovery Bottleneck}
&
\textbf{+14.0 percentage points}

100\% vs.\ 86\%
\\

&
\textbf{P4: Activity Bottleneck}
&
\textbf{+18.0 percentage points}

100\% vs.\ 82\%
\\

&
\textbf{P5: Balanced Maintainer}
&
\textbf{+11.1 percentage points}

96\% vs.\ 84\%
\\

\midrule


\multicolumn{3}{l}{
\textbf{Signal Synthesis$^*$ (Lifestyle Coach)}
\quad
\textit{$\Delta$ = Counterfactual $-$ Generic}
}
\\
\addlinespace[1mm]

&
\textbf{P1: Optimal Baseline}
&
\textbf{+10.0 percentage points}

56\% vs.\ 46\%
\\

&
\textbf{P2: Multi-Risk Challenger}
&
\textbf{$-2.0$ percentage points}

40\% vs.\ 42\%
\\

&
\textbf{P3: Recovery Bottleneck}
&
\textbf{+4.0 percentage points}

54\% vs.\ 50\%
\\

&
\textbf{P4: Activity Bottleneck}
&
\textbf{$-2.0$ percentage points}

46\% vs.\ 48\%
\\

&
\textbf{P5: Balanced Maintainer}
&
\textbf{$-8.9$ percentage points}

49\% vs.\ 58\%
\\

\midrule


\multicolumn{3}{l}{
\textbf{Easy to Understand (Lifestyle Coach)}
\quad
\textit{$\Delta$ = Counterfactual $-$ Generic}
}
\\
\addlinespace[1mm]

&
\textbf{P1: Optimal Baseline}
&
\textbf{0.0 percentage points}

94\% vs.\ 94\%
\\

&
\textbf{P2: Multi-Risk Challenger}
&
\textbf{$-8.0$ percentage points}

78\% vs.\ 86\%
\\

&
\textbf{P3: Recovery Bottleneck}
&
\textbf{$-10.0$ percentage points}

84\% vs.\ 94\%
\\

&
\textbf{P4: Activity Bottleneck}
&
\textbf{$-4.0$ percentage points}

78\% vs.\ 82\%
\\

&
\textbf{P5: Balanced Maintainer}
&
\textbf{$-2.2$ percentage points}

91\% vs.\ 93\%
\\

\midrule


\multicolumn{3}{l}{
\textbf{Baseline Personalization$^*$ (Clinician)}
\quad
\textit{$\Delta$ = Counterfactual $-$ Generic}
}
\\
\addlinespace[1mm]

&
\textbf{P1: Optimal Baseline}
&
\textbf{+16.0 percentage points}

76\% vs.\ 60\%
\\

&
\textbf{P2: Multi-Risk Challenger}
&
\textbf{+2.0 percentage points}

62\% vs.\ 60\%
\\

&
\textbf{P3: Recovery Bottleneck}
&
\textbf{+18.0 percentage points}

70\% vs.\ 52\%
\\

&
\textbf{P4: Activity Bottleneck}
&
\textbf{+38.0 percentage points}

76\% vs.\ 38\%
\\

&
\textbf{P5: Balanced Maintainer}
&
\textbf{+26.7 percentage points}

78\% vs.\ 51\%
\\

\midrule


\multicolumn{3}{l}{
\textbf{12-Week Biological Plausibility (Clinician)}
\quad
\textit{$\Delta$ = Counterfactual $-$ Generic}
}
\\
\addlinespace[1mm]

&
\textbf{P1: Optimal Baseline}
&
\textbf{+18.0 percentage points}

98\% vs.\ 80\%
\\

&
\textbf{P2: Multi-Risk Challenger}
&
\textbf{$-6.0$ percentage points}

76\% vs.\ 82\%
\\

&
\textbf{P3: Recovery Bottleneck}
&
\textbf{+6.0 percentage points}

92\% vs.\ 86\%
\\

&
\textbf{P4: Activity Bottleneck}
&
\textbf{+2.0 percentage points}

82\% vs.\ 80\%
\\

&
\textbf{P5: Balanced Maintainer}
&
\textbf{+2.2 percentage points}

82\% vs.\ 80\%
\\

\midrule


\multicolumn{3}{l}{
\textbf{Safe Intensity Limits (Clinician)}
\quad
\textit{$\Delta$ = Counterfactual $-$ Generic}
}
\\
\addlinespace[1mm]

&
\textbf{P1: Optimal Baseline}
&
\textbf{+2.0 percentage points}

100\% vs.\ 98\%
\\

&
\textbf{P2: Multi-Risk Challenger}
&
\textbf{+2.0 percentage points}

100\% vs.\ 98\%
\\

&
\textbf{P3: Recovery Bottleneck}
&
\textbf{+2.0 percentage points}

98\% vs.\ 96\%
\\

&
\textbf{P4: Activity Bottleneck}
&
\textbf{+2.0 percentage points}

100\% vs.\ 98\%
\\

&
\textbf{P5: Balanced Maintainer}
&
\textbf{0.0 percentage points}

98\% vs.\ 98\%
\\

\midrule


\multicolumn{3}{l}{
\textbf{Contrasts Benefits vs.\ Status Quo (Clinician)}
\quad
\textit{$\Delta$ = Counterfactual $-$ Generic}
}
\\
\addlinespace[1mm]

&
\textbf{P1: Optimal Baseline}
&
\textbf{+96.0 percentage points}

100\% vs.\ 4\%
\\

&
\textbf{P2: Multi-Risk Challenger}
&
\textbf{+84.0 percentage points}

90\% vs.\ 6\%
\\

&
\textbf{P3: Recovery Bottleneck}
&
\textbf{+84.0 percentage points}

90\% vs.\ 6\%
\\

&
\textbf{P4: Activity Bottleneck}
&
\textbf{+88.0 percentage points}

96\% vs.\ 8\%
\\

&
\textbf{P5: Balanced Maintainer}
&
\textbf{+68.9 percentage points}

84\% vs.\ 16\%
\\

\midrule


\multicolumn{3}{L{0.94\textwidth}}{
\footnotesize
$^*$ Rubric names marked with an asterisk denote 3-level ordinal measures for which the reported percentage corresponds to top-box achievement.
}
\\[1mm]

\multicolumn{3}{L{0.94\textwidth}}{
\footnotesize
For the side-by-side preference outcomes, the original analysis used $^{**}P<0.01$ and $^{***}P<0.001$ as significance thresholds. Exact $P$-values are reported directly here.
}
\\[1mm]

\multicolumn{3}{L{0.94\textwidth}}{
\footnotesize
CF, Counterfactual Coaching; Gen, Generic Coaching. Unless otherwise indicated, percentages within the Result column are reported as Counterfactual versus Generic.
}
\\

\end{longtable}

\clearpage

\begin{longtable}{
    L{0.24\textwidth}
    L{0.28\textwidth}
    L{0.40\textwidth}
}
\caption{\textbf{Individual Expert Rater Performance and Qualitative Feedback.} Detailed evaluation outcomes for individual lifestyle coaches and clinicians, accompanied by representative verbatim feedback highlighting expert perspectives on model performance, strengths, and limitations.}
\label{tab:supp_rater_breakdown}
\\

\toprule
\textbf{Expert Track \& Rater ID} &
\textbf{Counterfactual Win Rate / Key Rubric Highlights} &
\textbf{Representative Verbatim Qualitative Feedback (Unblinded Post-Analysis)} \\
\midrule
\endfirsthead

\multicolumn{3}{l}{
\textit{Supplementary Table~\ref{tab:supp_rater_breakdown} continued from previous page}
}
\\[1.5mm]

\toprule
\textbf{Expert Track \& Rater ID} &
\textbf{Counterfactual Win Rate / Key Rubric Highlights} &
\textbf{Representative Verbatim Qualitative Feedback (Unblinded Post-Analysis)} \\
\midrule
\endhead

\midrule
\multicolumn{3}{r}{\textit{Continued on next page}} \\
\endfoot

\bottomrule
\endlastfoot


\textbf{Lifestyle Coach 1}

Valid cases: $n=49$
&
\textbf{Best Overall:} 89.8\% (44/49)

\textbf{Adherence:} 98.0\% (48/49)

\textbf{Contrasts Benefits:} 98\% CF vs.\ 2\% Gen
&
\textit{``Although both plans are helpful, supportive, and use data and behavior change techniques, [the Counterfactual plan] provides much more detail and dives deeper into personalizing the plan and providing options for the user to choose, giving the user the autonomy to select the best option for them which will be stronger in long-term success.''}

(PIDs \texttt{186449}, \texttt{55807})
\\
\midrule


\textbf{Lifestyle Coach 2}

Valid cases: $n=49$
&
\textbf{Best Overall:} 53.1\% (26/49)

\textbf{Adherence:} 44.9\% (22/49)

\textbf{Trajectory Control:} 92\% CF vs.\ 67\% Gen
&
\textit{``[The Counterfactual plan] is by far more effective with the greatest likelihood to impact both short-term and long-term behavior. While both positively reinforce behavior at the outset, [the Generic plan] fails to contrast short- and long-term impacts nor does it explain the science-backed `why'. In P3 (PID \texttt{70621}), [Counterfactual] provides the best opportunity for user agency, whereas [Generic] misses sleep as the key factor for someone already at 10,000+ steps/day.''}
\\
\midrule


\textbf{Lifestyle Coach 3}

Valid cases: $n=49$
&
\textbf{Best Overall:} 98.0\% (48/49)

\textbf{Adherence:} 98.0\% (48/49)

\textbf{Compelling Reason:} 98\% CF vs.\ 4\% Gen
&
\textit{``I loved how thorough the coach was with detailing what would happen to the user by choosing the plan and following it to their long-term benefits... [The Counterfactual plan] would give me the motivation to start the plan I want to choose and know what benefits could come from doing so. [The Generic plan] does not do that at all.''}

\textbf{Constructive critique on PID \texttt{186449}:}

\textit{``Needs to clear up the phrasing of adding +1,000 steps per day so the user does not confuse a daily target with adding another 1,000 every single day.''}
\\
\midrule


\textbf{Lifestyle Coach 4}

Valid cases: $n=49$
&
\textbf{Best Overall:} 57.1\% (28/49)

\textbf{Adherence:} 53.1\% (26/49)

\textbf{Behavior Strategies:} 37\% CF vs.\ 39\% Gen
&
\textit{``In [the Generic plan], there is not an active contrast here for the user, and there could have been more tangible points to connect the dots on where their health is heading if they choose to adopt this plan. [The Counterfactual plan] connects the metrics directly to future outcomes and gives the user tangible options.''}
\\
\midrule


\textbf{Lifestyle Coach 5}

Valid cases: $n=49$
&
\textbf{Best Overall:} 59.2\% (29/49)

\textbf{Adherence:} 61.2\% (30/49)

\textbf{Trajectory Control:} 100\% CF vs.\ 98\% Gen
&
\textit{``[The Counterfactual plan] is more flexible with multiple pathways, while [the Generic plan] is more rigid and increasingly takes more time for the user to complete, which can be a barrier to success. In [the Generic plan], each metric is touched upon separately with nothing that addresses the bigger `why' for making these changes.''}
\\
\midrule


\textbf{Clinician 1}

Valid cases: $n=49$
&
\textbf{Safety Gate:} 100\% CF vs.\ 94\% Gen

\textbf{Contrasts Benefits:} 98\% CF vs.\ 2\% Gen

\textbf{Personalization:} 88\% CF vs.\ 82\% Gen
&
\textit{``[The Counterfactual plan] is much more thorough in explaining the physiological consequences of staying the same versus making changes, and gives options of paths to follow that are flexible and the user's choice.''}
\\
\midrule


\textbf{Clinician 2}

Valid cases: $n=49$
&
\textbf{Safety Gate:} 100\% CF vs.\ 100\% Gen

\textbf{Contrasts Benefits:} 98\% CF vs.\ 29\% Gen

\textbf{Personalization:} 73\% CF vs.\ 65\% Gen
&
\textit{``Overall [the Generic plan] is more vague; there are no suggestions of a path to choose and it does not give the user the numbers to motivate what the new habit will do for their health age. [The Counterfactual plan] clearly notes Future A and Future B and gives Options 1--3 as far as areas to choose focus.''}
\\
\midrule


\textbf{Clinician 3}

Valid cases: $n=49$
&
\textbf{Safety Gate:} 96\% CF vs.\ 96\% Gen

\textbf{Bio.\ Plausibility:} 88\% CF vs.\ 90\% Gen

\textbf{Personalization:} 73\% CF vs.\ 14\% Gen
&
\textit{``[The Generic plan] lacks contrast between active lifestyle adoption and current level of engagement. [The Counterfactual plan] understands where the patient starts, proposes gentle increments bounded by their history, and makes accurate claims across sleep, exercise, and steps; with perseverance and commitment, the improved metrics are achievable.''}
\\
\midrule


\textbf{Clinician 4}

Valid cases: $n=49$
&
\textbf{Safety Gate:} 100\% CF vs.\ 100\% Gen

\textbf{Bio.\ Plausibility:} 100\% CF vs.\ 100\% Gen

\textbf{Contrasts Benefits:} 98\% CF vs.\ 2\% Gen
&
\textit{``High clinical safety and physiological accuracy across both arms, with [the Counterfactual plan] providing superior quantification of expected RHR and Health Age improvements tailored to baseline bottlenecks.''}
\\
\midrule


\textbf{Clinician 5}

Valid cases: $n=49$
&
\textbf{Safety Gate:} 100\% CF vs.\ 98\% Gen

\textbf{Contrasts Benefits:} 94\% CF vs.\ 4\% Gen

\textbf{Bio.\ Plausibility:} 43\% CF vs.\ 20\% Gen
&
\textbf{Constructive clinical plausibility critique on Persona 2, PIDs \texttt{186449} and \texttt{56440}:}

\textit{``While the counterfactual options are motivating, dropping Health Age by 4--5 years in 12 weeks by only adding 30 minutes of sleep or modest steps is unlikely to happen that quickly in a severely deconditioned patient.''}
\\
\midrule


\multicolumn{3}{L{0.94\textwidth}}{
\footnotesize
CF, Counterfactual Coaching; Gen, Generic Coaching; SxS, side-by-side comparison; RHR, resting heart rate. Win rates and rubric highlights are reported at the individual expert-rater level.
}
\\[1mm]

\multicolumn{3}{L{0.94\textwidth}}{
\footnotesize
Qualitative comments are reproduced verbatim from the unblinded post-analysis feedback provided by lifestyle coaches and clinicians. Participant identifiers are shown only where they were included in the original qualitative feedback.
}
\\

\end{longtable}